\documentclass[fleqn,usenatbib]{mnras}

\usepackage{newtxtext,newtxmath}

\usepackage[T1]{fontenc}
\usepackage{ae,aecompl}

\usepackage{graphicx}	
\usepackage{amsmath}	
\usepackage{enumerate}
\usepackage{color}
\usepackage{graphicx}
\usepackage{multicol}
\usepackage{multirow}
\usepackage{savesym}
\usepackage{tipa}
\usepackage{csvsimple}
\usepackage{arydshln}
\usepackage{arydshln}
\usepackage{longtable}
\usepackage[utf8]{inputenc}

\title[{\it Chandra} survey of Virgo spirals]
  {A {\it Chandra} Virgo cluster survey of spiral galaxies. I. Introduction to the
survey and a new ULX sample}

\author[R. Soria et al.]
{Roberto Soria$^{1,2}$ \thanks{Email: rsoria@nao.cas.cn (RS)}, 
Mari Kolehmainen$^{3}$,
Alister W.~Graham$^{4}$,
Douglas A.~Swartz$^{5}$,\newauthor
Mihoko Yukita$^{6,7}$,
Christian Motch$^{3}$,
Thomas H.~Jarrett$^{8}$, James C.~A.~Miller-Jones$^{9}$, 
\newauthor
Richard M.~Plotkin$^{10}$, 
Thomas J.~Maccarone$^{11}$, 
Laura Ferrarese$^{12}$,
Alexander Guest$^{13}$,
\newauthor
Ariane Lan\c{c}on$^{3}$ 
\\
$^{1}$College of Astronomy and Space Sciences, University of the Chinese Academy of Sciences, Beijing 100049, China\\
$^{2}$Sydney Institute for Astronomy, School of Physics A28, The University of Sydney, Sydney, NSW 2006, Australia\\
$^{3}$Universit\'e de Strasbourg, CNRS, Observatoire Astronomique, CNRS, UMR 7550,F-67000, Strasbourg, France\\
$^{4}$Centre for Astrophysics and Supercomputing, Swinburne University of Technology, Hawthorn, VIC 3122, Australia\\
$^{5}$Astrophysics Office, NASA Marshall Space Flight Center, ST12, Huntsville, AL 35812, USA\\
$^{6}$Code 662, NASA Goddard Space Flight Center, Greenbelt, MD 20771, USA\\
$^{7}$Henry A.~Rowland Department of Physics and Astronomy, Johns Hopkins University, Baltimore, MD 21218, USA\\
$^{8}$Department of Astronomy, University of Cape Town, Private Bag X3, Rondebosch, 7701, South Africa\\
$^{9}$International Centre for Radio Astronomy Research, Curtin University, GPO Box U1987, Perth, WA 6845, Australia\\
$^{10}$Department of Physics, University of Nevada, Reno, NV 89557, USA\\
$^{11}$Department of Physics, Box 41051, Science Building, Texas Tech University, Lubbock, TX 79409-1051, USA\\
$^{12}$Herzberg Astronomy and Astrophysics Research Centre, National Research Council of Canada, Victoria, BC V9E 2E7, Canada\\
$^{13}$Department of Physics and Astronomy, University College London, Gower Street, London, WC1E 6BT, UK
\\
}

\date{Accepted 2022 January 17. Received 2022 January 16; in original form 2021 July 28}

\pubyear{2021}

\begin{document}
\label{firstpage}
\pagerange{\pageref{firstpage}--\pageref{lastpage}}
\maketitle

\begin{abstract}
We present an analysis of the ultraluminous X-ray source (ULX) population in 75 Virgo cluster late-type galaxies, including all those with a star formation rate $\gtrsim$1 $M_{\odot}$ yr$^{-1}$ and a representative sample of the less star-forming ones. This study is based on 110 observations obtained over 20 years with the {\it Chandra X-ray Observatory} Advanced Camera for Imaging Spectroscopy. As part of a Large {\it Chandra} Program, new observations were obtained for 52 of these 75 galaxies. The data are complete to a sensitivity of $\approx$10$^{39}$ erg s$^{-1}$, with a typical detection limit of $\approx$3 $\times 10^{38}$ erg s$^{-1}$ for the majority of the sources. The catalogue contains about 80 ULXs (0.3–-10 keV luminosity $>$10$^{39}$ erg s$^{-1}$), and provides their location, observed flux, de-absorbed luminosity, and (for the 25 most luminous ones) simple X-ray spectral properties. We discuss the ULX luminosity function in relation to the mass and star formation rate of the sample galaxies. We show that recent models of low-mass plus high-mass X-ray binary populations (scaling with stellar mass and star formation rate, respectively) are mostly consistent with our observational results. We tentatively identify the most luminous X-ray source in the sample (a source in IC\,3322A with $L_{\rm X} \approx 6 \times 10^{40}$ erg s$^{-1}$) as a recent supernova or its young remnant. The properties of the sample galaxies (morphologies, stellar masses, star formation rates, total X-ray luminosities from their point-source population) are also summarised.

\end{abstract}


\section{Introduction}
The population properties of compact X-ray sources are a tracer of star formation rate (SFR) and stellar mass ($M_{\ast}$) \citep{grimm03,gilfanov04,mineo12a,fragos13a,lehmer19}. On the one hand, accreting compact objects probe the last stages of stellar evolution and collapse. On the other hand, binary systems with compact objects include the progenitors of gravitational wave mergers. Thus, studying the number and distribution of X-ray binaries provides empirical constraints to population synthesis models and theoretical predictions of binary merger rates \citep{belczynski16,kruckow18,spera19}. Reliable scaling relations between galaxy properties and the total luminosity of the X-ray binary population are also essential to estimate the relative contribution of off-nuclear compact objects (compared with massive stars and quasars) to cosmic reionization and to the heating of the intergalactic and interstellar medium at high redshift \citep{mirabel11,fragos13b,kaaret14,artale15,lehmer16}.


More specifically, here we are interested in the high-luminosity end of the X-ray binary population, near and above the Eddington limit of typical stellar-mass accretors ($L_{\rm X} > 10^{39}$ erg s$^{-1}$). This class of ultraluminous X-ray sources (ULXs: see \citealt{kaaret17} for a review) is populated both by confirmed neutron stars and candidate stellar-mass black holes (BHs), but the relative contribution of the two types of accreting compact objects remains uncertain and widely debated \citep{bachetti14,walton18,wiktorowicz19,mushtukov19}. ULXs are found both in young and in old stellar environments, although the younger systems are more common, especially at highest luminosities ($\gtrsim 5 \times 10^{39}$ erg s$^{-1}$) \citep{swartz04,maccarone07,zhang12,lehmer19}. In young-population ULXs, the donor stars are usually consistent with massive blue or red supergiants \citep{lau19,lopez20}, but other types of donors with strong winds are also possible: for example, a Wolf-Rayet star, as seen in a ULX inside the  Circinus galaxy \citep{qiu19}, or a main-sequence B-type star, as in M82 X-2 \citep{fragos15}. Mass transfer from massive donors may occur via classical wind, focused wind (``wind Roche-lobe-overflow''), or Roche lobe overflow (if the mass ratio between donor and accretor is below the threshold for common-envelope formation); the relative contribution of the three processes is also still actively debated \citep{pavlovskii17,artale19,elmellah19,quast19,wiktorowicz21}. 
Instead, in old-population ULXs, the donor star may be a Roche-lobe-filling low-mass or intermediate-mass star; such systems would be analogous to transient low-mass X-ray binaries. If the Roche lobe of the compact object, and therefore its accretion disk, is very large, the thermal-viscous disk instability is predicted to cause long outburst phases and super-Eddington peak luminosities \citep{lasota15,hameury20}. Alternatively, even intermediate- and low-mass stars can go through evolutionary phases when they can transfer $>10^{-6} M_{\odot}$ yr$^{-1}$ onto their compact companion for a sustained period of time of $\sim$10$^5$--$10^6$ yr \citep{wiktorowicz17,misra20}, for example when they cross the Hertzsprung gap or as they ascend the red giant or asymptotic giant branch. Moreover, in at least one case, for an old-population ULX in the Virgo giant elliptical M49, the donor star is a white dwarf  \citep{maccarone07,maccarone10,steele14}.

Most of the empirical constraints to theoretical models of ULX formation come from X-ray surveys of accreting compact objects in nearby galaxies, followed by optical/IR studies of their stellar counterparts and environments. The {\it Chandra X-ray Observatory} is the best instrument available for such surveys, thanks to its still unsurpassed sub-arcsec spatial resolution and very low background noise. Already a few years after its launch, {\it Chandra}-based surveys showed the difference between the ULX luminosity function (LF) in spiral and elliptical galaxies. The early-type-galaxy LF is an indicator of $M_{\ast}$ \citep{kim04,gilfanov04,zhang12}, while the late-type-galaxy LF is mainly  proportional to SFR \citep{grimm03,mineo12a}; the young-population LF is a power-law with an exponential cut-off at $L_{\rm X} \approx 2 \times 10^{40}$ erg s$^{-1}$ \citep{swartz11,mineo12a}. More generally, the LF of any given galaxy is a mix of the two components \citep{lehmer19} and depends on the specific SFR (sSFR $\equiv$ SFR/$M_{\ast}$). There is an additional smaller dependence of the high mass X-ray binary (HMXB) population on the metal abundance of the host galaxy, in the sense that metal-poor galaxies contain more luminous HMXBs \citep{prestwich13,brorby14,kovlakas20,lehmer21}. 

There are now a few hundred nearby galaxies observed by {\it Chandra} deep enough to reach a completeness limit of about $10^{39}$ erg s$^{-1}$ or below, which can be used for ULX population studies (for example, a sample of about 300 galaxies within 40 Mpc was used by \citealt{kovlakas20} for their study of the ULX population). Surveys of nuclear X-ray sources suspected of being intermediate-mass BHs have also been conducted with archival {\it Chandra} data ({\it e.g.}, \citealt{chilingarian18}). A caveat is that archival {\it Chandra} targets are biased in favour of ``interesting'' galaxies, sometimes with higher-than-average X-ray luminosities or already known to contain particularly luminous ULXs discovered by earlier X-ray telescopes.
This may affect the inferred scalings of the LF with sSFR. For this work, we chose instead a physically selected sample of galaxies, representative of the spiral population in the Virgo Cluster. The methodology of our sample selection is described in Section~\ref{sub_select}.

\begin{figure}
\hspace{-0.5cm}
\includegraphics[height=0.49\textwidth, angle=270]{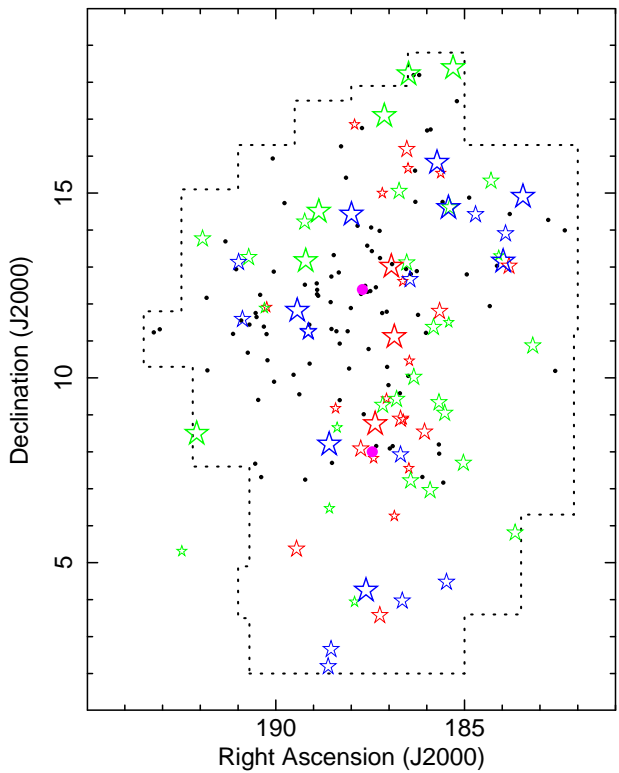}
 \caption{Sky map of the Virgo galaxies. Stars mark the positions of the 75 spirals in our study. Black dots mark the position of the 100 AMUSE-Virgo early-type galaxies; the two magenta dots represent M\,87 (further to the north) and M\,49 (further to the south). Small-sized stars represent spirals in our sample with low stellar masses ($\log (M_\ast/M_\odot) < 9.8$); intermediate-sized stars represent intermediate masses ($9.8 \leq \log (M_{\ast}/M_{\odot}) < 10.4$); and large-sized stars represent the most massive spirals ($\log (M_{\ast}/M_{\odot}) \geq 10.4$). Red-coloured stars denote galaxies with low SFR (SFR $<0.5 M_{\odot}$ yr$^{-1}$, according to the WISE 12-$\mu$m proxy); green-coloured stars are for intermediate rates ($0.5 M_{\odot}$ yr$^{-1} \leq$ SFR $< 1.5 M_{\odot}$ yr$^{-1}$); blue-coloured stars are for high rates (SFR $\geq 1.5 M_{\odot}$ yr$^{-1}$). (The choice of the three bands in stellar mass and SFR is arbitrary, for illustration purposes.) A dotted line marks the boundaries of the Virgo Cluster Survey \citep{binggeli84}. The only spiral in our sample that is outside the VCC region is included in the Extended VCC \citep{kim14} and is a probable member based on its redshift-independent distance and recession speed.} 
  \label{cluster_map}
\end{figure}

The Virgo Cluster is an interesting environment for a point-source X-ray population study in galaxies with a range of Hubble types and star-forming activity. It is the richest and biggest galaxy cluster in the local universe, with a total mass (gas, stars and dark matter) $M_{\rm tot} \approx 6$--8 $\times 10^{14} M_{\odot}$ \citep{kashibadze20,kashibadze18,shaya17,planck16}. 
It is close enough ($d = 16.1 \pm 1.0$ Mpc for its central galaxy, M\,87, in the weighted-average estimate of \citealt{degrijs20}) that {\it Chandra} imaging observations can reach $L_{\rm X} \approx$ a few $10^{38}$ erg s$^{-1}$ in $\lesssim$10 ks and are not affected by source confusion. Virgo is also just close enough to permit at least a detection and possibly a rough colour identification of the optical counterparts; at a distance modulus of 31 mag, the typical optical brightness of young-population ULXs ($M_V \sim -7$ mag to $-4$ mag: \citealt{ambrosi18,gladstone13,tao11}) corresponds to $V \sim 24$--27 mag. Moreover, Virgo already has a huge wealth of multiband datasets, which enable us to determine the SFR and $M_{\ast}$ of many of its galaxies.

A sample of 100 early-type Virgo galaxies (including E, dE and S0 morphological types) was observed by {\it Chandra} as part of the ``Active galactic nucleus MUltiwavelength Survey of Early-Type Galaxies'' (AMUSE) Virgo study \citep{gallo08,gallo10}. The sample was selected to coincide with the {\it Hubble Space Telescope} ({\it HST}) Virgo Cluster Survey \citep{cote04}. In turns, the {\it HST} sample was selected from early-type galaxies with blue brightness $B_T < 16$ mag, a published central velocity dispersion, and no peculiar properties such as mergers or close interactions. The total stellar mass ($M_{\ast}$) of the AMUSE-Virgo sample was $M_{\ast} \approx 6 \times 10^{12} M_{\odot}$ (derived by \citealt{gallo08,gallo10}, from the $g_0$ and $z_0$ luminosities, according to the relation of \citealt{bell03}). The {\it Chandra} coverage was designed to be complete down to $\approx 4 \times 10^{38}$ erg s$^{-1}$. The AMUSE X-ray survey provided an exceptional view of nuclear BH activity and Eddington ratio in ``normal'' early-type galaxies \citep{gallo10,miller12,gra-sor19}, as well as a sample of bright off-nuclear sources. 

From the AMUSE data, \cite{plotkin14} found 55 sources with an apparent 0.3--10 keV luminosity $L_{\rm X} > 10^{39}$ erg s$^{-1}$, in the Virgo ellipticals. However, only 13 of them had an apparent (projected) luminosity higher than $\approx 2 \times 10^{39}$ erg s$^{-1}$, and those brighter sources were statistically consistent with background AGN projected inside the $D_{25}$ area\footnote{The $D_{25}$ region of a galaxy is defined as the 25th magnitude B-band isophote.} of the Virgo galaxies. \cite{plotkin14} concluded that in the old stellar population, there is $\approx$ 1 ULX per $M_{\ast} \approx 1.6 \times 10^{11} M_{\odot}$. This is consistent with the expected LF of low mass X-ray binaries (LMXBs) in early-type galaxies \citep{zhang12}.

To have a more complete understanding of the nuclear and off-nuclear X-ray sources in Virgo, we must integrate the existing study of its elliptical galaxies with a new study of its spiral galaxies. Virgo is a spiral-rich cluster, a signature of an early evolutionary stage (clusters such as Coma, by contrast, are more evolved and spiral-poor). However, many of the Virgo Cluster's spirals have lost HI gas via ram pressure stripping and tidal stripping, and as a result, the sSFR of Virgo spirals tends to be lower than for field spirals of similar Jeans-Lundmark-Hubble type and mass (see \citealt{boselli06} for a detailed review and discussion of such issues, and \citealt{awgraham19} for a comprehensive discussion of morphological types).
For the off-nuclear X-ray sources, Virgo spirals contain two populations of stellar-mass objects. First, a sample of LMXBs in spirals; the number and luminosity of such sources as a function of stellar mass may not be the same as for the LMXBs in ellipticals. Second, a sample of young high mass X-ray binaries (HMXBs), in the galaxies with a higher SFR. Third, for centrally-located, {\it i.e.}, nuclear, BHs, a {\it Chandra} study can reveal, for example, the difference in the Eddington ratio distribution between galaxies dominated by hot gas (early-type) or cold gas (late-type). It can also reveal the presence of candidate X-ray-bright intermediate-mass BHs in the nuclei of spiral galaxies with a small bulge or no bulge (as predicted by galaxy/BH scaling relations), which can then be the subject of subsequent multiband investigations \citep{chilingarian18,mezcua18}. Furthermore, the unsurpassed sub-arcsec astrometric accuracy and sharp point-spread function of {\it Chandra} enables a search for slightly off-centre, ``wandering'' nuclear BHs \citep{khan20,ricarte21}. In this paper, we will mostly focus on off-nuclear sources, and we will leave the analysis of the nuclear BHs and candidate intermediate-mass BHs to separate work.

By contrast with the early-type galaxies, the archival {\it Chandra} coverage of late-type Virgo galaxies was, until recently, surprisingly scant. Less than 30 Virgo spirals had been observed before 2017, an insufficient number for population studies. Thus, we planned a larger survey and successfully applied for 550 ks of {\it Chandra} time (Cycle 18 Large Program, PI R. Soria). In the next Section, we describe the population properties of the {\it Chandra} sample. In Section 3, we outline our X-ray data analysis. In Section 4, we present the results for the off-nuclear source population, and place it in the context of the properties of their host galaxies. A census of the nuclear sources, a study of the diffuse X-ray emission in the most star-forming galaxies, and an investigation of the multiband counterparts will be presented in further work.

\begin{figure}
\centering
\includegraphics[height=0.475\textwidth, angle=270]{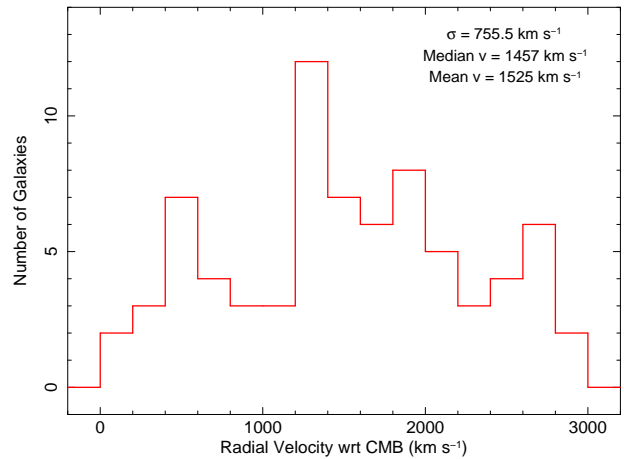}
 \caption{Distribution of radial velocities with respect to the cosmic microwave background, for the galaxies in our sample. The mean and median velocities, and the Gaussian standard deviation, are also labelled in the plot.} 
  \label{v3k_histo}
\end{figure}




\section{A Chandra sample of Virgo spirals}

\subsection{Selection criteria}\label{sub_select}
The selection of targets for our study resulted from a necessary compromise between various ideal requirements: a large number of galaxies, particularly with high SFR (because that is the other extreme of the range of galaxy activity compared to ellipticals; that is where we expect to find the most numerous and luminous X-ray binaries), but also a good sampling of all morphological types from early-type (large bulge) to late-type (small or no bulge) spirals and with a sufficiently large range of specific SFR (so as to differentiate between galaxies dominated by LMXBs, and those dominated by HMXBs). Each target should be observed deep enough to be complete at least down to $10^{39}$ erg s$^{-1}$ (in most cases, we required a detection limit of $\approx$3 $\times 10^{38}$ erg s$^{-1}$). Finally, the total exposure time has to be kept within plausible limits. 

There are about 70 Virgo spirals with SFR $\gtrsim$0.3 $M_{\odot}$ yr$^{-1}$, but the precise list depends on the various alternative proxies for the SFR (H$\alpha$ luminosity, or radio continuum, or far-IR, or far-UV plus 24-$\mu$m luminosity) as well as the adopted distance to the galaxies within Virgo. Moreover, 
some of the SFR studies we use for the analysis in this paper were not available at the time of our sample selection. With these caveats, we selected a representative sample\footnote{We do not claim that the sample is complete within a certain threshold; simply that it is well representative of most spirals with high SFR. All the $\approx$20 spirals with SFR $\gtrsim$1 $M_{\odot}$ yr$^{-1}$ are included. Among those with a SFR between $\approx$0.4--1$M_{\odot}$ yr$^{-1}$, we are aware of the following six Virgo galaxies missing from the sample: NGC\,4294, NGC\,4634, NGC\,4409, IC\,3476, NGC\,4630, and NGC\,4207. A case can be made also for NGC\,4383 (with an uncertain classification between S0 and Sa peculiar), and NGC\,4651 (located at the northern edge of Virgo, probably a member).} of about 60 galaxies with SFR $\gtrsim$0.3 $M_{\odot}$ yr$^{-1}$, plus about 15 galaxies with a lower SFR, but which were either already in the {\it Chandra} archives or were in the same field of view of another target with a higher SFR. In total, we compiled a list of 75 spirals: a sample large enough to be representative of the population of actively star-forming galaxies in Virgo, and to provide a sizeable population of ULXs. In particular, we included all the $\approx$20 spirals with SFR $\gtrsim$1 $M_{\odot}$ yr$^{-1}$. Of the 75 galaxies in the sample, 46 were observed for the first time by {\it Chandra} through our Large Program. For another six galaxies, new observations were combined with a previous snapshot observation in order to reach the required detection limit, though any detected sources were subsequently analyzed separately for each observation. Finally, the remaining 23 galaxies already had sufficiently deep archival observations.
In the rest of this Section we summarize the main properties of our sample. We will then describe our data analysis and main results, in particular for the ULX census. A more detailed analysis of the most interesting or peculiar individual sources will be presented in follow-up papers.

\subsection{Cluster membership}
Seventy-four out of our 75 galaxies (Figure 1) are included in the Virgo Cluster Catalog (VCC) of \cite{binggeli84}, which is often used as the standard reference for a galaxy census in this cluster. The only galaxy in our sample that falls outside the VCC footprint is NGC\,4713. However, it is included in the Extended Virgo Cluster Catalog \citep{kim14}, and it is a likely cluster member based on its recession speed and redshift-independent distance.

A VCC classification does not of course guarantee that a galaxy is physically associated with Virgo. This is especially true for spirals, which are more spread out around the edges of the cluster, while ellipticals are more concentrated in the central region, around M\,87 and M\,49.  As a test that all 75 galaxies are at least plausible cluster members, we inspected their radial velocity distribution with respect to the cosmic background, using the values of $v_{\rm {3K}}$ reported in HyperLEDA\footnote{http://leda.univ-lyon1.fr}. We obtain a broad, structured, tri-modal distribution (Figure 2), which spans $\approx$3000 km s$^{-1}$,  consistent with the findings of \cite{binggeli87} from the whole VCC. A few additional galaxies with {\it Chandra} archival data are physically located inside the VCC footprint (thus, they do have a VCC number) but their radial velocity is $>$3000 km s$^{-1}$, which locates them most likely behind the cluster: we did not include those outliers in our sample of 75 candidate members.

\begin{figure}
\hspace{-0.5cm}
\includegraphics[height=0.49\textwidth, angle=270]{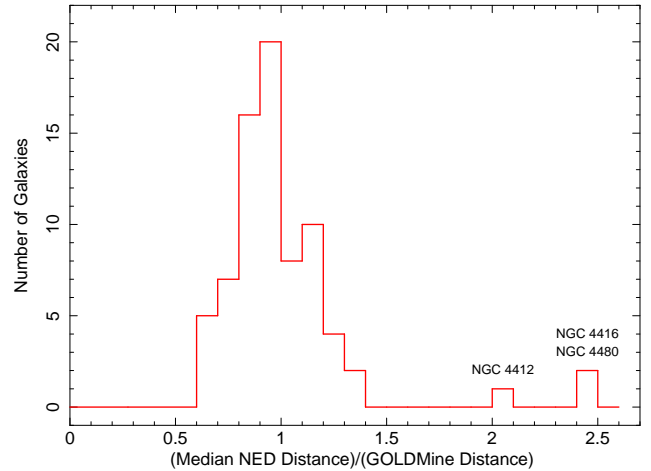}
 \caption{Ratio between the median redshift-independent distance (from NED) and the conventional Virgo sub-structure distance (17, 23 or 32 Mpc), for the 75 galaxies in our sample. Three galaxies with very discrepant distance measurements are labelled in the plot and discussed in Section 2.3.}
  \label{dist_ratio}
\end{figure}

\begin{figure}
\hspace{-0.5cm}
\includegraphics[height=0.49\textwidth, angle=270]{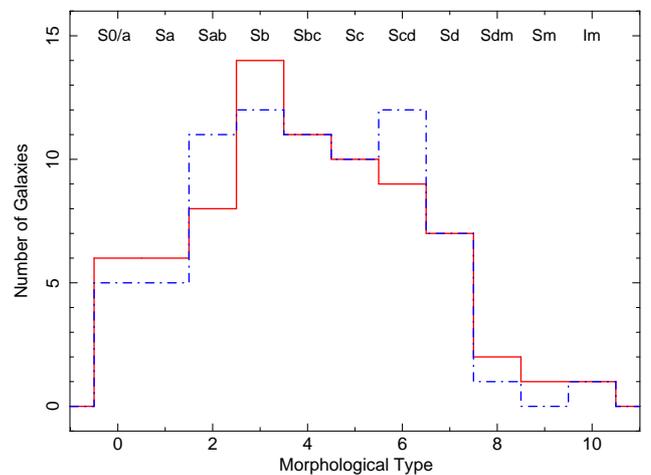}
 \caption{Morphological type distribution of the galaxies in our sample, expressed with the ``Hubble stage'' index $T$ (horizontal axis) \citep{devaucouleurs59,devaucouleurs91} and with the corresponding Hubble-Sandage-de Vaucouleurs classification (label over each histogram bin). The solid red histogram is for the galaxy morphologies reported in NED; the dash-dotted blue histogram is for the morphological classification of HyperLEDA.}
  \label{type_histo}
\end{figure}

\begin{figure}
\hspace{-0.5cm}
\includegraphics[height=0.49\textwidth, angle=270]{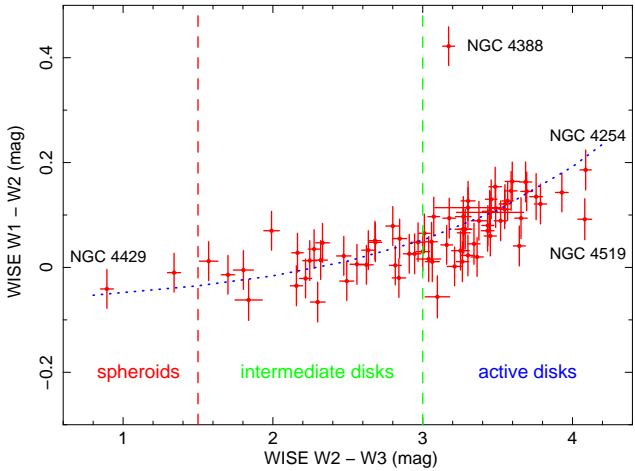}
 \caption{{\it {WISE}} classification of the galaxies in our sample, following \citet{jarrett19}. All the galaxies follow the ``star formation sequence'' (dotted blue line) of normal galaxies, except for NGC\,4388 (a Seyfert 2 galaxy), in which the {\it {W1}} colour is contaminated by the active nucleus. Spiral galaxies on the left of the sequence are more bulge-dominated, while those on the right have a younger and more active disk. The most bulge-dominated galaxy in our sample is NGC\,4429 (classified as S0/a in HyperLEDA), while the most disky are NGC\,4519 and NGC\,4254 = M\,99. }
  \label{wise_colors}
\end{figure}

\begin{figure}
\hspace{-0.5cm}
\includegraphics[height=0.49\textwidth, angle=270]{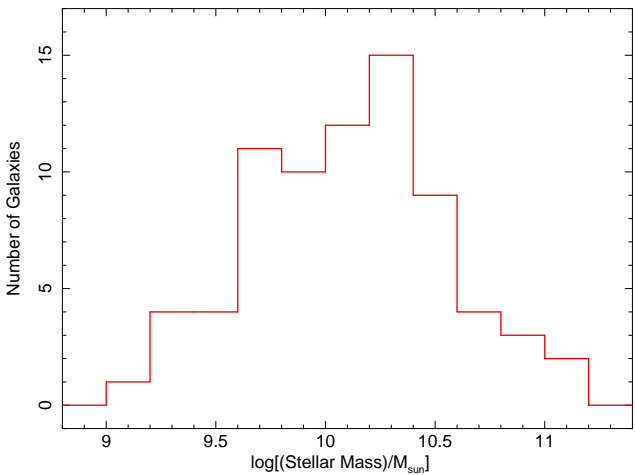}
 \caption{Stellar mass distribution, from the mid-infrared luminosity ({\it WISE} colours), assuming the median redshift-independent distances from NED (Section 2.3).}
  \label{mass_median}
\end{figure}





\begin{table*}
\caption{Main properties of the sample of 75 Virgo spirals}
\vspace{-0.2cm}
\scriptsize{
\begin{center}
\begin{tabular}{lccccccccccc}  
\hline \hline\\[-5pt]    
Galaxy  & Other Names &  Type & R.A.(J2000) & Dec.(J2000) & $v_{\rm 3K}$
& $d_1$ & $d_2$ & $\log (M_{\ast}/M_{\odot})$  
& SFR(12$\mu$m) & SFR(22$\mu$m) & SFR(B15) \\
   &   & &  &  & (km s$^{-1}$)  & (Mpc) & (Mpc) &  
   & ($M_{\odot}$ yr$^{-1}$) & ($M_{\odot}$ yr$^{-1}$) 
   & ($M_{\odot}$ yr$^{-1}$)\\
   \multicolumn{1}{c}{(1)} & (2) & (3) & (4) & (5) & (6) & (7) & (8) & (9) & (10) & (11) & (12)\\
\hline  \\[-5pt]
NGC\,4178 &  VCC\,0066 & SB(rs)dm  & 12 12 46.45 &  $+$10 51 57.5 & 714$\pm$2 &
17 & 13.20$\pm$3.00   &    9.84$\pm$0.09  &   0.59$\pm$0.19 &  
0.50$\pm$0.19  &   0.51  \\ [-1pt]
NGC\,4192 & M\,98, VCC\,0092 &  SAB(s)ab &  12 13 48.29 & $+$14 54 01.2  
& 212$\pm$12  & 17 &    13.55$\pm$2.93  &    10.63$\pm$0.09 &  
2.03$\pm$0.66  &    1.36$\pm$0.51   &   1.16 \\[-1pt]
NGC\,4197  &  VCC\,0120 &  Sd & 12 14 38.55 & $+$05 48 20.6    &         2427$\pm$10 &           32 &       26.40$\pm$3.92 &       9.92$\pm$0.10 &      1.00$\pm$0.33 &
     0.99$\pm$0.37 &     NA\\[-1pt]
NGC\,4206  & VCC\,0145  &  SA(s)bc & 12 15 16.81 & $+$13 01 26.3    &         1036$\pm$2 &           17 &       18.70$\pm$3.40 &       10.02$\pm$0.10 &      0.42$\pm$0.15 &  0.28$\pm$0.11 &      0.28\\[-1pt]
NGC\,4212  & NGC\,4208, VCC\,0157  & SAc  & 12 15 39.36 & $+$13 54 05.4   &          246$\pm$2 &           17 &       17.05$\pm$2.62 &   10.09$\pm$0.10 &       1.99$\pm$0.69 & 1.55$\pm$0.61 &       1.09\\[-1pt]
NGC\,4216  & VCC\,0167  &  SAB(s)b &  12 15 54.44 &  $+$13 08 57.8    &          468$\pm$3 &           17 &       15.50$\pm$2.08 &       11.00$\pm$0.09 &       1.66$\pm$0.56 & 0.93$\pm$0.36 &      0.66\\[-1pt]
NGC\,4222  & VCC\,0187  & Sd  & 12 16 22.52  & $+$13 18 25.4     &          563$\pm$2 &           17 &       21.30$\pm$5.16 &       9.61$\pm$0.10   &      0.49$\pm$0.18 &  0.39$\pm$0.16 &      0.26\\[-1pt]
NGC\,4237  & VCC\,0226  & SAB(rs)bc  &  12 17 11.42 & $+$15 19 26.3   &         1192$\pm$1 &           17 &       20.40$\pm$5.19 &       10.25$\pm$0.10    &       1.30$\pm$0.47 & 0.93$\pm$0.38 &      0.48\\[-1pt]
NGC\,4254  & M\,99, VCC\,0307  & SA(s)c  &  12 18 49.60 & $+$14 24 59.4   &         2737$\pm$8 &           17 &       14.40$\pm$2.05 &       10.29$\pm$0.10 &       7.37$\pm$2.45 & 5.51$\pm$2.09 &       5.17\\[-1pt]
NGC\,4276  & VCC\,0393  &  Sc &  12 20 07.48 & $+$07 41 30.7   &         2959$\pm$1 &           23 &       27.90$\pm \ast$ &       9.81$\pm$0.10 &      0.54$\pm$0.20 & 0.35$\pm$0.15 &      0.33\\[-1pt]
NGC\,4293  & VCC\,0460  &  (R)SB0/a(s) &  12 21 12.89 &  $+$18 22 56.6   &         1217$\pm$3 &           17 &       15.90$\pm$3.13 &       10.45$\pm$0.09 &      0.75$\pm$0.26 & 1.21$\pm$0.47 &  NA \\[-1pt]
NGC\,4298  & VCC\,0483  &  SA(rs)c &  12 21 32.76 & $+$14 36 22.2   &         1456$\pm$4 &           17 &       15.80$\pm$2.54 &       10.05$\pm$0.09 &       1.39$\pm$0.47 & 0.92$\pm$0.36 &      0.81\\[-1pt]
NGC\,4299  & VCC\,0491  & SAB(s)dm  &  12 21 40.55 & $+$11 29 59.7    &          576$\pm$2 &           17 &       21.80$\pm$2.96 &       9.51$\pm$0.10 &      0.65$\pm$0.27 & 0.86$\pm$0.36 &   NA\\[-1pt]
NGC\,4302  & VCC\,0497  & Sc  & 12 21 42.48 & $+$14 35 53.9    &         1457$\pm$16 &           17 &       17.40$\pm$8.60 &       10.40$\pm$0.09 &       1.58$\pm$0.55 & 0.96$\pm$0.38 &      0.86\\[-1pt]
NGC\,4303  & M\,61, VCC\,0508  & SAB(rs)bc & 12 21 54.90 & $+$04 28 25.1     &         1912$\pm$2 &           17 &       12.25$\pm$7.32 &       10.30$\pm$0.10   &       5.14$\pm$1.78 & 3.95$\pm$1.55 &       3.84\\[-1pt]
NGC\,4307  & VCC\,0524  & Sb  & 12 22 05.68 & $+$09 02 37.1    &         1397$\pm$4 &           23 &       21.95$\pm$3.39 &       10.38$\pm$0.09 &      0.64$\pm$0.22 & 0.48$\pm$0.19 &      0.22\\[-1pt]
NGC\,4312  & VCC\,0559  &  SA(rs)ab &  12 22 31.36  &  $+$15 32 16.5   &          475$\pm$3 &           17 &       10.90$\pm$0.96 &       9.68$\pm$0.09 &      0.29$\pm$0.10 & 0.18$\pm$0.07 &      0.13\\[-1pt]
NGC\,4313  & VCC\,0570  & SA(rs)ab  &  12 22 38.54 & $+$11 48 03.4    &         1767$\pm$3 &           17 &       14.15$\pm$2.92 &       10.01$\pm$0.09 &      0.33$\pm$0.11 & 0.20$\pm$0.08 &      0.11\\[-1pt]
NGC\,4316  & VCC\,0576  & Scd  & 12 22 42.24 & $+$09 19 56.9   &         1589$\pm$1 &           23 &       26.30$\pm$4.19 &       10.10$\pm$0.09 &       1.22$\pm$0.44 & 0.80$\pm$0.32 &      0.55\\[-1pt]
NGC\,4321  & M\,100, VCC\,0596  & SAB(s)bc  & 12 22 54.83 & $+$15 49 18.5    &         1898$\pm$2 &           17 &       16.10$\pm$1.57 &   10.71$\pm$0.10 &       7.45$\pm$2.54 & 6.05$\pm$2.35 &       5.46\\[-1pt]
NGC\,4330  & VCC\,0630  & Scd  & 12 23 17.25 & $+$11 22 04.7  &         1938$\pm$37 &           17 &       19.30$\pm$1.56 &       9.85$\pm$0.10 &      0.51$\pm$0.18 & 0.34$\pm$0.14 &      0.34\\[-1pt]
NGC\,4343  &  VCC\,0656 & SA(rs)b  &  12 23 38.70 & $+$06 57 14.7   &         1339$\pm$5 &           23 &       26.80$\pm$4.02 &       10.33$\pm$0.09 &       1.02$\pm$0.37 & 0.69$\pm$0.28 &      0.40\\[-1pt]
NGC\,4356  &  VCC\,0713 &  Scd & 12 24 14.53 & $+$08 32 09.1   &         1432$\pm$6 &           23 &       22.70$\pm$6.06 &       10.21$\pm$0.10  &      0.30$\pm$0.11 & 0.36$\pm$0.14 &      0.18\\[-1pt]
NGC\,4380  & VCC\,0792  & SA(rs)b  & 12 25 22.17 & $+$10 01 00.5  &         1287$\pm$4 &           23 &       19.30$\pm$3.56 &       10.41$\pm$0.10   &      0.63$\pm$0.21 & 0.35$\pm$0.13 &      0.27\\[-1pt]
IC\,3322A  & VCC\,0827  & SB(s)cd & 12 25 42.56 & $+$07 13 00.0   &          1335$\pm$3 &           23 &       25.30$\pm$4.12 &       9.85$\pm$0.10 &       1.10$\pm$0.39 & 
     0.77$\pm$0.31 &      0.73\\[-1pt]
NGC\,4388  & VCC\,0836  & SA(s)b  & 12 25 46.75 & $+$12 39 43.5   &         2845$\pm$3 &           17 &       18.10$\pm$4.96 &       10.10$\pm$0.09 &       3.01$\pm$1.06 & 5.11$\pm$2.03 &       3.01\\[-1pt]
NGC\,4390  & VCC\,0849  &  SAB(s)c &  12 25 50.67 & $+$10 27 32.6   &         1433$\pm$3 &           23 &       23.90$\pm$7.81 &       9.55$\pm$0.10 &      0.42$\pm$0.15 & 0.31$\pm$0.12 &      0.37\\[-1pt]
IC\,3322  & VCC\,0851  &  SAB(s)cd  & 12 25 54.10 & $+$07 33 17.2   &       1526$\pm$5 &           23 &       20.15$\pm$2.58 &       9.36$\pm$0.10 &      0.33$\pm$0.11 & 
     0.19$\pm$0.07 &      0.15\\[-1pt]
NGC\,4394  &  VCC\,0857 &  (R)SB(r)b & 12 25 55.53 & $+$18 12 50.6  &         1232$\pm$4 &           17 &       16.80$\pm$3.13 &       10.52$\pm$0.10  &      0.54$\pm$0.19 & 0.33$\pm$0.13 &       NA \\[-1pt]
NGC\,4396  & VCC\,0865  & SAd  & 12 25 58.82 & $+$15 40 17.3   &          208$\pm$6 &           17 &       13.00$\pm$1.97 &       9.12$\pm$0.10 &      0.25$\pm$0.09 & 0.18$\pm$0.07 &      0.24\\[-1pt]
NGC\,4402  & VCC\,0873  & Sb  &  12 26 07.65 & $+$13 06 48.0   &          563$\pm$2 &           17 &       14.50$\pm$3.83 &       9.87$\pm$0.09 &       1.33$\pm$0.44 & 0.96$\pm$0.36 &      0.65\\[-1pt]
NGC\,4405  & VCC\,0874  &  SA0/a & 12 26 07.15  & $+$16 10 51.6   &         2052$\pm$5 &           17 &       17.85$\pm$3.32 &       9.86$\pm$0.10   &      0.44$\pm$0.16 & 0.30$\pm$0.12 &       NA \\[-1pt]
NGC\,4411A  & VCC\,0905  & SB(rs)c  & 12 26 30.10  & $+$08 52 20.0   &         1616$\pm$2 &           17 &       16.80$\pm$3.15 &       9.27$\pm$0.11 &      0.14$\pm$0.05 &  0.08$\pm$0.04 &   NA \\[-1pt]
NGC\,4407  &  NGC\,4413, VCC\,0912 & (R')SB(rs)ab  & 12 26 32.25 & $+$12 36 39.6  &          426$\pm$4 &           17 &       16.05$\pm$1.40 &       9.79$\pm$0.10  &      0.39$\pm$0.13 & 0.27$\pm$0.11 &      0.18\\[-1pt]
NGC\,4412  &  VCC\,0921 & SB(r)b pec &  12 26 36.08 & $+$03 57 52.9   &         2603$\pm$18 &           17 &       35.60$\pm \ast$ &       10.18$\pm$0.10 &       2.34$\pm$0.81 & 3.04$\pm$1.19 &       2.40\\[-1pt]
NGC\,4416  & VCC\,0938  & SB(rs)cd  & 12 26 46.72 & $+$07 55 08.4   &         1729$\pm$2 &           17 &       42.10$\pm \ast$ &       10.29$\pm$0.10   &       2.09$\pm$0.72 & 1.34$\pm$0.53 &       1.12\\[-1pt]
NGC\,4411B  &  VCC\,0939 & SAB(s)cd  & 12 26 47.23 & $+$08 53 04.6    &         1608$\pm$2 &           23 &       22.40$\pm$7.92 &       9.94$\pm$0.10 &      0.37$\pm$0.13 & 0.21$\pm$0.08 &   NA\\[-1pt]
NGC\,4419  & VCC\,0958  & SB(s)a & 12 26 56.44 & $+$15 02 50.6   &           96$\pm$23 &           17 &       16.10$\pm$3.62 &       10.31$\pm$0.09 &       1.47$\pm$0.50 &  2.05$\pm$0.79 &       1.21\\[-1pt]
NGC\,4424  &  VCC\,0979 &  SB(s)a & 12 27 11.61 & $+$09 25 14.4   &          776$\pm$3 &           23 &       16.40$\pm$0.80 &       9.93$\pm$0.09 &      0.55$\pm$0.18 & 0.59$\pm$0.22 &    0.32\\[-1pt]
NGC\,4430  & VCC\,1002  &  SB(rs)b & 12 27 26.41 & $+$06 15 46.0   &         1788$\pm$2 &           23 &       15.85$\pm$2.77 &       9.66$\pm$0.10 &      0.51$\pm$0.16 &  0.32$\pm$0.12 &      0.25\\[-1pt]
NGC\,4429  & VCC\,1003  &  S0/a & 12 27 26.51 & $+$11 06 27.8   &         1325$\pm$67 &           17 &       15.80$\pm$4.18 &       10.97$\pm$0.10  &      0.16$\pm$0.06 & 0.15$\pm$0.06 &      NA\\[-1pt]
NGC\,4438  & VCC\,1043  & SA0/a(s) pec  & 12 27 45.59 & $+$13 00 31.8   &          427$\pm$7 &           17 &       11.30$\pm$3.54 &       10.49$\pm$0.10    &      0.35$\pm$0.11 & 0.24$\pm$0.09 &      0.20\\[-1pt]
NGC\,4445  & VCC\,1086  & Sab  & 12 28 15.93 & $+$09 26 10.3    &          684$\pm$3 &           23 &       18.50$\pm$3.13 &       9.69$\pm$0.10  &      0.20$\pm$0.07 &  0.11$\pm$0.04 &      0.11\\[-1pt]
NGC\,4450  &  VCC\,1110 & 	SA(s)ab  &  12 28 29.63  & $+$17 05 05.8   &         2273$\pm$3 &           17 &       15.25$\pm$3.36 &       10.75$\pm$0.10   &      0.58$\pm$0.20 &   0.40$\pm$0.16 &      0.27\\[-1pt]
NGC\,4451  & VCC\,1118  & Sbc  & 12 28 40.55 & $+$09 15 31.7   &         1195$\pm$2 &           23 &       27.00$\pm$3.95 &       9.95$\pm$0.10    &       1.01$\pm$0.36 &  0.77$\pm$0.31 &      0.63\\[-1pt]
IC\,3392  & VCC\,1126  & SAb  & 12 28 43.26 & $+$14 59 58.2    &         1992$\pm$2 &           17 &       13.65$\pm$3.86 &       9.75$\pm$0.09 &      0.28$\pm$0.09 & 
     0.18$\pm$0.07 &     0.10\\[-1pt]
NGC\,4457  &  VCC\,1145 & (R)SAB0/a(s)  & 12 28 59.01 & $+$03 34 14.1   &         1228$\pm$2 &           17 &       10.70$\pm$3.15 &       10.20$\pm$0.10 &      0.41$\pm$0.13 &  0.34$\pm$0.12 &      0.29\\[-1pt]
NGC\,4469  & VCC\,1190  & SB(s)0/a  & 12 29 28.03 & $+$08 44 59.7   &          918$\pm$11 &           23 &       16.75$\pm$0.07 &       10.42$\pm$0.10   &      0.18$\pm$0.06 & 0.11$\pm$0.04 &      NA\\[-1pt]
NGC\,4470  & NGC\,4610, VCC\,1205  & Sa  & 12 29 37.78 & $+$07 49 27.1   &         2674$\pm$2 &           17 &       16.40$\pm$6.60 &       9.40$\pm$0.10   &      0.42$\pm$0.14 &   0.29$\pm$0.11 &      0.34\\[-1pt]
NGC\,4480  &  VCC\,1290 & SAB(s)c  & 12 30 26.78 & $+$04 14 47.7   &         2774$\pm$3 &           17 &       42.20$\pm$3.80 &       10.42$\pm$0.10    &       2.14$\pm$0.73 &   1.68$\pm$0.65 &       1.42\\[-1pt]
NGC\,4492  &  VCC\,1330 & 	SA(s)a  & 12 30 59.71 & $+$08 04 40.3   &         2079$\pm$10 &           17 &       19.30$\pm$3.54 &       10.35$\pm$0.10 &      0.24$\pm$0.09 &  0.09$\pm$0.04 &     0.07\\[-1pt]
NGC\,4496A  & NGC\,4505, VCC\,1375  & SB(rs)m  &  12 31 39.21 & $+$03 56 22.1   &         2072$\pm$2 &           17 &       15.70$\pm$1.14 &       9.62$\pm$0.10 &      0.72$\pm$0.24 &  0.62$\pm$0.21 &      0.80\\[-1pt]
%
NGC\,4498  & VCC\,1379  &  SAB(s)d &  12 31 39.56  & $+$16 51 10.0   &         1824$\pm$4 &           17 &       14.55$\pm$3.62 &       9.46$\pm$0.10 &      0.35$\pm$0.12 & 0.23$\pm$0.09 &      0.21\\[-1pt]
NGC\,4501  & M\,88, VCC\,1401  &  SA(rs)b & 12 31 59.16 &$+$14 25 13.4   &         2605$\pm$2 &           17 &       17.10$\pm$4.58 &       11.03$\pm$0.09 &       6.00$\pm$2.08 &   4.19$\pm$1.65 &       3.98\\[-1pt]
NGC\,4519  &  VCC\,1508 &  SB(rs)d & 12 33 30.25 & $+$08 39 17.1   &         1555$\pm$2 &           17 &       19.60$\pm$8.48 &       9.69$\pm$0.10 &       1.26$\pm$0.45 &   1.33$\pm$0.54 &       1.05\\[-1pt]
NGC\,4522  & VCC\,1516  & SB(s)cd  &  12 33 39.71 &$+$09 10 30.1   &         2653$\pm$4 &           17 &       16.05$\pm$4.73 &       9.75$\pm$0.10 &      0.43$\pm$0.15 &  0.28$\pm$0.11 &      0.34\\[-1pt]
NGC\,4527  & VCC\,1540  &  SAB(s)bc &  12 34 08.42 &$+$02 39 13.2   &         2077$\pm$1 &           17 &       14.05$\pm$0.62 &       10.37$\pm$0.09 &       4.53$\pm$1.57 &  3.92$\pm$1.53 &       2.89\\[-1pt]
NGC\,4532  &  VCC\,1554 &  IBm & 12 34 19.33  &$+$06 28 03.7   &         2350$\pm$4 &           17 &       13.00$\pm$3.45 &       9.28$\pm$0.10 &      0.60$\pm$0.19 &  0.93$\pm$0.35 &       1.18\\[-1pt]
NGC\,4535  & VCC\,1555  & SAB(s)c  & 12 34 20.31  &$+$08 11 51.9   &         2296$\pm$1 &           17 &       15.95$\pm$1.42 &       10.43$\pm$0.10 &       3.82$\pm$1.30 &  3.34$\pm$1.30 &       2.71\\[-1pt]
NGC\,4536  &  VCC\,1562 & SAB(rs)bc  &  12 34 27.05 &$+$02 11 17.3   &         2149$\pm$1 &           17 &       14.95$\pm$1.73 &       10.22$\pm$0.09 &       3.64$\pm$1.29 &  4.74$\pm$1.86 &       3.49\\[-1pt]
NGC\,4548  & M\,91, VCC\,1615  & SB(rs)b  &  12 35 26.45 &$+$14 29 46.8   &          802$\pm$2 &           17 &       15.60$\pm$1.06 &       10.66$\pm$0.10   &       1.41$\pm$0.48 &  0.88$\pm$0.34 &      0.62\\[-1pt]
NGC\,4567  & VCC\,1673  &  SA(rs)bc &  12 36 32.71 &$+$11 15 28.8   &         2588$\pm$8 &           17 &       22.45$\pm$5.79 &       10.28$\pm$0.10  &       2.18$\pm$0.75 &   1.43$\pm$0.56 &      0.65\\[-1pt]
NGC\,4568  & VCC\,1676  & SA(rs)bc  & 12 36 34.26 &$+$11 14 20.0   &         2563$\pm$6 &           17 &       17.40$\pm$5.96 &       10.25$\pm$0.09 &       3.94$\pm$1.37 &   3.07$\pm$1.21 &       2.83\\[-1pt]
NGC\,4569  & M\,90, VCC\,1690  & SAB(rs)ab  &  12 36 49.79 &$+$13 09 46.6   &          103$\pm$2 &           17 &       11.45$\pm$3.16 &       10.53$\pm$0.09 &       1.58$\pm$0.54 &    1.31$\pm$0.52 &      0.89\\[-1pt]
NGC\,4571  &  VCC\,1696 & SA(r)d  & 12 36 56.38 &$+$14 13 02.5   &          657$\pm$3 &           17 &       15.10$\pm$2.33 &       10.13$\pm$0.10  &       1.04$\pm$0.35 &  0.65$\pm$0.25 &     NA\\[-1pt]
NGC\,4579  & M\,58, VCC\,1727  & SAB(rs)b  &  12 37 43.52 &$+$11 49 05.5   &         1842$\pm$2 &           17 &       16.80$\pm$3.07 &       10.96$\pm$0.10  &       1.93$\pm$0.67 &    1.29$\pm$0.50 &       1.55\\[-1pt]
NGC\,4580  & VCC\,1730  &  SAB(rs)a pec &  12 37 48.39 &$+$05 22 06.7  &         1372$\pm$2 &           17 &       17.35$\pm$3.15 &       10.11$\pm$0.10  &      0.48$\pm$0.17 &  0.34$\pm$0.13 &      0.20\\[-1pt]
NGC\,4606  & VCC\,1859  & SB(s)a  & 12 40 57.54 &$+$11 54 44.0   &         1969$\pm$7 &           17 &       14.10$\pm$4.31 &       9.77$\pm$0.10  &      0.23$\pm$0.07 &  0.14$\pm$0.05 &     0.06\\[-1pt]
NGC\,4607  & VCC\,1868  & SBb  & 12 41 12.40  &$+$11 53 11.9   &         2580$\pm$8 &           17 &       19.70$\pm$6.55 &       9.63$\pm$0.10 &      0.96$\pm$0.34 &  0.73$\pm$0.29 &      0.62\\[-1pt]
NGC\,4639  & VCC\,1943  & SAB(rs)bc  & 12 42 52.39 &$+$13 15 26.6   &         1302$\pm$7 &           17 &       21.95$\pm$1.97 &       10.29$\pm$0.10 &      0.81$\pm$0.27 &  0.58$\pm$0.22 &      0.57\\[-1pt]
NGC\,4647  & VCC\,1972  & SAB(rs)c  & 12 43 32.31  &$+$11 34 54.7   &         1733$\pm$2 &           17 &       16.80$\pm$3.04 &       10.01$\pm$0.10    &       1.91$\pm$0.66 &   1.36$\pm$0.53 &       1.07\\[-1pt]
NGC\,4654  & VCC\,1987  & SAB(rs)cd  & 12 43 56.58 &$+$13 07 36.1   &         1362$\pm$4 &           17 &       14.40$\pm$3.15 &       10.12$\pm$0.11 &       2.93$\pm$1.01 &  2.43$\pm$0.95 &       1.96\\[-1pt]
NGC\,4689  & VCC\,2058  & SA(rs)bc  & 12 47 45.56 &$+$13 45 46.1   &         1924$\pm$5 &           17 &       15.70$\pm$3.33 &       10.36$\pm$0.10 &       1.34$\pm$0.46 &   0.86$\pm$0.33 &      0.54\\[-1pt]
NGC\,4698  & VCC\,2070  &  SA(s)ab &  12 48 22.91 &$+$08 29 14.6   &         1333$\pm$2 &           17 &       22.15$\pm$6.77 &       10.87$\pm$0.10  &      0.88$\pm$0.31 &  0.39$\pm$0.16 &     NA\\[-1pt]
NGC\,4713  & EVCC\,1189  & SAB(rs)d  & 12 49 57.87 &$+$05 18 41.1   &          980$\pm$2 &           17 &       14.80$\pm$3.55 &       9.46$\pm$0.10 &      0.73$\pm$0.24 & 
     0.63$\pm$0.24 &      0.82\\
\hline\\ [-5pt]
\end{tabular} 
\label{tab1}
\end{center}
\begin{flushleft} 
{\footnotesize{
Col.~(3): from NED. Col.~(6): radial velocity with respect to the cosmic microwave background (from HyperLeda). Col.~(7): galaxy distance assuming the conventional structure of the Virgo Cluster, that is: 17 Mpc for the Virgo A subclump and the N, S and E clouds; 23 Mpc for the Virgo B subclump; 32 Mpc for the W and M clouds (as adopted, for example, by \citealt{gavazzi99}, \citealt{boselli15}, and the GOLDMine database). Col.~(8): median redshift-independent distance from NED, and standard deviation of the published values. When a galaxy has Cepheid distance measurements, we used the median and standard deviation of those values only. Col.~(9): stellar mass, derived from WISE W1 and W2 fluxes, as calibrated by \cite{cluver14} (see also \citealt{jarrett19}). The median NED distances were adopted. Col.~(10): SFR from the {\it WISE} W3 flux, derived with the calibration from \cite{cluver17} (see Eq.~\ref{Eq2}), and assuming median NED distances. Col.~(11): SFR from the {\it WISE} W4 flux, derived with the calibration from \cite{cluver17} (see Eq.~\ref{Eq3}). Col~(12): average SFR computed by \cite{boselli15}, from four different proxies. We have rescaled their published values (originally based on sub-structure distances) to our median NED distances. The 1-$\sigma$ errors for the SFRs in Col~(12) are $\approx$20\% of the values.
}}
\end{flushleft}
}
\end{table*}

\subsection{Distances}
Recession speed is not a useful proxy for the distance of individual Virgo galaxies, because it is comparable to the proper motion inside the cluster. There are two common alternative ways of allocating distances in Virgo cluster surveys. One method is to divide the cluster into sub-clusters and ``clouds'', determine an average distance for each substructure, and attribute that distance to all the galaxies in the substructure. The most commonly used subdivision \citep{gavazzi99,gavazzi02,boselli15} consists of a cluster A around M\,87, located at an assumed (partly conventional) distance of 17 Mpc; a cluster B around the position of M\,49, at 23 Mpc (but M\,49 itself is only projected in front of cluster B, and is allocated a distance of 17 Mpc); clouds N, S, E at 17 Mpc; clouds W, M at 32 Mpc (see Fig.~8 in \citealt{gavazzi99} for a definition of the conventional sky boundaries of those regions). This is also the convention adopted in the Galaxy On-Line Database MIlano NEtwork (GOLDMine\footnote{http://goldmine.mib.infn.it}, \citealt{gavazzi03}). 

In terms of our sample, 29 out of our 75 spirals belong to cluster A, 15 to cluster B, another 15 to cloud S, 8 to cloud N, 7 to cloud E and 1 to cloud W. Thus, in this classification, 59 of our galaxies have a distance of 17 Mpc, 15 are at 23 Mpc, and one (NGC\,4197) is at 32 Mpc (see Table 1, Col.~7). The relative high number of targets in clouds S (in particular, at its southern end, for Declination $\lesssim$5$^{\circ}$) and N is not surprising, as those two infalling subgroups consist of spirals for $\approx$80\% of their members \citep{gavazzi99}, while the ellipticals are more concentrated near the centre.

While attractive for its simplicity, the main shortcoming of this classification is that the projected sky location inside a subgroup boundary is no guarantee that a particular galaxy is not in the foreground or background: the clouds and sub-clusters themselves may be several Mpc thick. An example of this issue is discussed in \cite{boselli02}, who note the presence of gas-rich spirals in cluster A alongside the more common (and expected) gas-deficient ones. They suggest that the gas-rich galaxies are located $\approx$10 Mpc behind the core of Virgo, at the outskirts of the cluster, where they have not yet been ram-pressure stripped of their gas by the intracluster medium. One of the best examples of such gas-rich galaxies is NGC\,4480.

The alternative choice, adopted here, for the galaxy distances is to use the database of redshift-independent measurements in the literature, as reported in the NASA/IPAC Extragalactic Database (NED\footnote{https://ned.ipac.caltech.edu}). Such distances come from a variety of methods, and it is usually impossible to check the various assumptions or limitations that affected the original analyses. When Cepheid distances were available (for 9 of the 75 galaxies, namely NGC\,4321, NGC\,4424, NGC\,4496A, NGC\,4527, NGC\,4535, NGC\,4536, NGC\,4548, NGC\,4571 and NGC\,4639), we used the median and standard deviation of those measurements. Otherwise, we adopted the median and standard
deviation of all NED values (mostly Tully-Fisher measurements)
after first inspecting the histogram of redshift-independent distances, removing outliers and re-deriving the median and its uncertainty. These are recorded in Table 1, Col.~8. Finally, we compared the median NED distances with the substructure distance (Figure 3). The scatter around a ratio of 1 is expected, given the typical standard deviation of several Mpc in the measured distances for each galaxy, and the intrinsic thickness of the substructures (also several Mpc). There are only three outliers (NGC\,4412, NGC\,4416, NGC\,4480, labelled in Figure 3) which have Tully-Fisher distances ($\sim$40 Mpc) much larger than their substructure distances (17 Mpc). We do not have elements to decide at this stage which set of distances is more reliable, and whether those three galaxies are Virgo cluster members. The uncertainty in the distance to NGC\,4480, in particular, affects our census of the most luminous ULXs, because the most luminous point-like source in that galaxy has either an X-ray luminosity $\sim$10$^{40}$ erg s$^{-1}$ or barely above 10$^{39}$ erg s$^{-1}$ depending on the choice of distance.

\subsection{Morphology and stellar masses}

Our sample includes a good balance of early- and late-type spirals (Figure 4). About half of the galaxies are of type S0/a to Sbc, and the other half from Sc to Im (with a little discrepancy between the classification in NED and  HyperLEDA). There are eight Messier galaxies \citep{messier1781} in our sample: M\,58, M\,61, M\,88, M\,90, M\,91, M\,98, M\,99, M\,100. 

An alternative classification of spiral galaxies is based on their mid-infrared colours, as measured by the {\it {Wide-field Infrared Survey Explorer}} ({\it {WISE}}; \citealt{wright10}). All 75 galaxies in our sample have been imaged and characterized with {\it WISE}: this is part of a larger project called the WISE Extended Source Catalog\footnote{https://vislab.idia.ac.za/research-wxsc} (WXSC). {\it WISE} photometry and global properties for the 100 largest galaxies in the sky (first data release of the WXSC) was published by \cite{jarrett19}: that sample already included a few of our 75 Virgo targets. For all other galaxies, we used {\it WISE} results that will be published in follow-up WXSC data releases.
For a detailed technical description of how the total {\it WISE} luminosity of each galaxy was measured, see \cite{jarrett19}. In short, the steps followed by the WSXC team were the following. First, Galactic foreground stars projected onto a galaxy were identified and removed. Then, a local background was computed from an elliptical annulus, with inner and outer radii located $\approx$30\% and $\approx$50\% beyond the maximum extent of the galactic disk. Foreground stars were also masked from the background annulus. Each galaxy was assumed to have a constant ellipticity and to be axisymmetric. 
To estimate the total flux of the galaxies, \cite{jarrett19} simultaneously fit two S\'ersic functions (bulge $+$ disk) to the surface brightness profile.
See \cite{jarrett19} for details.

{\it WISE} has four channels: W1, W2, W3 and W4, corresponding to the 3.4-$\mu$m, 4.6-$\mu$m, 12-$\mu$m and 22-$\mu$m bands, respectively. 
Plotting the $W1 - W2$ colours against the $W2 - W3$ colours defines a ``star formation sequence'' of normal galaxies \citep{jarrett17,jarrett19}. All our galaxies are located along that sequence (Figure 5), except for NGC\,4388, which is $\approx$0.3 mag above the sequence, in a region typically occupied by Seyfert and low-luminosity AGN \citep{mingo16,huang17}, which is indeed the case for NGC\,4388. The reddest galaxy along the mid-infrared star formation sequence is NGC\,4429 (classified as S0/a in HyperLEDA); the bluest one is NGC\,4254 = M\,99 (an Sc galaxy).

\begin{figure}
\hspace{-0.5cm}
\includegraphics[height=0.49\textwidth, angle=270]{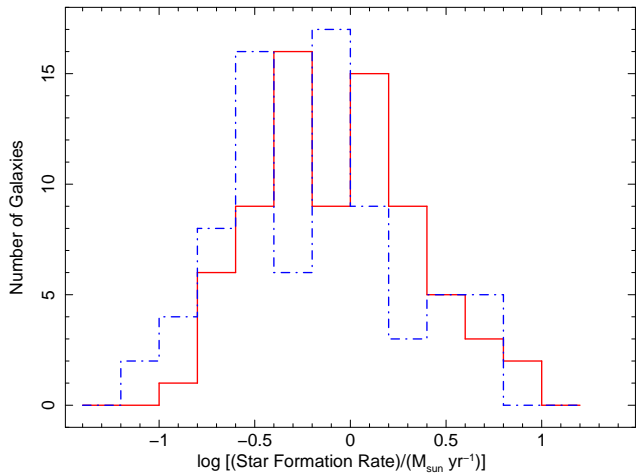}
 \caption{Solid red histogram: SFR distribution, inferred from the WISE 12-$\mu$m luminosity \citep{jarrett19}. Dash-dotted blue histogram: SFR distribution from the WISE 22-$\mu$m luminosity with the scaling of Jarrett et al.~(2019). }
  \label{sfr_median}
\end{figure}

\begin{figure}
\hspace{-0.5cm}
\includegraphics[height=0.49\textwidth,angle=270]{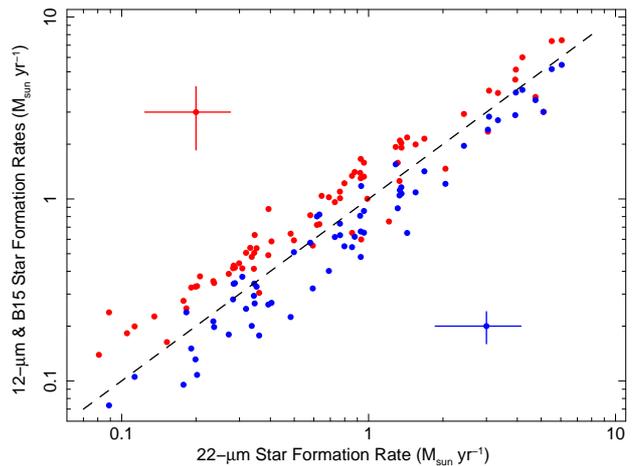}
 \caption{Comparison between the SFRs derived from the 12-$\mu$m luminosities, the 22-$\mu$m luminosities, and the average SFRs from Boselli et al.~(2015), B15.  The dashed line is the one-to-one relation.  On average, the 22-$\mu$m SFRs are $\approx$20\% higher than the B15 values (plotted in blue) and $\approx$20\% lower than the 12-$\mu$m values (plotted in red). The difference between the three estimates is relatively more prominent for low SFRs. Typical uncertainties on the {\it WISE} SFRs from Equations~\ref{Eq2} and \ref{Eq3} are around 33--38 percent (see Table~\ref{tab1}), and around 20 percent for the B15 values. The red datapoint in the top left sector of the plot and the blue datapoint in the bottom right sector are labels, displayed to represent the typical size of the error bars. }
  \label{sfr_12_22_boselli_median}
\end{figure}

Stellar masses ($M_{\ast}$) were derived from the mid-infrared luminosity. Specifically, we used the relation from \cite{cluver14}, based on the {\it {WISE}} photometry; see also \cite{jarrett13,jarrett19}:
\begin{equation}
   \log \left(M_{\ast}/L_{\rm{W1}}\right) = -2.54({\rm {W1}} - {\rm{W2}}) - 0.17, 
\end{equation}
where $M_{\ast}$ is in solar units, W1 and W2 are the magnitudes in the 3.4-$\mu$m and 4.6-$\mu$m bands, respectively, and $L_{W1}$ is the luminosity in the W1 band in units of solar luminosity. (The absolute magnitude of the Sun in the W1 band is $M_{\rm{W1},\odot} = 3.24$ mag). This definition of $M_{\ast}$ is in excellent agreement (better than 10\%) with the mass relation used by \cite{graham19} in their study of Virgo spirals, namely $M_{\ast}/L_K = 0.6$, where $L_K$ is the $K$-band luminosity. (The good agreement between the $K$-band and {\it WISE} mass values holds also for NGC\,4388, the galaxy where the {\it WISE} photometry was moderately affected by an AGN, as noted earlier). The main reason we adopt the {\it WISE} scaling rather than the $K$-band scaling in this paper is because we have also used the {\it {WISE}} band luminosities as a proxy for the SFR (Section~\ref{SecSFR}). 

From the mass values listed in Table 1 (Column 9), we see that $\approx$3/4 of our sample galaxies have masses between $\approx$5 $\times 10^9 M_{\odot}$ and $\approx$5 $\times 10^{10} M_{\odot}$ (Figure 6). The four most massive galaxies are NGC\,4216, NGC\,4429, NGC\,4501 (= M\,88) and NGC\,4579 (= M\,58): all four of them have $M_{\ast} \approx 1 \times 10^{11} M_{\odot}$. The mass values scale with the square of the assumed distances. The values listed in Table~\ref{tab1} and adopted for subsequent analysis are for the median NED distances, after correcting for outliers. The error range for each galaxy includes only the photometric error in {\it WISE} W1 and W1 $-$ W2, not the systematic uncertainty in the distance; it also does not include the scatter around the best-fitting relation, which we estimate as $\pm 0.15$ dex, from Fig.~6 of \cite{cluver14}. In total, the stellar mass of our galaxy sample is $M_{\ast,\rm{tot}} \approx 1.5 \times 10^{12} M_{\odot}$, if we assume the redshift-independent NED distances, or $M_{\ast,\rm{tot}} \approx 1.7 \times 10^{12} M_{\odot}$ if we use the substructure distances.  
The main reason for computing a total stellar mass is that we shall estimate the number of LMXB ULXs per unit stellar mass in our sample of star-forming galaxies (Section 4.4, Section 5), and compare it both with the value found in the early-type galaxies of the AMUSE sample \citep{plotkin14} and with theoretical predictions of X-ray binary populations.


\subsection{Star formation rates}\label{SecSFR}

To characterize the star formation properties of the sample, we looked for SFR proxies that were available for all 75 galaxies, for consistency.
The luminosity in the {\it {WISE}} bands W3 (12 $\mu$m) and W4 (22 $\mu$m) can be used as a proxy for the total infrared luminosity and, therefore, for the SFR \citep{jarrett13,cluver14,cluver17,jarrett19}. The relations calibrated by \cite{cluver17} are:
\begin{eqnarray}
    \log \left(\frac{{\mathrm{SFR}}}{M_{\odot} {\mathrm{yr}}^{-1}}\right) 
    & = & (0.873 \pm 0.021) \, \log \left(\frac{\nu L_{12\mu\mathrm{m}}}{L_{\odot}}\right) \label{Eq2} \\ 
    && - (7.62 \pm 0.18) \nonumber \\
    \log \left(\frac{{\mathrm{SFR}}}{M_{\odot} {\mathrm{yr}}^{-1}}\right) 
    & = & (0.900 \pm 0.027) \, \log \left(\frac{\nu L_{22\mu\mathrm{m}}}{L_{\odot}}\right) \label{Eq3} \\  
    && - (7.87 \pm 0.24) \nonumber 
\end{eqnarray}
for the W3 and W4 bands, respectively. In Equations (2--3), $\nu$ is the central frequency of the band, and the luminosity $\nu L_{\nu}$ is normalized to the bolometric luminosity of the Sun ($L_{\odot} = 3.83 \times 10^{33}$ erg s$^{-1}$), not its band-limited luminosity. The contribution of the stellar continuum was subtracted from  $L_{12\mu\mathrm{m}}$ and $L_{22\mu\mathrm{m}}$ as described in \cite{cluver17}. A Kroupa initial mass function (IMF) was adopted \citep{kroupa01}.

\begin{figure}
\hspace{-0.5cm}
\includegraphics[height=0.49\textwidth, angle=270]{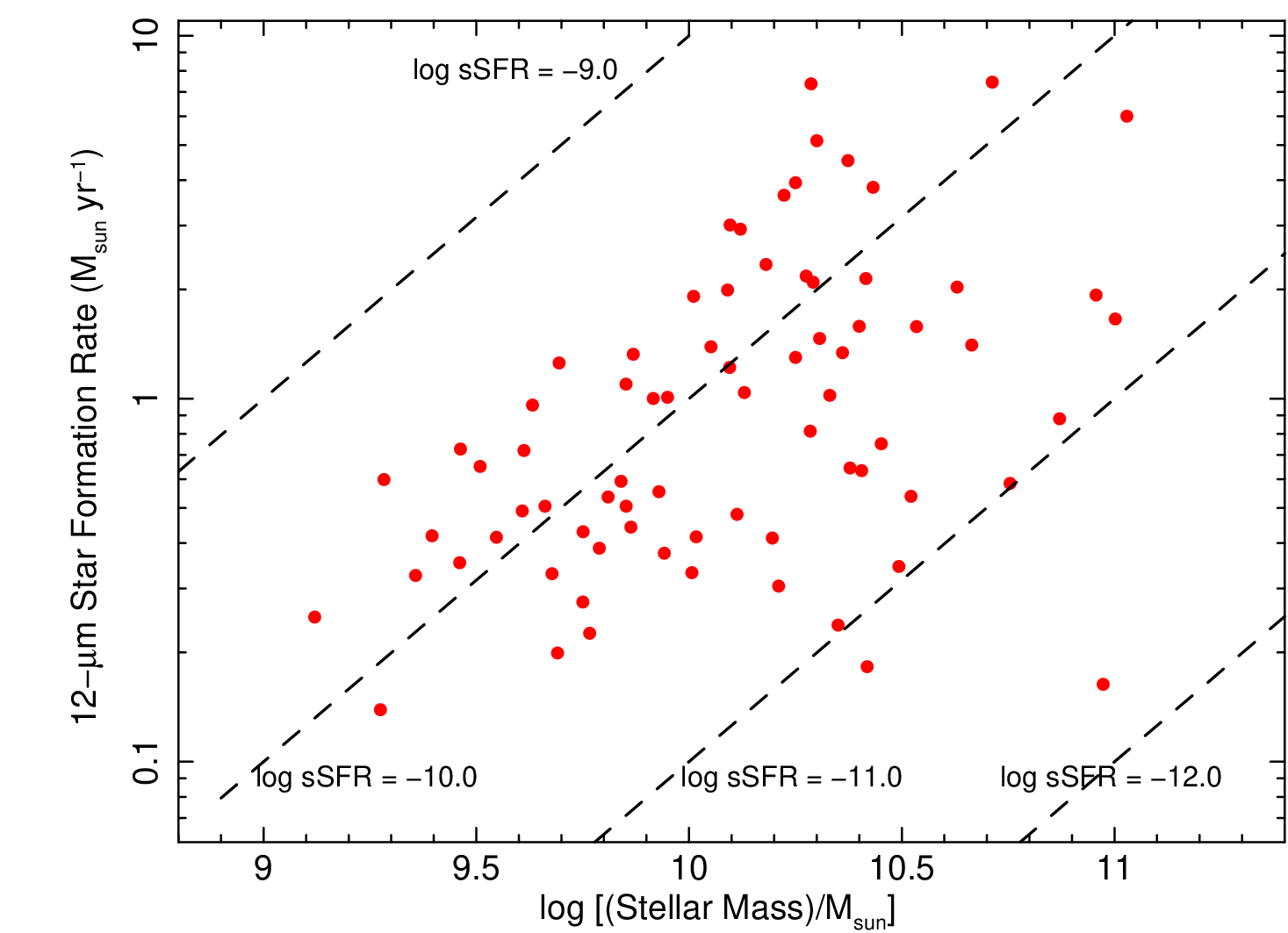}\\
\vspace{0.2cm}
\hspace{-0.5cm}
\includegraphics[height=0.49\textwidth, angle=270]{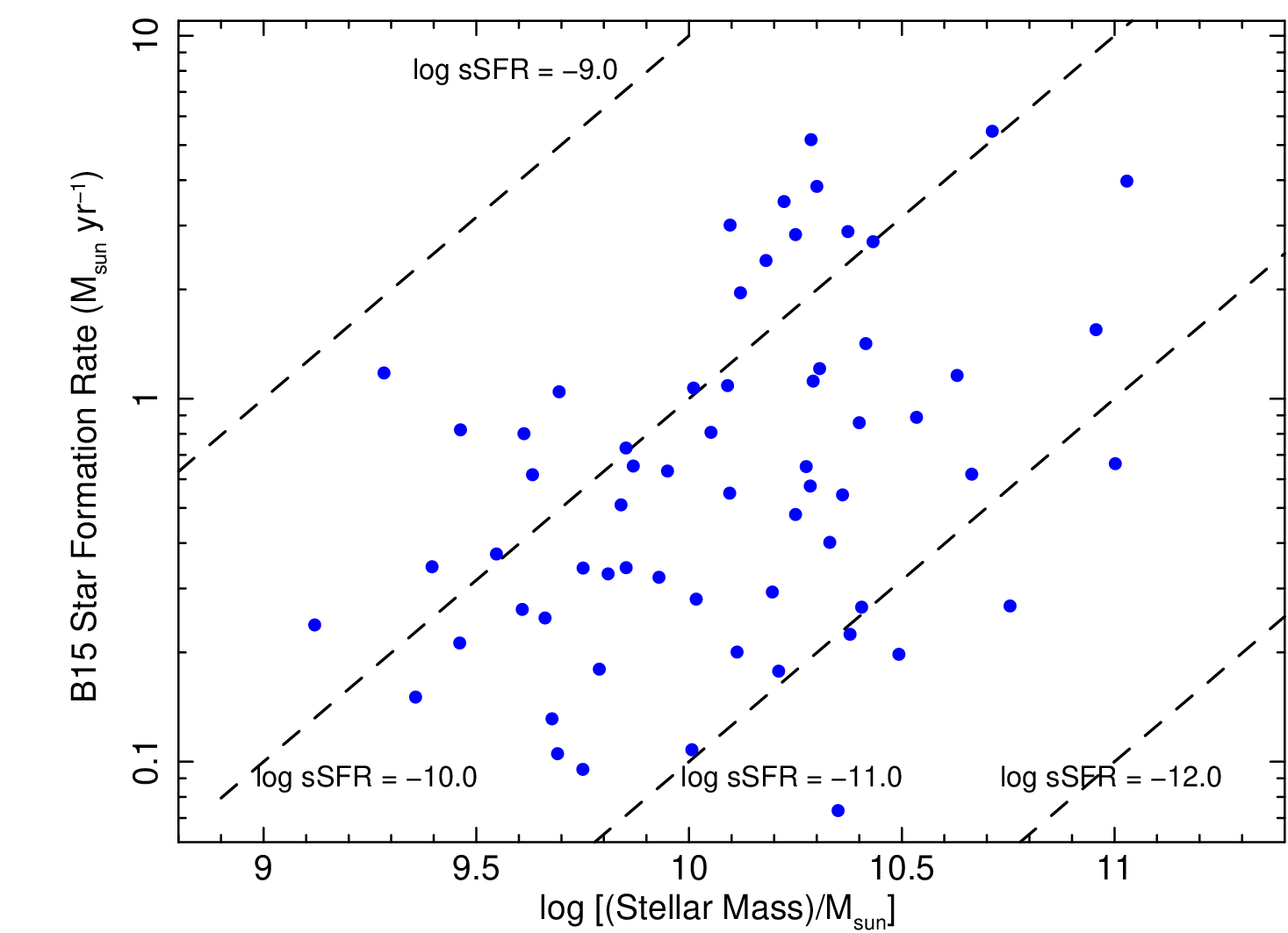}
 \caption{Top panel: 12-$\mu$m SFR  versus $M_{\ast}$. Loci of constant sSFR are marked with dashed lines. Bottom panel: as in the top panel, for the average B15 SFR.}
  \label{mass_sfr}
\end{figure}

The {\it WISE} SFRs are presented in Table~\ref{tab1} (columns~10 and 11) and are graphically shown in the upper panel of Figure~\ref{mass_sfr}. The total SFR$_{12\mu\mathrm{m}}$ of the sample is $(104 \pm 6) M_{\odot}$ yr$^{-1}$, if we adopt redshift-independent distances, or $(115 \pm 8) M_{\odot}$ yr$^{-1}$ for substructure distances. The total SFR$_{22\mu\mathrm{m}}$ is $(86 \pm 6) M_{\odot}$ yr$^{-1}$ or $(95 \pm 7) M_{\odot}$ yr$^{-1}$ in the two cases. (Note that the two sets of WISE-derived SFRs scale as $d^{1.75}$ and $d^{1.80}$, respectively, as a function of assumed galaxy distances, while stellar masses scale as $d^2$). It is clear (Figure 7) that there is a systematic offset in the distribution of SFR$_{12\mu\mathrm{m}}$ compared with the corresponding values of SFR$_{22\mu\mathrm{m}}$. This is evident in particular at lower SFRs ($\lesssim$0.3 $M_{\odot}$ yr$^{-1}$), where SFR$_{12\mu\mathrm{m}}$ tends to be $\sim$1.5--2 times higher than SFR$_{22\mu\mathrm{m}}$; at higher SFRs ($\gtrsim$2 $M_{\odot}$ yr$^{-1}$), the excess of SFR$_{12\mu\mathrm{m}}$ over SFR$_{22\mu\mathrm{m}}$ is only $\sim$10\%.
\cite{cluver17} and \cite{jarrett19} argue that, for normal galaxies (excluding extreme dust-obscured starbursts, which are not a concern in our sample), SFR$_{12\mu\mathrm{m}}$ is a better proxy of total infrared luminosity and SFR than SFR$_{22\mu\mathrm{m}}$, because it is less contaminated by polycyclic aromatic hydrocarbon emission and less dependent on metallicity.

In addition to the far-infrared luminosity, the most common empirical proxies for the SFR in the literature are the H$\alpha$ luminosity (corrected for extinction), the far-UV plus 24-$\mu$m luminosity, and the 20-cm radio continuum luminosity; see \cite{kennicutt98} for a well-known review. \cite{boselli15} (B15) collected a large database of SFR measurements in nearby spirals, determined with those three methods, as part of their {\it Herschel} Reference Survey; some of the SFR values in B15 came from their own H$\alpha$ observations and data analysis, others were taken from published literature. When more than one proxy was available for a galaxy, B15 also included the average between the different values. Their SFR catalog includes 64 of our 75 galaxies (Table 1, Column 12). 

The list of average SFRs in \cite{boselli15} does not include errors. We estimated an approximate error from the dispersion of the SFR values reported for the individual proxies around their respective average SFRs, for all galaxies with more than one SFR proxy. The SFRs from the different proxies have an approximately Gaussian distribution around the (normalized) average values, with a standard deviation $\sigma$ of about 20\% of the mean. Thus, we take this standard deviation as an estimate of the 1-$\sigma$ error for each B15 average SFR value.

We compared the B15 average values with the {\it WISE} values (scaled to the same redshift-independent distance). The B15 values are lower than both the 12-$\mu$m and 22-$\mu$m values (Figure~\ref{sfr_12_22_boselli_median}). For the 64 galaxies that are in common between the {\it WISE} catalog and the B15 catalog, we determine a total SFR$_{\mathrm{B15}} \approx 66 M_{\odot}$ yr$^{-1}$, compared with SFR$_{22\mu\mathrm{m}} \approx 81 M_{\odot}$ yr$^{-1}$ and SFR$_{12\mu\mathrm{m}} \approx 98 M_{\odot}$ yr$^{-1}$. The discrepancy occurs at all values of SFRs, although it is more accentuated at the low end (Figure 7). With the plausible assumption that the same fractional difference would be found in the remaining 11 galaxies, we conclude that the three SFR proxies used by B15 suggest a total SFR $\approx 70 M_{\odot}$ yr$^{-1}$ for our full sample. The large difference cannot be explained with the different choice of IMF (a Salpeter IMF in B15, \citealt{salpeter55}); a conversion to a Kroupa IMF in B15 would reduce rather than increase the amount of stellar mass required to produce the same ionizing flux. An investigation into the different calibrations of the various SFR proxies is beyond the scope of this paper.

\begin{figure}
\hspace{-0.5cm}
\includegraphics[height=0.49\textwidth, angle=270]{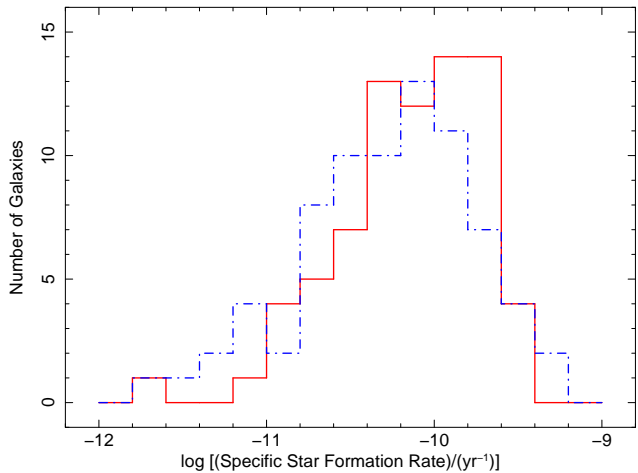}
 \caption{Solid red histogram: sSFR distribution, inferred from the WISE 12-$\mu$m luminosity. Dash-dotted blue histogram: sSFR distribution from the WISE 22-$\mu$m luminosity. }
  \label{ssfr_median}
\end{figure}

Knowing the mass and the SFR, we can then determine the distribution of our sample galaxies in terms of their sSFR. This is a parameter that will be particularly useful later, when we interpret the X-ray luminosity of the galaxies.
In the mass versus SFR$_{12\mu\mathrm{m}}$ plane (Figures 9,10), about half of our galaxies have sSFR $< -10$, which characterizes the ``blue cloud'' of star-forming galaxies (compare with Fig.~2 in \citealt{lehmer19}), and the other half can be described as ``green valley'' galaxies. If we adopt the B15 SFR, about two thirds of the galaxies are in the green valley. The sSFR distribution inferred from  SFR$_{22\mu\mathrm{m}}$ is of course intermediate between the other two. 
From this rough classification, we can already speculate that on average, we should find comparable contributions to the X-ray luminosity from galaxies dominated by young HMXBs and those dominated by intermediate-age or old LMXBs (see Fig.~6 in \citealt{lehmer19} for the predicted transition between the two classes). We investigate this point with our {\it Chandra} study.

\section{X-ray data analysis}

We selected and downloaded from the {\it Chandra} archives all the Advanced CCD Imaging Spectrometer (ACIS) observations that included our sample galaxies. That is a total of 110 observations from 2000 February to 2020 March: 106 observations placed the target on the back-illuminated ACIS-S3 chip, and the other 4 on the front-illuminated ACIS-I chips. Fifty-two of the observations were specifically obtained for this project (Proposal 18620568, PI R.~Soria, observing cycle 18), with exposure times between 8 and 15 ks, for a total exposure time of 551 ks; they all placed the target on S3. The other 58 were gathered from the archives, with various durations, from as short as 1.4 ks to as long as 148 ks, adding another 1,413 ks (Table 2, Columns 7--10).

We reprocessed and analysed each dataset with the Chandra Interactive Analysis of Observations ({\sc ciao}) software version 4.12 \citep{fruscione06}, with calibration database version 4.9.1. In particular, we created new level-2 event files with the {\it ciao} tasks {\it chandra\_repro}. For galaxies with multiple observations, we applied {\it reproject\_obs} to align and create stacked images. We created individual and stacked images in the soft band (0.3--1.0 keV), medium band (1.0--2.0 keV), hard band (2.0--7.0) and full band (0.3--7.0 keV). We used the {\sc ds9} imaging tool \citep{joye03} for displaying images, defining source and background regions, and performing various other tasks ({\it e.g.}, smoothing, creating contour plots, aligning and overplotting X-ray contours over optical images, finding centroids of point-like sources, etc.).

\subsection{Integrated galaxy luminosities}
First, we measured or constrained the total X-ray luminosity of the X-ray binary population of their host galaxies (inside their $D_{25}$), including the contribution from faint, unresolved sources. The main hurdles to overcome for such analysis were the choice of suitable background regions (especially in cases when the $D_{25}$ of a galaxy occupied most of the S3 chip) and the potential additional contribution to the X-ray emission from diffuse hot gas and/or from an active nucleus. 

We have taken the $B$-band 25 mag arcsec$^{-2}$ isophotal radii, and thus
diameters ($D_{25}$), from \citet{devaucouleurs91}. 
We drew corresponding elliptical regions with {\sc ds9}, in a format suitable to {\sc ciao} analysis. We then ran the {\sc ciao} task {\it specextract} to create spectra and associated response and ancillary response files. More specifically, because we were extracting spectra of extended regions, we built spatially weighted ancillary response functions ({\it specextract} parameter ``weight = yes'') without a point-source aperture correction (``correctpsf = no''). 
For each target galaxy, background regions on the same CCD were chosen as large as possible, outside the $D_{25}$ of the galaxy, but we avoided regions of low transmission (caused by molecular contamination) near the edges of the CCD array.  Each of the six ACIS-S chips is $8\arcmin \times 8\arcmin$, providing enough galaxy-free field-of-view. As noted previously, the bulk of the galaxy sample had their centre positioned on the S3 chip.  We also built response and ancillary response files for the background regions with {\it specextract}.

We adopted two methods to subtract the background and model the net galaxy emission. The first method consists of first modelling the background spectrum, scaling that model to the area of the galaxy, then holding this background model component fixed and adding one or more spectral components representing the emission from the galaxy. The second method simply subtracts the background spectrum (suitably scaled) directly from the galaxy-plus-background spectrum. 

In the first method, the background model consists of a celestial X-ray background (which includes unresolved cosmic background from distant AGN, and possible intracluster emission from diffuse hot gas in Virgo) and a particle-induced instrumental background. The celestial X-ray background is modelled as a two-temperature thermal plasma plus a power law with spectral parameter values given by the {\it Athena} Wide Field Imager \citep{meidinger20} background preparation document\footnote{https://www.mpe.mpg.de/ATHENA-WFI/public/resources/background/WFI-MPE-ANA-0010\_i7.1\_Preparation-of-Background-Files.pdf}, provided by Arne Rau (Max Planck Institute for Extraterrestrial Physics).  
We scaled the normalization of this component by the area of the background region on the sky. The instrumental background was modelled as a blackbody and three power-law continua superposed with three Gaussians. 
All parameters for the black body and the power-law components were allowed to vary in the fitting; however, the Gaussian energies and widths were fixed following the analytical particle background model of \cite{bartalucci14}, and only their normalizations were allowed to vary. The best-fitting background model was then applied to the source spectrum, with all its parameters frozen, and a global normalization factor scaled by the ratio of the source area to the background area.

Models were fitted to the spectra with the {\sc xspec} \citep{arnaud96} package, version 12.11.0. The Cash statistics \citep{cash79} was used as the fit statistics ({\it cstat} in {\sc xspec}). An absorbed power-law component was introduced to represent the galactic emission from (resolved and unresolved) point-like sources. Initially, the absorption column density $N_{\rm H}$ of the galactic power-law component was fixed to the Milky Way line-of-sight value, and the power-law photon index was fixed to $\Gamma = 1.7$. In the event of an unacceptable fit, $N_{\rm H}$ was allowed to vary first, and then, if necessary, the power-law photon index was also thawed.

The second method was a standard, direct subtraction of the background photon spectrum from the source spectrum ({\it i.e.}, with the source and background spectra loaded in {\sc xspec} as data file and bkg file, respectively). In this case, we did not need to assume any model for the background. We simply modelled the residual emission with an absorbed power-law, with free $N_{\rm H}$ and $\Gamma$ when we had $\gtrsim$100 net counts, or a column density fixed to the Galactic line-of-sight value and $\Gamma = 1.7$ for galaxies with very few net counts.

We compared the results of the two methods, and found that they were consistent within the uncertainties. In this paper, we have reported the results from the first method. All fits spanned the 0.5--6~keV energy range, because the detected photons are strongly dominated by background above those energies. Observed (model) fluxes and de-absorbed emitted luminosities were then extrapolated and quoted over the 0.5--8~keV range, for consistency with the range chosen by \cite{lehmer19}, which we will use extensively in comparison to our results. 

\subsubsection{Caveat 1: galaxies larger than the S3 chip}
In three cases, a sector at the outskirts of the $D_{25}$ region was outside the detector's field of view (namely, about half of the $D_{25}$ of NGC\,4178 and NGC\,4579, and about 10\% of NGC\,4501). In those cases, the net galaxy flux and luminosity were scaled under the assumption of a uniform surface brightness. The resulting flux may be overestimated by this assumption, because the surface brightness typically decreases towards larger galactic radii. However, given the small number of galaxies affected by this issue, this likely over-estimation in flux does not significantly affect our population results; in fact, the uncertainty caused by the rescaling is within the overall model flux uncertainties for each of those galaxies.  In another four galaxies with a large apparent size, the central part of the $D_{25}$ falls onto the ACIS-S3 chip but a small sector falls onto either of the two adjoining chips (S2 or S4). In those cases, we defined local background regions that also spanned two chips, with the same fractional area as the source regions. We compared this method with the alternative possibility of computing the galaxy emission only over the fraction of the $D_{25}$ inside the S3 chip, and then rescaling the total flux as explained above. The two approaches give the same result within the model uncertainties.

\subsubsection{Caveat 2: galaxies with a nuclear X-ray source}
A few of our galaxies contain a bright AGN, with 0.5--8 keV luminosity $\sim$10$^{39}$--$10^{41}$ erg s$^{-1}$: they are classified in the literature as Seyferts (NGC\,4388, NGC\,4569, NGC\,4579, NGC\,4639, NGC\,4698) or LINERs (NGC\,4438, NGC\,4450, NGC\,4548). Those strong nuclear X-ray sources were immediately excluded from the extraction of the $D_{25}$ spectra of their host galaxies, with the use of ``exclude'' circular regions around the nucleus. In few cases, bright AGN readout streaks were visible on a chip and were also excluded from our source extraction. In all other cases, when we detected candidate nuclear sources but with an X-ray luminosity $\lesssim$10$^{39}$ erg s$^{-1}$, we included the nuclear region in the $D_{25}$ source extraction and subsequent model fitting. The emission of those faint, candidate nuclear point sources was extracted and modelled separately, and subtracted from the total $D_{25}$ luminosity of their host galaxies {\it a posteriori}. This is because modelling a weak nuclear source on its own and then subtracting its luminosity gives a more accurate result than simply placing an exclusion circle around the nuclear position. In total, 39 of the 75 galaxies (including the 8 mentioned above) have a point-like {\it Chandra} source within $\approx$1$^{\prime\prime}$.5 of the position of the optical nucleus reported in the literature. A separate paper on the nuclear X-ray properties of those galaxies suspected of hosting an intermediate-mass black hole can be found in \citet{grahamIMBH21}, and we have already presented preliminary results for candidate intermediate-mass nuclear BHs (based on archival {\it Chandra} data prior to our Large Program observations) in \cite{graham19}. 

\subsubsection{Caveat 3: galaxies with strong thermal-plasma emission} 
In a few cases, a simple power law model was statistically unacceptable. Inspection of the fit residuals (and, sometimes, direct inspection of the images) indicated that a soft thermal component was also present in the $D_{25}$ emission, with typical emission features around 1 keV, 1.3 keV and 1.8 keV; not surprisingly, this happened in galaxies with a high SFR (a beautiful example is NGC\,4303). For those galaxies, an additional {\it apec} model component was added. The flux from this thermal component was excluded from the galaxy flux reported here, because for this work we are specifically looking at the X-ray binary contribution (both the individually detected sources and the faint, unresolved ones). We are aware that part of the unresolved emission from a galaxy can be in the form of thermal plasma but come from point-like sources; for example, from accretion columns in magnetic Cataclysmic Variables ({\it e.g.}, \citealt{revnivtsev07,li07,li11,hong16}). However, such contribution is negligible compared with the power-law component produced by the X-ray binary population; it is also typically characterized by a temperature $\gtrsim$10 keV, so that its spectrum is indistinguishable from a hard power-law component, given the small number of counts for most of our galaxies, and the low {\it Chandra} sensitivity above 5 keV. It is only the thermal plasma component at temperatures $\sim$0.3--1 keV, mostly from gas shock-heated by young SNRs, that (occasionally) stands out in a galaxy spectrum and can be separated from the power-law component. We leave a discussion of the hot gas emission to follow-up work.  

\subsubsection{Caveat 4: background AGN projected onto the $D_{25}$}
Area-scaling for background subtraction assumes that the cosmic background emission is uniform between source and background regions. This applies in principle both to unresolved emission and (on a statistical level) to point-like sources (background AGN), leaving aside the issue of cosmic variance. This may not be true for individual galaxies, if their $D_{25}$ ellipse includes one or few bright AGN projected through the galaxy, which may not be exactly balanced by a scaled number of similar AGN in the background region, given the small-number statistics. We will illustrate the expected relative contribution of resolved background AGN in Section 4.2, when we discuss the ULX LF. Here we can anticipate that such a contamination is not a significant problem. We will show that most galaxies have 0.5--8 keV X-ray luminosities $\gtrsim$10$^{39}$, and we expect to find a total of only $\sim$10 AGN projected inside the $D_{25}$ of the 75 sample galaxies, at a flux greater than this apparent luminosity (using the median NED distance for each galaxy).
 

\subsection{Point-source X-ray luminosities}
For this part of the analysis, we used again standard {\sc ciao} tasks. In particular, we used {\it wavdetect} to identify point-like sources. We then checked and improved the centroid position with {\it dmstat}. In a few cases, especially when a source was observed far from the aimpoint, we further refined its position using a 2-D Gaussian fit with the {\sc sherpa} package \citep{freeman01}. For regions with significant diffuse emission from hot gas, we used the hard band or the 1.5--7 keV band to filter out the thermal plasma emission and identify accreting point-like sources. This includes also the candidate nuclear sources, which will be discussed in a follow-up paper. In this work, we present the results for the off-nuclear ULX population.   

We measured absorbed and absorption-corrected fluxes of every moderately bright point-like source ({\i.e.}, every source with an observed flux $\gtrsim$10$^{-14}$ erg cm$^{-2}$ s$^{-1}$, corresponding to a luminosity of about a few $\times 10^{38}$ erg s$^{-1}$) with {\it srcflux}, over the 0.5--8 keV band. In doing so, we used the ancillary response function to determine the shape of the point spread function (PSF) at the location of the source ({\it i.e.}, ``PSF Method = arfcorr'' in {\it srcflux}). As usual for {\it Chandra}/ACIS studies, the size of the source extraction region depended on how far from the aimpoint each particular source was. For isolated sources near the aimpoint, the typical source extraction radius was 2$^{\prime\prime}$.5, or 2$^{\prime\prime}$.0 for sources in more crowded regions. For background subtraction, we defined local background regions at least four times larger that the corresponding source regions. 

We made two assumptions in our initial {\it srcflux} analysis of the ULX population: a) that the spectrum was a power law with a photon index $\Gamma = 1.8$; b) that the absorbing column density $N_{\rm H}$ was limited to the line-of-sight Galactic value, obtained from \cite{hi4pi16} via NASA's High Energy Astrophysics Science Archive Research Center (HEASARC). The rationale for the first assumption is that $\Gamma \approx 1.8$ is the mean photon index over the 0.5--8 keV band determined from the \cite{swartz04}'s {\it Chandra} survey of ULXs in nearby galaxies. It is also the slope used for the conversion from count rates to luminosities in the follow-up work of \cite{swartz11}. The effect of different choices of $\Gamma$ is small, over a plausible range of values. For example, for Cycle-18 observations, with $N_{\rm H} = 3 \times 10^{20}$ cm$^{-2}$, the choice of $\Gamma = 1.6$ would increase the inferred unabsorbed luminosity by about 6\%; the choice of $\Gamma = 2.0$ would decrease it by about 3\%. The reason for the second assumption (line-of-sight Galactic absorption) is that we wanted to start from a ULX list as simple and model-independent as possible. 

In the second step of our point-source analysis (Section 4.2.1), we selected all sources with $\gtrsim$50 counts.
We used {\it specextract} to create spectra and associated response and ancillary response files for those sources. We then rebinned the spectra with the 
{\sc ftools} \citep{blackburn95} task {\it grppha}, to 1 count per bin, and modelled them with {\sc xspec} \citep{arnaud96}, using the Cash statistics \citep{cash79}. We assumed a power-law model whenever possible, but unlike for our initial {\it srcflux} estimate, this time we left the photon index and the intrinsic $N_{\rm H}$ as free parameters. We used the {\it cflux} convolution model in {\sc xspec} to measure absorbed and de-absorbed fluxes in the 0.5--8 keV and 0.3--10 keV bands.

\begin{figure}
\hspace{-0.5cm}
\includegraphics[height=0.49\textwidth, angle=270]{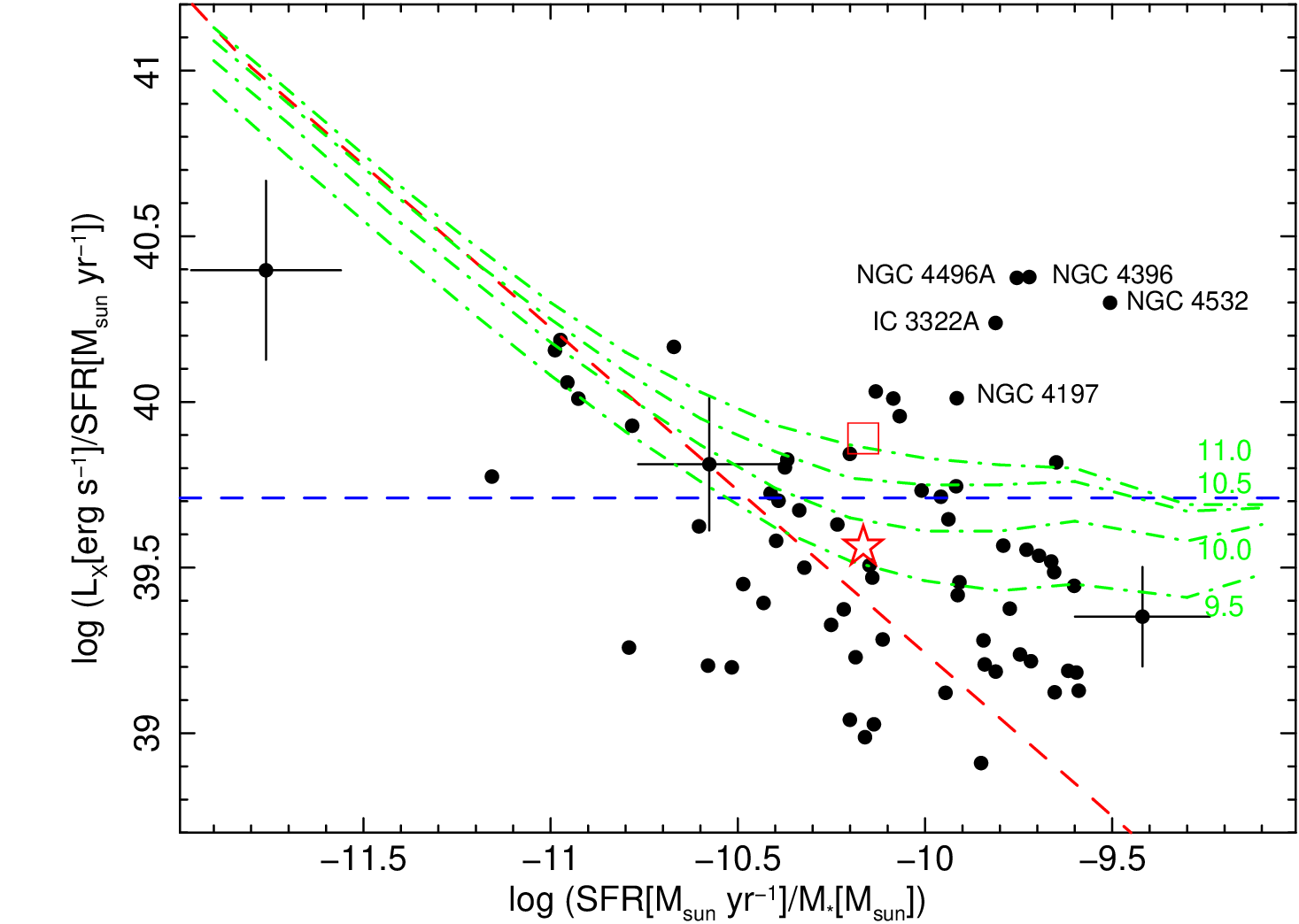}\\
\vspace{0.2cm}
\hspace{-0.5cm}
\includegraphics[height=0.49\textwidth, angle=270]{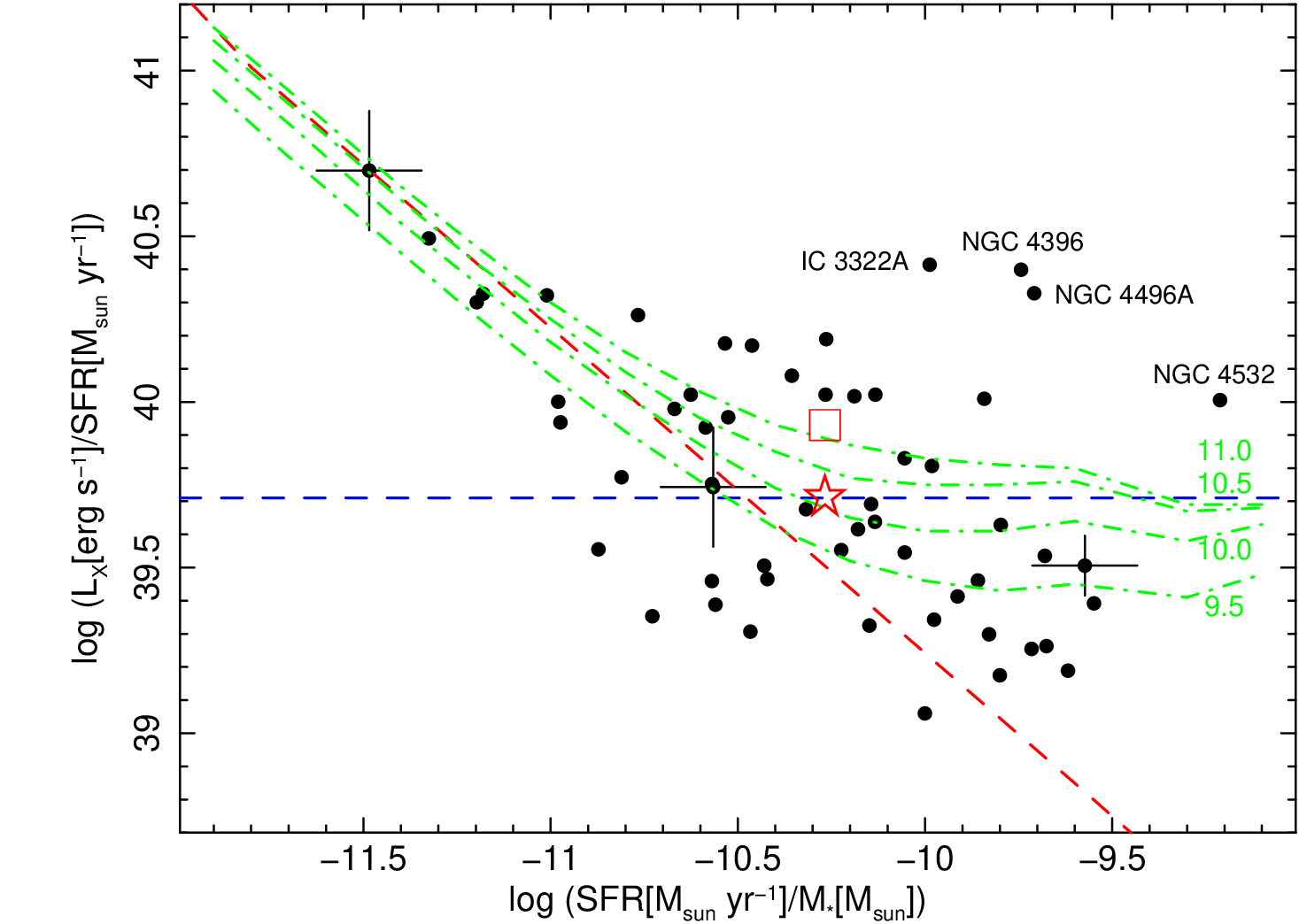}
 \caption{Luminosity of the X-ray binary population (0.5--8 keV band) of each galaxy, normalized to the SFR of the galaxy, plotted versus the specific SFR. Error bars are omitted for clarity, except for three galaxies in each plot, which visualize representative values of the error bars for low, intermediate and high sSFRs. Errors for all individual galaxies can be derived from the values of $M_{\ast}$, SFR and $L_{\rm X}$ listed in Table 1, Columns 9--12, and Table 2, Column 2.
 Top panel: SFR values from the {\it WISE} 12-$\mu$m band; bottom panel: SFR values from B15 (hence, not all 75 galaxies have a datapoint in the bottom panel). The dashed diagonal red line is the theoretical luminosity expected for a pure LMXB population ({\it i.e.}, scaling with the stellar mass), in the limit of perfect sampling of the LF (\citealt{lehmer19}). The dashed horizontal blue line is the theoretical luminosity for a pure HMXB population, again in the limit of perfect sampling (\citealt{lehmer19}). The red star marks the total value of $L_{\rm X}$/SFR versus sSFR for all the sample galaxies. The red square is the expected value of $L_{\rm X}$/SFR versus sSFR for the whole sample ({\it i.e.}, the sum of the LMXB and HMXB luminosities predicted by \citealt{lehmer19} at that value of sSFR). The green dash-dotted curves are the predicted median values of the observed luminosity, for a population of galaxies of various SFRs and stellar masses, based on the Monte Carlo simulations of \citet{lehmer19}; the number next to each curve represents the stellar mass along that track. The green curves differ from a simple sum of LMXB plus HMXB luminosity, because of the incomplete sampling of the LF in each galaxy, especially those of lower masses. A few outliers with an exceptionally high $L_{\rm X}$ in relation to their SFR and stellar mass are labelled in both panels and will be discussed individually in follow-up work. }
  \label{lehmer_boselli}
\end{figure}

\section{Main Results}

We summarize our main results for the de-absorbed X-ray luminosity of all the sample galaxies, and the fluxes and luminosities of their ULXs, in Table 2.  

\subsection{Galaxy luminosities}
The galaxy luminosities (Table 2, Column 2) include the ULXs but exclude the thermal plasma emission and the nuclear sources. This is because here we want to study the relation between the galaxy properties (stellar mass and SFR) and their X-ray binary populations. For galaxies with multiple observations of comparable duration, the luminosity values provided in Column 2 are an average luminosity. For galaxies with multiple observations of very uneven duration, only the deepest observations were used. When a galaxy contains ULXs, the ULX luminosity itself is a significant fraction of the total luminosity ({\it i.e.}, the total point-source luminosity is usually dominated by the sources at the upper end of the LF).

We stress that for this paper, our main interest is in the population properties of the sample rather than in individual galaxies: specifically, whether the Virgo spirals are dominated by LMXBs or HMXBs\footnote{The possibility of off-centre, wandering intermediate-mass BHs among the point-source population will be discussed in further work.}. The combined X-ray luminosity of LMXBs is proportional to the stellar mass \citep{gilfanov04,lehmer10,boroson11,zhang12,lehmer19}, although other factors such as population age and specific frequency of globular clusters also affect the correlation \citep{zhang12}. The combined HMXB luminosity is proportional to SFR \citep{grimm03,ranalli03,lehmer10,mineo12a,lehmer19}. Regardless of the precise value of the normalization constants, we expect a total X-ray luminosity $L_{\rm X} = \alpha_{\rm{LMXB}}M_{\ast} + \beta_{\rm{HMXB}}\,{\rm{SFR}}$ \citep{lehmer10,lehmer19}. Hence, if a population of galaxies are dominated by LMXBs, we expect $L_{\rm X}/{\rm{SFR}} \propto M_{\ast}/{\rm{SFR}} = 1/{\rm{sSFR}}$, that is we expect to find them along a line of slope $-1$ in the $\log \left(L_{\rm X}/{\rm{SFR}}\right)$ versus $\log$ sSFR plane. Instead, if the galaxies are dominated by HMXBs, we expect them scattered along a horizontal line, because $L_{\rm X}/{\rm{SFR}} \approx $ constant. In the analysis of \cite{lehmer19}, the transition from LMXB to HMXB domination occurs at sSFR $\approx 10^{-10.5}$ yr$^{-1}$. Their best-fitting relation (which we will use for comparison with our data) is 
\begin{equation}
    L_{\rm X} = \left( 5.1^{+2.0}_{-0.9} \times {\mathrm{SFR}} + 1.8^{+0.3}_{-0.3} \times M_{\ast,10} \right) \times 10^{39} \ \ {\mathrm{erg~s}}^{-1},
\end{equation} 
where the SFR is in units of $M_{\odot}$ yr$^{-1}$ and $M_{\ast,10}$ is the stellar mass in units of $10^{10} M_{\odot}$. We have ignored here the small dependence of the SFR coefficient on the metal abundance \citep{lehmer21}; we leave this discussion to further work.

Our analysis shows (Figure 11) that there is indeed a (1/sSFR) trend in the distribution of our galaxies at the lower end of the sSFR scale (where we expect LMXBs to dominate). This is more evident when we use the B15 values of SFR (Figure 11, bottom panel): in this case, about 1/3 of the 75 galaxies are in the region of the diagram in which the X-ray luminosity is dominated by LMXBs. This does not mean that 1/3 of the total X-ray luminosity of the Virgo spirals comes from LMXBs, because the galaxies dominated by HMXBs are also the most luminous and the most likely to contain ULXs. 

For sSFR $\gtrsim 10^{-10.5}$ yr$^{-1}$, the galaxy distribution is very scattered and flatter than the (1/sSFR) trend, but most galaxies fall below the median tracks predicted by the Monte Carlo simulations of \cite{lehmer19} (despite a few exceptions with higher-than-expected X-ray luminosity, namely NGC\,4532, NGC\,4396A, NGC\,4396 and NGC\,4197). We had previously shown (Figure 6) that about two thirds of our sample galaxies have a stellar mass in the range $9.5 \lesssim \log \left(M_{\ast}/M_{\odot} \right) \lesssim 10.5$. However, the median point-source luminosity of our sample of galaxies in the high sSFR region (Figure 11) falls below the simulated median line for $\log \left(M_{\ast}/M_{\odot}\right) = -9.5$ in \cite{lehmer19}.
Thus, the X-ray luminosity of several of the most actively star-forming disks appears underestimated, compared with the scaling relations of \cite{lehmer19}. Those scaling relations were derived from galaxies with discs orientated close to face-on, with low or intermediate inclination; however, our sample contains several galaxies seen at high inclination, with much higher column densities through their discs (particularly in the regions where HMXBs are preferentially located).

On the same $\log \left(L_{\rm X}/{\rm{SFR}}\right)$ versus $\log$ sSFR plane we can also plot the total values from the whole sample, knowing that $L_{\rm {X,tot}} \approx 3.6 \times 10^{41}$ erg s$^{-1}$ (from a simple addition of the 0.5--8 keV luminosities of each galaxy) and the total SFR $\approx$70--104 $M_{\odot}$ yr$^{-1}$. The total observed value of $L_{\rm X}/{\rm{SFR}}$ (marked by a red star in the two panels of Figure 11) is lower than predicted by the model of \cite{lehmer19} (red squares in the two panels of Figure 11), when we include both the predicted HMXB contribution (dashed blue line) and the predicted LMXB contribution (dashed red line). In particular, the observed luminosity is a factor of 2 lower than predicted if we adopt the 12$\mu$m set of SFRs (Figure 11, top panel), or a factor of 1.6 lower if we use the B15 set of SFRs (Figure 11, bottom panel).  

At this stage of our analysis, there is no glaring single reason for this discrepancy between inferred and predicted luminosities. We have already hinted at an underestimation of the de-absorbed luminosity from galaxies seen at higher inclination as the most likely cause. We will investigate this possibility further when we fit the individual spectra of the most luminous ULXs and model their luminosity distribution (Sections 4.2.1, 4.3).  
We will show there that the median absorption column density of the most luminous ULXs is $N_{\rm H} \approx 3 \times 10^{21}$ cm$^{-2}$. If we assume this column density, instead of line-of-sight Galactic absorption, our inferred luminosities would be increased by a factor of $\approx$1.5 (for observations in Cycle 4 or earlier), a factor $\approx$1.4 (observations in Cycles 5--14) and a smaller factor for more recent observations ($\approx$1.2 for Cycle 18). This would bring our total $L_{\rm X}/SFR$ in line with the prediction from \cite{lehmer19}, within the errors. (For comparison, \citealt{lehmer19} found a median intrinsic column density $N_{\rm H} = 2 \times 10^{21}$ for their brighter sources, and applied that value to convert count rates to luminosities for the fainter sources).
Even the choice of a higher but constant $N_{\rm H}$ would be an over-simplification (which is why we avoided it here): galaxies with a high sSFR contain more cold gas and should have an absorption column density above the median value (thus requiring a larger luminosity correction in Figure 11), while galaxies with low sSFR may be better approximated with $N_{\rm H}$ close to the line-of-sight Galactic value.


Moreover, the relation between $L_{\rm X}$, SFR and $M_{\ast}$ may be different for the galaxies in the Virgo sample and those in the \cite{lehmer19} sample. 
There are only 5 Virgo spirals (NGC\,4254, NGC\,4321, NGC\,4450, NGC\,4536, NGC\,4569) in the sample of 38 galaxies used by \cite{lehmer19} to fit their scaling parameters. Even for those 5 galaxies, we cannot directly compare our and their adopted values of $L_{\rm X}$, SFR and $M_{\ast}$, because \cite{lehmer19} use only the inner part of the $D_{25}$ of those spirals (about half of the projected $D_{25}$ area). More specifically, \cite{lehmer19} consider only the portion of a galaxy inside the $K_s = 20$ mag (arcsec)$^{-2}$ surface brightness isophote \citep{jarrett03}, while we have used the traditional $D_{25}$ definition of galactic sizes. This may introduce a small bias, for example if the outskirts of spiral disks have a lower number of X-ray binaries and lower $L_{\rm X}$ for a given stellar mass or SFR, compared with the denser inner regions. Furthermore, we used different proxies for SFR and different proxies for stellar mass, compared with \cite{lehmer19}. Finally, we used a different method for determining the total point-source luminosity of a galaxy. We estimated $L_{\rm X}$ from a spectral analysis (Section 3.1). Instead, \cite{lehmer19} estimate $L_{\rm X}$ by integrating the X-ray binary luminosity functions. Considering all these methodological differences, the small discrepancy between our empirical estimates and the predicted scaling relations of \cite{lehmer19} is not surprising.
A more detailed discussion of this issue is beyond the scope of this work.

\begin{table*}
\caption{Host-galaxy X-ray binary luminosities and ULX census}
\vspace{-0.2cm}
\scriptsize{
\begin{center}
\begin{tabular}{lccccccccccc}  
\hline \hline\\[-5pt]    
Galaxy  & $L_{{\rm {XRB,}}0.5-8}$ &   R.A.(J2000) & Dec.(J2000) & ULX $F_{0.5-8}$ 
& ULX $L_{0.3-10}$ & ULX $L_{0.3-10}^c$
& ObsID & Obs.~Date  & Detector & Exp.~Time & Notes \\[4pt]
   & ($10^{39}$ erg s$^{-1}$)  & &  & ($10^{-14}$ erg cm$^{-2}$ s$^{-1}$) 
   & ($10^{39}$ erg s$^{-1}$) & ($10^{39}$ erg s$^{-1}$) & &  &  & (ks) & \\
   \multicolumn{1}{c}{(1)} & (2) & (3) & (4) & (5) & (6) & (7) & (8) & (9) & (10)&(11) & (12)\\
\hline  \\[-5pt]
NGC\,4178 &  5.4[4.6--6.1]$^*$ & 12 12 44.51 & $+$10 51 13.6  &
    14.7[13.7--15.8] &  3.95[3.67--4.21] &  6.0[5.5--6.6] & 12748  & 2011-02-19 & ACIS-S & 36.3 & \\[-1pt]
NGC\,4192  & 6.4[4.5--8.8]
  & 12 13 48.92 & $+$14 55 01.1  & 11.2[9.6--12.8]  &  3.21[2.74--5.57] & 4.1[3.5--4.7] & 19390  & 2017-11-04 & ACIS-S &  14.9 &  \\[-1pt]
NGC\,4197  &       10.3[8.0--12.7]
  &  12 14 40.29 &  $+$05 49 00.0  &  2.62[1.69--3.82]  & 2.76[1.78--4.03]  & 5.2[2.8--11.1] &
  19420 & 2017-07-26 & ACIS-S & 8.0 &   \\[-1pt]
  &   
  & 12 14 41.31 & $+$05 49 03.9  & 2.99[2.00--4.25]	&	3.15[2.10--4.48] & 4.9[3.2--8.7] &
  19420 & 2017-07-26 & ACIS-S & 8.0 &   \\[-1pt]
NGC\,4206  &       1.59[0.53--2.75]
  & 12 15 17.48 & $+$13 00 15.5  & 2.72[1.79--3.93] &	1.46[0.96--2.11]  & 2.0[1.3--2.8] &
  19431  & 2017-02-06 & ACIS-S & 8.0 &   \\[-1pt]
NGC\,4212  &       7.3[6.2--9.3]
  & 12 15 35.53 & $+$13 53 49.1  & 2.76[2.06--3.60]	&	1.19[0.92--1.61] & 1.7[1.2--2.2] &
    19395 & 2017-02-14 & ACIS-S & 14.9 &   \\[-1pt]
    &
  & 12 15 39.59 & $+$13 53 42.2  & 13.0[11.2--14.7]	&	5.77[5.00--6.58]  & 7.1[6.0--8.9] &
    19395 & 2017-02-14 & ACIS-S & 14.9 &   \\[-1pt]
NGC\,4216  &       14.0[11.5--17.0]
  & 12 15 50.05 & $+$13 07 13.0  & 6.69[4.60--9.34]  & 2.47[1.70--3.44]  &  3.2[2.2--4.5] &
  19391 & 2017-07-24 & ACIS-S  & 9.5 &  \\[-1pt]   
  &   
  & 12 15 50.12  & $+$13 06 18.3   & 7.07[5.46--8.96] &   2.61[2.01--3.31] & 4.8[2.8--21.0] &
  19391 & 2017-07-24 & ACIS-S  & 9.5 &  \\[-1pt] 
NGC\,4222  &       2.7[1.0--5.3]
  &  12 16 20.55 & $+$13 18 06.5 &  2.70[1.73--3.95]	 &	1.88[1.21--2.76] & 2.4[1.6--3.5] &
  19432  & 2018-05-10   & ACIS-S & 8.0  &   \\[-1pt] 
NGC\,4237  &       1.4[0.7--8.1]
  &   12 17 09.86  & $+$15 18 53.8  & 1.88[1.14--2.91]	&  1.20[0.73--1.86]  &  1.6[1.0--2.5] &
  19404  & 2017-02-12  & ACIS-S  & 8.0 &   \\[-1pt] 
NGC\,4254  &       16.6[15.5--17.7]
  & 12 18 51.87  &  $+$14 24 21.6 & 3.23[2.17--4.59]	&	1.03[0.69--1.46] &  1.7[1.1--2.4] &
  7863 & 2007-11-21  & ACIS-S &  5.1 &   \\[-1pt]   
  & & & & 3.17[2.65--3.71]	&	1.01[0.84--1.18] & 1.3[1.1--1.5] &
  17462  & 2015-02-16 & ACIS-I & 44.5 & \\[-1pt]
  &
  & 12 18 56.10  & $+$14 24 19.3  &  46.0[41.6--50.5]	&	14.6[13.3--16.1] &  21.6[19.0--24.8] &
  7863 & 2007-11-21  & ACIS-S &  5.1 &   \\[-1pt]   
  & & & & 16.2[15.0--17.3]	&	7.15[6.64--7.71]  & 6.6[5.9--7.4] &
  17462  & 2015-02-16 & ACIS-I & 44.5 & \\[-1pt]
NGC\,4276  &      $<$1.5
  &  -- & --  & --  & --  & -- &
  19427 & 2017-07-30 & ACIS-S & 15.2 &  \\[-1pt] 
NGC\,4293  &       4.9[3.4--6.4]
  & --  & --  & --  &  -- & -- &
  19409 & 2018-04-03 & ACIS-S & 8.0 &  \\[-1pt] 
NGC\,4298  &       4.0[2.0--6.6] 
  & 12 21 36.10 & $+$14 35 45.1  & 2.37[1.68--3.23]	& 0.91[0.65,1.24] 
  &  1.2[0.9--1.7] &
  19392 & 2018-04-09 & ACIS-S & 14.2 &  \\[-1pt]
  & 
  &  &  & 3.51[2.38--4.96]	&	1.35[0.91--1.90] &  1.8[1.2--2.5] &
  19397 & 2018-04-09 & ACIS-S & 7.8 &  \\[-1pt]
NGC\,4299  &       2.2[1.4--3.3]
  & --  & --  & --  & --  & -- &
  7834 & 2007-11-18  & ACIS-S & 5.1 &  \\[-1pt] 
NGC\,4302  &       1.74[1.06--2.66]
  & 12 21 42.04 &  $+$14 36 54.5  & 2.46[1.53--3.68] &		1.10[0.69--1.66]  & 1.2[0.7--2.1] &
  19392 & 2018-04-09 & ACIS-S & 14.2 &  \\[-1pt]
  & 
  &  &  & 3.00[1.59--4.97] &		1.34[0.72--2.24] & 2.0[1.0--3.5] &
  19397 & 2018-04-09 & ACIS-S & 7.8 &  \\[-1pt]
NGC\,4303  &       6.9[6.0--7.9]
  &  12 22 01.24  & $+$04 29 37.9 &  4.96[4.36--5.56]  & 1.13[0.99--1.26]	 &  1.9[1.7--2.1] &
  2149  &  2001-08-07 &  ACIS-S & 28.0 &  \\[-1pt] 
  & & & & 4.08[3.30--4.86]	& 	0.93[0.75--1.11] & 1.3[1.1--1.6] &
  17550 & 2014-11-16 & ACIS-S & 19.8 & \\[-1pt]
NGC\,4307  &     $<$2.7
  & -- & -- & -- & -- & -- &
  19405 & 2018-05-14  & ACIS-S  &  14.7  &  \\[-1pt]     
NGC\,4312  &      0.32[0.12--0.55]
  & --  &  -- & --  & --  & -- &
  7083 & 2005-11-07 & ACIS-S & 1.9 &  \\[-1pt] 
  &&&&&&&
  19414 & 2017-02-15 & ACIS-S & 8.0 &  \\[-1pt]
NGC\,4313  &      0.9[0.1--1.9]
  & --  & --  & --  & --  & -- &
  19416 & 2018-04-14 & ACIS-S & 8.0 &  \\[-1pt] 
NGC\,4316  &       3.9[2.8--5.4]
  & 12 22 39.54  & $+$09 20 21.0  & 3.44[2.61--4.43]	&	3.60[2.73--4.63]  & 5.4[3.0--22.0] &
  19400 & 2018-04-20  &  ACIS-S & 14.9  &   \\[-1pt]  
NGC\,4321  &       12.0[10.9--13.1]
  & 12 22 46.20 & $+$15 48 50.0  & 1.45[1.16--1.74]	&	0.57[0.46--0.69]  & 0.9[0.7--1.1] &
  6727 &  2006-02-18 & ACIS-S & 37.9 &   \\[-1pt]
  &&&& 0.97[0.62--1.43]	&	0.39[0.24--0.57]  & 0.6[0.4--0.9] &
  9121 &  2008-04-20 & ACIS-S & 14.9 &   \\[-1pt]
  &&&& 2.73[2.11--3.47]	&	1.08[0.83--1.36]  & 1.7[1.3--2.1] &
  12696 &  2011-02-24 & ACIS-S & 14.9 &   \\[-1pt]
  &&&& 1.27[1.07--1.48]	&	0.50[0.42--0.58]  & 0.7[0.6--0.8] &
  14230 &  2012-02-16 & ACIS-S & 7.9 &   \\[-1pt]
  &&&&  1.35[0.72--2.25] &	0.53[0.28--0.88] & 0.7[0.4--1.1] &
  23140 &  2020-02-15 & ACIS-S & 10.0 &   \\[-1pt]
  &&&&  1.87[1.11--2.88]	&	0.74[0.44--1.13] & 0.9[0.5--1.4] &
  23141 &  2020-03-13 & ACIS-S & 10.0 &   \\[-1pt]
  &
  & 12 22 47.37  & $+$15 49 11.5  & $<$0.07	& $<$0.03   & $<$0.05 &
  6727 &  2006-02-18 & ACIS-S & 37.9 &   \\[-1pt] 
  &&&& 5.93[4.94--6.93]	&	2.33[1.95--2.73]  & 3.6[3.0--4.2] &
  9121 &  2008-04-20 & ACIS-S & 14.9 &   \\[-1pt]
  &&&&  $<$0.23	& $<$0.09 & $<$0.1 &
  12696 &  2011-02-24 & ACIS-S & 14.9 &   \\[-1pt]
  &&&&  $<$0.08		&	$<$0.03 & $<$0.05 &
  14230 &  2012-02-16 & ACIS-S & 7.9 &   \\[-1pt]
  &&&&  $<$0.3	& $<$0.1 & $<$0.1 &
  23140 &  2020-02-15 & ACIS-S & 10.0 &   \\[-1pt]
  &&&&  $<$0.3	& $<$0.1 & $<$0.1 &
  23141 &  2020-03-13 & ACIS-S & 10.0 &   \\[-1pt]
%
%
  &
  & 12 22 51.29 & $+$15 51 23.3 & $<$0.14  & $<$0.06  & $<$0.1  &
  6727 &  2006-02-18 & ACIS-S & 37.9 &   \\[-1pt] 
  &&&& $<$0.03	 &	$<$0.02  & $<$0.03 &
  9121 &  2008-04-20 & ACIS-S & 14.9 &   \\[-1pt]
  &&&&  0.55[0.25--0.99]	&	0.22[0.10--0.39] & 0.3[0.1--0.5] &
  12696 &  2011-02-24 & ACIS-S & 14.9 &   \\[-1pt]
  &&&&  4.75[4.31--5.19]	&	1.87[1.70--2.04] & 2.6[2.4--2.9] &
  14230 &  2012-02-16 & ACIS-S & 7.9 &   \\[-1pt]
  &
  &  12 22 53.71  & $+$15 48 46.2 & $<$0.11  & $<$0.05  & $<$0.08  &
  6727 &  2006-02-18 & ACIS-S & 37.9 &   \\[-1pt] 
  &&&& 3.53[2.82--4.25]	&	1.39[1.11--1.67]  & 2.1[1.7--2.6] &
  9121 &  2008-04-20 & ACIS-S & 14.9 &   \\[-1pt]
  &&&&  $<$0.19  & $<$0.07 & $<$0.1 &
  12696 &  2011-02-24 & ACIS-S & 14.9 &   \\[-1pt]
  &&&&  $<$0.05  & $<$0.02   & $<$0.03 &
  14230 &  2012-02-16 & ACIS-S & 7.9 &   \\[-1pt]
  &&&&  0.98[0.45--1.78]	&	0.39[0.18--0.72] & 0.55[0.28--0.95] &
  23140 &  2020-02-15 & ACIS-S & 10.0 &   \\[-1pt]
  &&&&  0.34[0.05--0.88]	&	0.13[0.02--0.34] & 0.12[0.08--0.35] &
  23141 &  2020-03-13 & ACIS-S & 10.0 &   \\[-1pt]
  &
  &  12 22 54.14 & $+$15 49 12.6 & 4.02[3.55--4.49]	&	1.58[1.40--1.77] & 2.4[2.1--2.7] &
  6727 &  2006-02-18 & ACIS-S & 37.9 &   \\[-1pt] 
  &&&& 13.8[12.4--15.3]	&	5.30[5.88--5.99] & 14.4[8.1--31.1] &
  9121 &  2008-04-20 & ACIS-S & 14.9 &   \\[-1pt]
  &&&&  4.16[3.36--4.96] &		1.63[1.32--1.96] & 2.5[2.1--3.0] &
  12696 &  2011-02-24 & ACIS-S & 14.9 &   \\[-1pt]
  &&&&  3.00[2.69--3.31] &		1.18[1.06--1.30]   & 1.8[1.6--2.0] &
  14230 &  2012-02-16 & ACIS-S & 7.9 &   \\[-1pt]
  &&&&  6.88[5.26--8.50] &		2.70[2.07--3.35] & 3.5[2.7--4.4] &
  23140 &  2020-02-15 & ACIS-S & 10.0 &   \\[-1pt]
  &&&&  14.1[11.8--16.3]	&	5.53[4.65--6.41] & 14.0[6.5--44.6] &
  23141 &  2020-03-13 & ACIS-S & 10.0 &   \\[-1pt]
  & 
  & 12 22 54.26  & $+$15 49 44.4  & 1.87[1.37--2.49] &	0.74[0.54--0.98]  & 1.0[0.8--1.4] &
  6727 &  2006-02-18 & ACIS-S & 37.9 &   \\[-1pt] 
  &&&& 2.12[1.57--2.78]	&	0.83[0.62--1.09] & 1.3[1.0--1.7] &
  9121 &  2008-04-20 & ACIS-S & 14.9 &   \\[-1pt]
  &&&&  2.43[1.55--3.56]	&	0.96[0.61--1.40] & 1.3[0.8--1.9] &
  12696 &  2011-02-24 & ACIS-S & 14.9 &   \\[-1pt]
  &&&&  2.47[2.09--2.85]	&	0.97[0.82--1.12]   & 1.3[1.1--1.5] &
  14230 &  2012-02-16 & ACIS-S & 7.9 &   \\[-1pt]
  &&&& 2.57[1.69--3.71]	&	1.01[0.67--1.46]  & 1.3[0.9--1.9] &
  23140 &  2020-02-15 & ACIS-S & 10.0 &   \\[-1pt]
  &&&& 1.69[0.99--2.64]	&	0.66[0.39--1.04]  & 0.9[0.5--1.4] &
  23141 &  2020-03-13 & ACIS-S & 10.0 &   \\[-1pt]
  & 
  & 12 22 54.78  & $+$15 49 16.3 & 6.79[6.15--7.43]	&	2.67[2.42--2.92] &  4.2[3.8--4.6] &
  6727 &  2006-02-18 & ACIS-S & 37.9 &   \\[-1pt] 
  &&&& 11.1[9.72--12.4]	&	4.34[3.82--4.88] & 6.9[5.4--10.4] &
  9121 &  2008-04-20 & ACIS-S & 14.9 &   \\[-1pt]
  &&&&  4.07[3.22--4.92]	&	1.61[1.26--1.94] & 2.6[2.1--3.1] &
  12696 &  2011-02-24 & ACIS-S & 14.9 &   \\[-1pt]
  &&&&   5.85[5.42--6.28]	&	2.30[2.13--2.48]  & 3.5[3.3--3.8] &
  14230 &  2012-02-16 & ACIS-S & 7.9 &   \\[-1pt]
  &&&&  6.33[4.86--8.06]	&	2.49[1.91--3.18] & 3.3[2.5--4.1] &
  23140 &  2020-02-15 & ACIS-S & 10.0 &   \\[-1pt]
  &&&&  4.55[3.33--6.02]	&	1.75[1.31--2.37] & 2.3[1.7--3.0] &
  23141 &  2020-03-13 & ACIS-S & 10.0 &   \\[-1pt]
  &
  & 12 22 54.83 & $+$15 49 18.4  &  3.04[2.60--3.49]	&	1.20[1.02--1.37] &  2.0[1.7--2.5] &
  6727 &  2006-02-18 & ACIS-S & 37.9 &   \\[-1pt] 
  &&&& 4.53[3.63--5.43]	 &	1.79[1.39--2.08] & 2.9[2.3--3.4] &
  9121 &  2008-04-20 & ACIS-S & 14.9 &   \\[-1pt]
  &&&&  5.11[4.19--6.03]	&	2.01[1.65--2.38] & 3.0[2.5--3.6] &
  12696 &  2011-02-24 & ACIS-S & 14.9 &   \\[-1pt]
  &&&&  3.63[3.28--3.98]	&	1.43[1.29--1.57]  & 2.0[1.8--2.2] &
  14230 &  2012-02-16 & ACIS-S & 7.9 &   \\[-1pt]
  &&&&  4.14[2.98--5.54]	&	1.62[1.18--2.18] & 2.1[1.5--2.8] &
  23140 &  2020-02-15 & ACIS-S & 10.0 &   \\[-1pt]
  &&&& 2.90[1.94--4.13]	&	1.14[0.76--1.62] & 1.6[1.1--2.2] &
  23141 &  2020-03-13 & ACIS-S & 10.0 &   \\[-1pt]
  &
  & 12 22 55.69 &  $+$15 49 25.2 & 1.84[1.49--2.19]	&	0.72[0.58--0.86] & 1.1[0.9--1.3] &
  6727 &  2006-02-18 & ACIS-S & 37.9 &   \\[-1pt] 
  &&&& 2.58[1.97--3.20]	&	1.01[0.77--1.26] & 1.6[1.2--2.0] &
  9121 &  2008-04-20 & ACIS-S & 14.9 &   \\[-1pt]
  &&&&  2.88[2.21--3.57]	&	1.13[0.87--1.40] & 1.8[1.4--2.2] &
  12696 &  2011-02-24 & ACIS-S & 14.9 &   \\[-1pt]
  &&&&  1.93[1.62--2.25]	&	0.76[0.64--0.88]  & 1.2[1.0--1.4] &
  14230 &  2012-02-16 & ACIS-S & 7.9 &   \\[-1pt]
  &&&&  2.49[1.63--3.59]	&	0.98[0.65--1.42] & 1.3[0.9--1.9] &
  23140 &  2020-02-15 & ACIS-S & 10.0 &   \\[-1pt]
  &&&& 1.91[1.16--2.92]	&	0.75[0.46--1.15] & 0.9[0.6--1.4] &
  23141 &  2020-03-13 & ACIS-S & 10.0 &   \\[-1pt]
&
  & 12 22 57.47 & $+$15 49 16.5  & $<$0.07	& $<$0.03  &$<$0.05  &
  6727 &  2006-02-18 & ACIS-S & 37.9 &   \\[-1pt] 
  &&&& $<$0.34	& $<$0.13  & $<$0.20  &
  9121 &  2008-04-20 & ACIS-S & 14.9 &   \\[-1pt]
  &&&&  $<$0.13	& $<$0.05 & $<$0.08 &
  12696 &  2011-02-24 & ACIS-S & 14.9 &   \\[-1pt]
  &&&&  3.93[3.59--4.28]	&	1.55[1.42--1.69] & 2.4[2.2--2.6] &
  14230 &  2012-02-16 & ACIS-S & 7.9 &   \\[-1pt]
  &&&&  $<$0.75	& $<$0.25 & $<$0.30 &
  23140 &  2020-02-15 & ACIS-S & 10.0 &   \\[-1pt]
  &&&&  $<$0.46	& $<$0.20 & $<$0.25 &
  23141 &  2020-03-13 & ACIS-S & 10.0 &   \\[-1pt]
\hline\\[-5pt]
\end{tabular} 
\label{tab2a}
\end{center}
}
\end{table*}

\begin{table*}
\contcaption{Host-galaxy X-ray binary luminosities and ULX census}
\vspace{-0.2cm}
\scriptsize{
\begin{center}
\begin{tabular}{lccccccccccc}  
\hline \hline\\[-5pt]    
Galaxy  & $L_{{\rm {XRB,}}0.5-8}$ &   R.A.(J2000) & Dec.(J2000) & ULX $F_{0.5-8}$ 
& ULX $L_{0.3-10}$ & ULX $L^c_{0.3-10}$ & ObsID & Obs.~Date  & Detector & Exp.~Time & Notes \\[4pt]
   & ($10^{39}$ erg s$^{-1}$)  & &  & ($10^{-14}$ erg cm$^{-2}$ s$^{-1}$) 
   & ($10^{39}$ erg s$^{-1}$)  &($10^{39}$ erg s$^{-1}$)  & &  &  & (ks) & \\
   \multicolumn{1}{c}{(1)} & (2) & (3) & (4) & (5) & (6) & (7) & (8) & (9) & (10)&(11) & (12)\\
\hline  \\[-5pt]
NGC\,4321  &       12.0[10.9--13.1]
%
  & 12 22 57.52 &  $+$15 48 47.5  & $<$0.08	& $<$0.03  &$<$0.05  &
  6727 &  2006-02-18 & ACIS-S & 37.9 &   \\[-1pt] 
  &&&& $<$0.13	& $<$0.05  & $<$0.08  &
  9121 &  2008-04-20 & ACIS-S & 14.9 &   \\[-1pt]
  &&&&  $<$0.13	& $<$0.05 & $<$0.08  &
  12696 &  2011-02-24 & ACIS-S & 14.9 &   \\[-1pt]
  &&&&  $<$0.06	& $<$0.03 & $<$0.05  &
  14230 &  2012-02-16 & ACIS-S & 7.9 &   \\[-1pt]
  &&&&  4.78[3.55--6.27]	& 1.88[1.40--2.47] & 2.4[1.8--3.2] &
  23140 &  2020-02-15 & ACIS-S & 10.0 &   \\[-1pt]
  &&&&  5.01[3.74--6.54]	&	1.97[1.47--2.56] & 2.5[1.9--3.3] &
  23141 &  2020-03-13 & ACIS-S & 10.0 &   \\[-1pt]
  & 
  & 12 22 58.64 & $+$15 51 24.0 & $<$0.13	& $<$0.05 & $<$0.08  &
  6727 &  2006-02-18 & ACIS-S & 37.9 &   \\[-1pt] 
  &&&& $<$0.02	& $<$0.01  & $<$0.02  &
  9121 &  2008-04-20 & ACIS-S & 14.9 &   \\[-1pt]
  &&&&  $<$0.1	& $<$0.05 & $<$0.08  &
  12696 &  2011-02-24 & ACIS-S & 14.9 &   \\[-1pt]
  &&&&  3.28[2.92--3.65]	&	1.26[1.15--1.44] & 1.8[1.6--2.0] &
  14230 &  2012-02-16 & ACIS-S & 7.9 &   \\[-1pt]
  &&&&   $<$0.38	& $<$0.15 & $<$0.20  &
  23141 &  2020-03-13 & ACIS-S & 10.0 &   \\[-1pt]
  &
  & 12 22 58.68 & $+$15 47 51.9 & 3.20[2.79--3.60]	& 1.29[1.13--1.46]  & 2.1[1.8--2.4] &
  6727 &  2006-02-18 & ACIS-S & 37.9 &  A \\[-1pt] 
  &&&&  3.70[2.98--4.42]	&	1.46[1.18--1.74] & 2.4[2.0--2.8] &
  9121 &  2008-04-20 & ACIS-S & 14.9 &  A \\[-1pt]
  &&&&  2.40[1.84--3.06]	&	0.94[0.73--1.20] & 1.5[1.2--1.9] &
  12696 &  2011-02-24 & ACIS-S & 14.9 & A  \\[-1pt]
  &&&&  3.17[2.86--3.47]	&	1.25[1.12--1.36] & 1.9[1.7--2.1] &
  14230 &  2012-02-16 & ACIS-S & 7.9 &  A \\[-1pt]
  &&&&  1.98[1.21--3.01]	&	0.78[0.48--1.18] & 1.0[0.6--1.5] &
  23140 &  2020-02-15 & ACIS-S & 10.0 & A  \\[-1pt]
  &&&&  1.62[0.93--2.58]	&	0.64[0.37--1.01] & 0.7[0.4--1.2] &
  23141 &  2020-03-13 & ACIS-S & 10.0 & A  \\[-1pt]
NGC\,4330  &       1.62[0.46--3.66]
  &  12 23 13.26 & $+$11 21 26.0 & 3.63[2.51--5.05]	&	2.06[1.42--2.86]  & 2.8[1.9--3.8] &
  19419 &  2018-04-16 & ACIS-S & 8.0 &  \\[-1pt] 
NGC\,4343  &      0.5[0.3--1.6]
  &  -- & --  & --  & --  & -- &
  4687 & 2005-02-11 & ACIS-S & 38.3 &  \\[-1pt]
  &&&&&&&
  7129 & 2006-07-16  & ACIS-S &  4.6 & \\[-1pt]
  &&&&&&&
  12955 & 2011-02-17 & ACIS-S & 72.8 & \\[-1pt]
NGC\,4356  &     $<$1.0
  & --  & --  & --  & --  & -- &
  19435 & 2017-04-16 & ACIS-S & 14.9 &  \\[-1pt] 
NGC\,4380  &       2.7[2.0--4.0]
  & 12 25 22.21  & $+$10 01 03.4  &  2.92[2.16--3.83]	& 1.65[1.21--2.16] & 2.2[1.6--2.9] &
  19406 &  2018-05-11 & ACIS-S & 14.7 &   \\[-1pt] 
IC\,3322A  &       6.4[4.6--8.5]
  & 12 25 41.70  & $+$07 13 38.9  &  5.46[4.26--6.67]	&	5.23[4.01--6.39] &   62[23--132] &
  19401 &  2018-04-24 & ACIS-S & 13.9 &  \\[-1pt] 
  &
  & 12 25 42.25  & $+$07 13 19.8  & 1.19[0.71--1.84]	&	1.15[0.68--1.78] &  1.6[0.9--2.4] &
  19401 &  2018-04-24 & ACIS-S & 13.9 &  \\[-1pt] 
NGC\,4388  &       4.6[3.7--5.6]
  & 12 25 38.98 & $+$12 40 00.8  & 2.89[2.37--3.41]	&	(1.45[1.19--1.71])  & (1.7[1.3--2.2]) &
  1619 &  2001-06-08 & ACIS-S & 20.0 &  B \\[-1pt]
  &&&& $<$0.073 &	($<$0.037) & ($<$0.05) & 
  12291  & 2011-12-07 & ACIS-S & 27.6 & B\\[-1pt]
  &
  & 12 25 39.14 & $+$12 40 09.4	& 4.05[3.44--4.67]	& 2.03[1.72--2.35]	 &   3.4[2.9--3.9] &
  1619 &  2001-06-08 & ACIS-S & 20.0 &  \\[-1pt]
  &&&& 1.48[1.13--1.88]	&	0.75[0.57--0.94] & 1.1[0.8--1.4] &
  12291  & 2011-12-07 & ACIS-S & 27.6 & \\[-1pt]
  &
  & 12 25 54.06  & $+$12 40 04.2  & 1.67[1.29--2.07] &	0.86[0.66--1.06]	 & 1.5[1.1--1.8] &
  1619 &  2001-06-08 & ACIS-S & 20.0 & C \\[-1pt]
  &&&& 2.98[2.48--3.48]	&	1.53[1.28--1.79] & 2.4[2.0--2.8] &
  12291  & 2011-12-07 & ACIS-S & 27.6 & C \\[-1pt]
NGC\,4390  &      0.3[0.1--1.1]
  & --  & --  & --  & --  & -- &
  19425 & 2018-04-14 & ACIS-S & 15.4 &  \\[-1pt] 
IC\,3322  &      0.6[0.16--1.6]
  & --  &  -- & --  &  -- & -- &
  19424 & 2018-04-22 & ACIS-S & 14.9 & \\[-1pt] 
NGC\,4394  &      1.0[0.5--2.0]
  & --  & --  & --  & --  & -- &
  7864 & 2007-11-11 & ACIS-S & 5.1 &  \\[-1pt]
NGC\,4396  &       3.8[2.9--4.7]
  & 12 26 01.54  & $+$15 39 41.4  & 4.17[2.97--5.66]	&	1.08[0.77--1.46]  & 1.5[1.1--2.0] &
  19417 & 2018-05-25  & ACIS-S & 8.1 &   \\[-1pt] 
  &
  & 12 26 01.81  & $+$15 39 41.8  & 12.3[10.1--14.6] &	3.18[2.58--3.76]  & 9.4[4.2--40.4] &
  19417 & 2018-05-25  & ACIS-S & 8.1 &   \\[-1pt] 
%
NGC\,4402  &       2.3[1.7--3.4]
  & --  & --  & --  & --  & -- &
  15149 & 2013-07-24 & ACIS-S & 148.1 &  \\[-1pt]    
NGC\,4405  &       1.0[0.6--2.2]
  & --  & --  & --  & --  & -- &
  19410 & 2018-04-09 & ACIS-S & 8.0 &  \\[-1pt] 
NGC\,4411A  &       1.49[0.88--2.25]
  & 12 26 27.86 & $+$08 51 45.03  &  3.34[2.29--4.67]	&	1.43[0.98--2.00] &  2.3[1.6--3.2] &
  7837 &  2007-11-13 & ACIS-S & 5.1 &   \\[-1pt] 
  &
  & 12 26 31.62 & $+$08 52 28.1  & 3.84[2.55--5.50]	&	1.63[1.09--2.34] & 2.5[1.7--3.6] &
  7840  & 2007-02-13 & ACIS-S & 4.9 & \\[-1pt]
  &&&&
  1.06[0.51--1.90]	&	0.45[0.21--0.81] & 0.7[0.3--1.3] &
  7837 &  2007-11-13 & ACIS-S & 5.1 &   \\[-1pt] 
  &&&& 3.61[2.49--5.02]	&	1.54[1.06--2.13] & 1.9[1.3--2.6] &
  19429  & 2017-02-09 & ACIS-S & 9.9 & \\[-1pt]
NGC\,4412  &       3.6[1.8--6.6]
  &  12 26 35.16 & $+$03 58 15.0 & 0.85[0.36--1.62]  & 1.67[0.70--3.20]  &  2.4[1.1--4.5] &
  19418 & 2017-11-23 & ACIS-S  &  8.0 & D  \\[-1pt]
%
NGC\,4411B  &       2.4[1.7--3.6]
  & 12 26 48.76 & $+$08 53 38.1  & 1.68[0.96--2.70]	&	1.27[0.72--2.05]  & 2.0[1.1--3.2] &
  7840  & 2007-02-13 & ACIS-S & 4.9 & \\[-1pt]
  &&&& 2.84[1.98--3.91]	&	2.12[1.50--2.01] & 2.8[2.0--3.9] &
  19429  & 2017-02-09 & ACIS-S & 9.9 & \\[-1pt]
  &
  & 12 26 49.48 & $+$08 52 18.9 & 2.83[1.85--4.10]	&	2.14[1.40--3.10]&  3.4[2.2--4.8] & 
  7840  & 2007-02-13 & ACIS-S & 4.9 & \\[-1pt]
  &&&& 3.80[2.79--5.03]	&	2.88[2.08--3.74] & 5.5[2.8--29.3] &
  19429  & 2017-02-09 & ACIS-S & 9.9 & \\[-1pt]
NGC\,4407  &      2.7[0.9--4.9]
  & --  & --  & --  & --  & -- &
  19434 & 2018-05-10 & ACIS-S & 7.8 &  \\[-1pt] 
NGC\,4416  &  $<$1
  &  -- & --  & --  & --  & -- &
  19436 & 2018-05-06 & ACIS-S & 7.8 &  \\[-1pt]    
NGC\,4419  &       4.3[3.2--6.5]
  &  -- &  -- & --  & --  & -- &
  19394 & 2016-12-24 & ACIS-S & 9.9 &  \\[-1pt]    
NGC\,4424  &      0.9[0.5--1.5]
  & --  & --  & --  & --  & -- &
  19408 & 2017-04-17 & ACIS-S & 14.9 &  \\[-1pt]    
NGC\,4429  &       4.1[2.0--6.3]
  & --  &  -- & --  & --  & -- &
  19430 & 2018-05-10 & ACIS-S & 7.8 &  \\[-1pt] 
NGC\,4430  &       2.6[1.9--3.7]
  & 12 27 27.09 & $+$06 15 53.9  & 5.53[4.04--7.35]	&	2.10[1.53--2.79]  & 2.9[2.1--3.8] &
  19426 & 2018-05-05  & ACIS-S & 7.8 &  \\[-1pt] 
NGC\,4438  &       3.9[2.9--5.1]
  & --  & --  & --  & --  & -- &
  2883 & 2002-01-29 & ACIS-S & 25.1 &  \\[-1pt] 
  &&&&&&&
  8042 & 2008-02-11 & ACIS-S & 4.9 &  \\[-1pt] 
  &&&&&&&
  21376 & 2020-03-20 & ACIS-S & 29.7 &  \\[-1pt] 
  &&&&&&&
  23189 & 2020-03-20 & ACIS-S & 19.8 &  \\[-1pt] 
  &&&&&&&
  23037 & 2020-03-27 & ACIS-S & 19.8 &  \\[-1pt] 
  &&&&&&&
  23200 & 2020-03-28 & ACIS-S & 25.7 &  \\[-1pt] 
NGC\,4445  &      1.0[0.4--2.2]
  & 12 28 16.38 & $+$09 26 08.4  &  2.86[2.12--3.76]	&	1.48[1.09--1.94] &  1.9[1.4--2.5] &
  19433 &  2018-05-13 & ACIS-S & 14.9 &   \\[-1pt] 
%
NGC\,4450  &       8.4[5.6--11.3]
  & 12 28 31.37 & $+$17 04 26.8  &  3.68[2.58--5.04]	&	1.30[0.92--1.79] & 1.8[1.3--2.5] &
  19399 & 2017-04-06  & ACIS-S & 8.0 &   \\[-1pt] 
NGC\,4451  &       1.3[0.7--2.1]
  & --  & --  & --  & --  & -- &
  19412 & 2018-05-13 & ACIS-S & 14.7 &  \\[-1pt] 
IC\,3392  &       $<$0.6
  & --  & --  & --  & --  & -- &
  19421 & 2018-04-11 & ACIS-S  & 7.6 & \\[-1pt] 
NGC\,4457  &      0.7[0.4--0.9]
  & --  & --  & --  & --  & -- &
  3150 & 2002-12-03 & ACIS-S & 38.9 &  \\[-1pt]
NGC\,4469  &       1.1[0.6--1.6]
  &  -- & --  & --  & --  & -- &
  19415 & 2018-06-23 & ACIS-S & 14.9 &  \\[-1pt] 
NGC\,4470  &     1.0[0.8--1.2]
  &  12 29 38.03 & $+$07 49 32.3 &  3.11[2.46--3.78] &  1.30[1.02--1.58]  & 2.0[1.6--2.4] &
  12978 & 2010-11-20  & ACIS-S & 19.8 &   \\[-1pt] 
  &&&& 4.23[3.12--5.35] & 1.74[1.28--2.20] & 2.7[1.9--3.5] &
  15756 & 2014-04-16  & ACIS-I & 32.1 & \\[-1pt]
  &&&& 2.57[1.84--3.44] & 1.06[0.76--1.42] & 1.5[1.1--1.9] &
  15760 & 2014-04-26  & ACIS-I & 29.4 &  \\[-1pt]
  &&&& 1.86[1.35--2.47]  & 0.76[0.56--1.02] & 1.1[0.8--1.5] &
  16260 & 2014-08-04 & ACIS-S & 24.7 & \\[-1pt]
  &&&& 2.10[1.57--2.71]  & 0.87[0.65--1.11] & 1.2[0.9--1.6] &
  16261 & 2015-02-24 & ACIS-S & 22.8 & \\[-1pt]
  &&&& 2.27[1.51--3.24]  & 0.94[0.62--1.34] & 1.4[1.0--1.8] &
  16262 & 2016-04-30T & ACIS-S & 24.7 & \\[-1pt]
  &&&& 2.78[2.11--3.56]	&	1.15[0.87--1.46] & 1.4[1.1--1.8] &
  21647 & 2019-04-17  & ACIS-S & 29.7 &   \\[-1pt] 
NGC\,4480  &       11.0[6.8--16.6]
  & 12 30 27.86 & $+$04 14 59.9  & 1.14[5.89--1.96]	&	3.08[1.60--5.30]  & 4.3[2.2--7.4] & 
  19423 & 2016-12-06  & ACIS-S & 8.0 & D  \\[-1pt] 
    &
    & 12 30 28.14 & $+$04 14 40.9 & 4.02[2.89--5.42] &		10.9[7.9--14.7] & 13.0[8.3--35.4] &
  19423 & 2016-12-06  & ACIS-S & 8.0 & D  \\[-1pt] 
NGC\,4492  &       3.7[2.6--5.2]
  & 12 30 57.83 & $+$08 04 35.2 & 0.83[0.33--1.65]	&	0.46[0.18--0.93] & 0.6[0.2--1.3] &
  7845 & 2007-02-22 & ACIS-S & 4.9 & E\\[-1pt]
  &&&& 7.01[6.05--7.99]	&	3.94[3.40--4.50] & 8.3[6.2--18.1] &
  15759 & 2014-04-25 & ACIS-I & 29.7 & E \\[-1pt]
  &
  & 12 30 58.88 & $+$08 04 52.2  &  4.18[2.98--5.67] &	2.35[1.68--3.18] & 3.7[2.6--5.0] &
  7845 & 2007-02-22 & ACIS-S & 4.9 & \\[-1pt]
  &&&& 2.54[1.94--3.15]	&	1.43[1.10--1.77] & 2.0[1.5--2.4] &
  15759 & 2014-04-25 & ACIS-I & 29.7 & \\[-1pt]
NGC\,4496A  &     17.1[12.0--23.3]
  & 12 31 38.08 & $+$03 56 42.2  &  35.5[29.3--41.8] &		13.4[10.9--15.8] & 16.6[13.0--21.8] &
  16995 & 2015-07-26  & ACIS-S & 2.7 &   \\[-1pt] 
%
NGC\,4498  &      0.9[0.5--1.4]
  &  -- &  -- & --  & --  & -- & 
  19422 & 2018-04-07 & ACIS-S & 8.1 &  \\[-1pt] 
NGC\,4501  &       12.7[9.5--15.9]$^*$
  & 12 32 00.39 & $+$14 24 42.4  & 8.05[7.09--9.02]	&	3.61[3.18--4.04]  &  12.9[6.1--35.4] &
  2922 & 2002-12-09  & ACIS-S & 17.9 &   \\[-1pt] 
  &
  & 12 32 00.94 & $+$14 25 02.6 & 9.16[8.15--10.2]	&	4.10[3.64--4.59] &  6.0[4.9--8.1] &
  2922 & 2002-12-09  & ACIS-S & 17.9 &   \\[-1pt] 
  &
  & 12 32 06.19 & $+$14 23 21.5	& 2.36[1.76--3.05]	&	1.06[0.79--1.36]&  1.6[1.2--2.1] &
  2922 & 2002-12-09  & ACIS-S & 17.9 &   \\[-1pt] 
NGC\,4519  &       1.9[0.9--2.9]
  &  -- &  -- & --  & --  & -- &   
  19411 & 2018-05-05 & ACIS-S & 8.5 &  \\[-1pt]
%
NGC\,4522  &      $<$0.9
  & --  &  -- & --  & --  & -- &
  19428 & 2018-05-05 & ACIS-S & 7.8 &  \\[-1pt]

  %
  %
%
%
%
\hline\\[-5pt]
\end{tabular} 
\end{center}
\begin{flushleft} 
\end{flushleft}
}
\end{table*}

\begin{table*}
\contcaption{Host-galaxy X-ray binary luminosities and ULX census}
\vspace{-0.2cm}
\scriptsize{
\begin{center}
\begin{tabular}{lccccccccccc}  
\hline \hline\\[-5pt]    
Galaxy  & $L_{{\rm {XRB,}}0.5-8}$ &   R.A.(J2000) & Dec.(J2000) & ULX $F_{0.5-8}$ 
& ULX $L_{0.3-10}$ & ULX $L^c_{0.3-10}$ & ObsID & Obs.~Date  & Detector & Exp.~Time & Notes \\[4pt]
   & ($10^{39}$ erg s$^{-1}$)  & &  & ($10^{-14}$ erg cm$^{-2}$ s$^{-1}$) 
   & ($10^{39}$ erg s$^{-1}$)  & ($10^{39}$ erg s$^{-1}$)  & &  &  & (ks) & \\
   \multicolumn{1}{c}{(1)} & (2) & (3) & (4) & (5) & (6) & (7) & (8) & (9) & (10)&(11) & (12)\\
\hline  \\[-5pt]
NGC\,4527  &       7.5[5.6--9.6]
  & 12 34 10.94 & $+$02 39 25.2  & 24.7[21.6--27.9]	&	7.4[6.5--8.4]  & 21.4[18.1--25.7] &
  4017 &  2003-03-09 & ACIS-S & 4.9 &   \\[-1pt] 
  &&&& 0.43[0.13--0.95]	&	0.13[0.04--0.29] &  0.18[0.06--0.40] &
  19386 & 2016-12-18  & ACIS-S & 9.9 &  \\[-1pt]
  &
  & 12 34 11.43  & $+$02 39 28.9  & 18.0[15.4--20.7] &	5.4[4.6--6.2] & 13.3[11.1--16.1] &
  4017 &  2003-03-09 & ACIS-S & 4.9 &   \\[-1pt] 
  &&&& 3.11[2.20--4.23]	&	0.94[0.66--1.27]  & 1.3[0.9--1.8] &
  19386 & 2016-12-18  & ACIS-S & 9.9 &  \\[-1pt]
  &
  & 12 34 11.76  & $+$02 39 57.7 & $<$0.34 & $<$0.10  & $<$0.17 &
  4017 &  2003-03-09 & ACIS-S & 4.9 &   \\[-1pt] 
  &&&& 3.94[2.91--5.18]	&	1.18[0.87--1.56] & 1.6[1.2--2.1] &
  19386 & 2016-12-18  & ACIS-S & 9.9 &  \\[-1pt]
  &
  & 12 34 14.87  & $+$02 40 05.3  &  $<$0.34 & $<$0.10  & $<$0.17 &
  4017 &  2003-03-09 & ACIS-S & 4.9 &   \\[-1pt] 
  &&&& 5.28[4.08--6.72]	&	1.58[1.22--2.01] & 2.2[1.7--2.7] &
  19386 & 2016-12-18  & ACIS-S & 9.9 &  \\[-1pt]
NGC\,4532  &       11.9[10.6--13.4]
  &  12 34 18.89  & $+$06 28 15.4 &  20.0[17.1--23.0]	& 5.13[4.38--5.91] & 7.3[5.9--10.5] &
  19407 & 2018-05-15  & ACIS-S & 8.0 &   \\[-1pt] 
  &
  & 12 34 20.34  & $+$06 27 37.5  & 11.4[9.22--13.7] &	2.93[2.36--3.51] & 7.3[3.3--27.9] &
  19407 & 2018-05-15  & ACIS-S & 8.0 &   \\[-1pt] 
NGC\,4535  &       3.1[1.2--7.5]
  & 12 34 29.94  & $+$08 10 35.4  & 7.68[6.35--9.02]  &  (2.98[2.47--3.50])  & (4.0[3.3--4.7]) &
  19388 & 2017-04-21 & ACIS-S & 14.9 & F  \\[-1pt] 
NGC\,4536  &       12.0[10.8--13.1]
  & 12 34 26.16  & $+$02 11 22.3  & 9.31[7.89--10.7]	&	3.15[2.67--3.64]  & 5.1[4.1--7.3] &
  19387 & 2017-07-01  & ACIS-S & 14.9 &   \\[-1pt] 
  &
  & 12 34 26.83  & $+$02 11 18.8 &  3.08[2.30--4.02]	&	1.08[0.80--1.40] & 1.4[1.0--1.9] &
  19387 & 2017-07-01  & ACIS-S & 14.9 &   \\[-1pt] 
  &
  & 12 34 31.81 & $+$02 11 21.9  & 3.99[3.05--4.93]	&	1.35[1.04--1.66]  & 1.8[1.4--2.3] &
  19387 & 2017-07-01  & ACIS-S & 14.9 &   \\[-1pt] 
  &
  & 12 34 32.56  &  $+$02 10 54.5  & 	7.89[6.57--9.21]	&	2.67[2.23--3.12] &  3.7[3.1--4.3] &
  19387 & 2017-07-01  & ACIS-S & 14.9 &   \\[-1pt] 
  &
  & 12 34 36.93 & $+$02 08 51.4  & 13.1[11.3--15.0]	&	4.44[3.80--5.06]   &  9.8[5.2--24.0] &
  19387 & 2017-07-01  & ACIS-S & 14.9 &   \\[-1pt] 
  &
  & 12 34 38.81  & $+$02 10 28.5  & 5.57[4.41--6.74]	& 1.89[1.49--2.28]  &  2.5[2.0--3.0] &
  19387 & 2017-07-01  & ACIS-S & 14.9 &   \\[-1pt] 
NGC\,4548  &       2.2[1.7--3.4]
  & 12 35 31.33  & $+$14 30 48.8  &  4.72[3.37--6.37]	&	1.77[1.25--2.38] &  2.4[1.7--3.2] &
  19402 & 2018-04-11  & ACIS-S & 7.6 &   \\[-1pt] 
NGC\,4567  &       9.6[7.2--13.4]
  & 12 36 33.75 & $+$11 16 10.6  &  4.96[3.62--6.59]	&	3.87[2.83--5.14] & (4.9[3.4--7.4]) &
  19396 & 2018-05-10  & ACIS-S & 7.8 & E \\[-1pt] 
NGC\,4568  &       12.1[9.0--15.1]
  & 12 36 32.32  & $+$11 13 06.3  &  3.18[2.10--4.54]	&	1.44[0.95--2.05] & 1.6[1.1--2.1] &
  19389 & 2018-03-01  & ACIS-S & 13.7 &   \\[-1pt]
  &&&& 3.41[1.74--5.82]	&	1.58[0.81--2.72]  & 2.0[1.0--3.5] &
  19396 & 2018-05-10  & ACIS-S & 7.8 &  \\[-1pt] 
  &
  & 12 36 33.04 & $+$11 14 05.6  & 12.1[9.72--14.4]	&	5.43[4.38--6.50] & 6.7[5.2--8.6] &
  19389 & 2018-03-01  & ACIS-S & 13.7 &   \\[-1pt]
  &&&&  17.3[13.6--20.8]	&	7.81[6.14--9.43] & 9.0[7.0--13.6] &
  19396 & 2018-05-10  & ACIS-S & 7.8 &  \\[-1pt] 
  &
  & 12 36 33.22 & $+$11 14 37.1  & 9.87[7.82--11.9]	&	4.45[3.43--5.36] & 4.9[3.9--6.2] &
  19389 & 2018-03-01  & ACIS-S & 13.7 &   \\[-1pt]
  &&&&  1.90[0.92--3.42]	&	0.86[0.42--1.55] & 1.6[1.0--2.3] &
  19396 & 2018-05-10  & ACIS-S & 7.8 &  \\[-1pt] 
  &
  & 12 36 34.48  & $+$11 14 40.2 & 2.20[1.54--3.03]	&	1.03[0.72--1.14]& 1.4[0.9--1.9] &
  19389 & 2018-03-01  & ACIS-S & 13.7 &   \\[-1pt]
  &&&&  1.86[1.08--2.93]	&	0.87[0.51--1.37] & 1.2[0.7--1.8] &
  19396 & 2018-05-10  & ACIS-S & 7.8 &  \\[-1pt] 
NGC\,4569  &       7.4[6.3--8.5]
  & 12 36 53.70  & $+$13 11 54.0  &  4.93[3.03--7.50]	&	1.00[0.61--1.52] & 1.8[1.1--2.7] &
  405 & 2000-02-17  & ACIS-S & 1.7 &   \\[-1pt]
  &&&& 2.53[2.16--2.89]	&	0.51[0.44--0.59] & 0.82[0.70--0.94] &
  5911 & 2005-11-13 & ACIS-S & 39.1 & \\[-1pt]
NGC\,4571  &   2.0[0.3--5.6]
  & --  & --  & --  & --  & -- &
  7858 & 2008-02-14 & ACIS-S & 4.7 &  \\[-1pt]
%
NGC\,4579  &       28.3[26.5--30.2]$^*$
  & 12 37 40.31  & $+$11 47 27.8  & 28.9[27.8--30.1] &	12.6[12.1--13.1]  & 17.1[16.2--18.1] &
  807 & 2000-05-02  & ACIS-S & 33.9 &   \\[-1pt] 
  &
  & 12 37 43.21 &  $+$11 49 01.5  & 3.34[2.53--4.15] &	1.41[1.07--1.82] & 2.4[2.1--2.7] &
  807 & 2000-05-02  & ACIS-S & 33.9 &   \\[-1pt] 
  &
  & 12 37 53.88  & $+$11 50 20.4 & 9.47[8.14--10.8]	&	4.12[3.53--4.74]& 4.0[2.8--4.7] &
  807 & 2000-05-02  & ACIS-S & 33.9 &   \\[-1pt] 
NGC\,4580  &       1.2[0.6--2.1]
  & --  & --  & --  & --  & -- &
  10413 & 2017-12-01 & ACIS-S & 8.0 &  \\[-1pt]  
NGC\,4606  &  1.2[0.4--2.1]
  &  -- & --  & --  & --  & -- &
   19437  & 2017-04-22 & ACIS-S & 7.7 &  \\[-1pt]
%
NGC\,4607  &       3.2[2.4--4.7]
  & 12 41 12.61  & $+$11 53 50.2  &   1.49[0.75--2.60]	&	0.89[0.44--1.54] & 1.2[0.6--2.2] &
  19437  & 2017-04-22 & ACIS-S & 7.7 &  \\[-1pt]
  &&&& 7.98[6.12--9.85]	&	4.74[3.64--5.87] & 10.5[5.5--50.6] &
  19403 &  2018-05-09 & ACIS-S & 8.0 &   \\[-1pt] 
NGC\,4639  &       5.2[3.8--7.8]
  & 12 42 51.20  & $+$13 14 40.3  & 3.52[1.82--6.05]	& 2.58[1.33--4.45]  & 4.8[2.5--16.8] &
  408 &  2000-02-05 &  ACIS-S & 1.4 &   \\[-1pt] 
  &&&& 4.70[3.66--5.75]	&	3.45[2.68--4.22] & 4.5[3.5--5.5] &
  19398 &  2018-04-14 &  ACIS-S & 14.7 &   \\[-1pt] 
NGC\,4647  &       6.9[5.1--8.8]
  &  -- &  -- & --  & --  & -- &
  785 &  2000-04-20 & ACIS-S & 38.1 &  \\[-1pt] 
  &&&&&&&
  8182 & 2007-01-30 & ACIS-S & 52.3 & \\[-1pt]
  &&&&&&&
  8507 & 2007-02-01 & ACIS-S & 17.5  &  \\[-1pt]
  &&&&&&&
  12975 & 2011-08-08 & ACIS-S & 84.9 & \\[-1pt]
  &&&&&&&
  12976 & 2011-02-24 & ACIS-S & 101.0 &  \\[-1pt]
  &&&&&&&
  14328 & 2011-08-12 & ACIS-S  & 14.0  &  \\[-1pt]
NGC\,4654  &       3.9[3.0--4.9]
  &  -- & --  & --  & --  & -- &
  3325 & 2002-05-03 & ACIS-S & 4.9 &  \\[-1pt] 
  &&&&&&&
  19393 & 2016-12-14 & ACIS-S & 9.9 & \\[-1pt]
NGC\,4689  &       5.7[4.0--7.4]
  & 12 47 47.71  & $+$13 46 17.3  & 7.93[6.09--9.79]	&	2.99[2.29--3.68]  & 4.7[3.6--5.8] &
  7865 &  2007-05-07 & ACIS-S  & 4.9 &   \\[-1pt] 
NGC\,4698  &       9.0[6.8--11.5] 
  & 12 48 25.87 &  $+$08 30 20.8  &  14.7[13.8--15.7] &  NA & NA & 
  3008 & 2002-06-16 & ACIS-S & 29.7 & G  \\[-1pt] 
NGC\,4713  &       2.0[1.5--3.4]
  &  --  &  -- & --  & -- & -- &
  4019 &  2003-01-28 & ACIS-S  & 4.9  &     \\[-1pt] 
\hline\\[-5pt]
\end{tabular} 
\end{center}
}
\begin{flushleft} 
{\footnotesize{
Col.~(2): total unabsorbed 0.5--8 keV luminosity of the point-source (X-ray binaries) population (resolved plus unresolved) in each galaxy, estimated with a power-law model fit. The range of values in brackets represents the 90\% confidence range. When a galaxy was observed multiple times, the luminosity was determined from a stack of all the observations (combined with the {\it epicspeccombine} task). An asterisk (on 3 galaxies) denotes that they are larger than the detector field of view in one direction. See Section 3.1 for details. 
Cols.~(3),(4): coordinates of each ULX, defined as each source that exceeded a 0.3--10 keV luminosity of $10^{39}$ erg s$^{-1}$ in at least one of the observations. 
Col.~(5): absorbed 0.5--8 keV flux (with 90\% confidence range) of each ULX in each observation, measured with {\it srcflux}; here, we assumed a power-law model, with fixed photon index $\Gamma = 1.8$ and only Galactic line-of-sight column density. 
Col.~(6): unabsorbed 0.3--10 keV luminosity (with 90\% confidence range) of each ULX in each observation (including those in which they were in a faint state), assuming $\Gamma = 1.8$ and Galactic line-of-sight column density. The conversion from fluxes to luminosities is based on the median redshift-independent NED distances. The error range in the luminosity includes only the flux measurement uncertainty, not the systematic uncertainty in the distance to each galaxy. 
Col.~(7): ``corrected'' 0.3--10 keV ULX luminosity (with 90\% confidence range), based on individual spectral models for the most luminous sources, and, for all the others, adopting the median values of photon index and column density inferred from the modelled subsample ($\Gamma = 1.8$ and total $N_{\rm H} = 3.0 \times 10^{21}$, respectively). We regard these luminosities as our best estimate, and use them for further analysis.
Col.~(12): Notes. A = SN\,1979C; B = likely background galaxy at $z\approx 0.156$; C = likely foreground star; D = NED Tully-Fisher distance possibly overestimated; E = candidate nuclear source of a small (satellite?) galaxy; F = likely quasar; G = known BL Lac at $z \approx 0.43$.
}}
\end{flushleft}
\end{table*}

\begin{figure}
\hspace{-0.5cm}
\includegraphics[height=0.49\textwidth, angle=270]{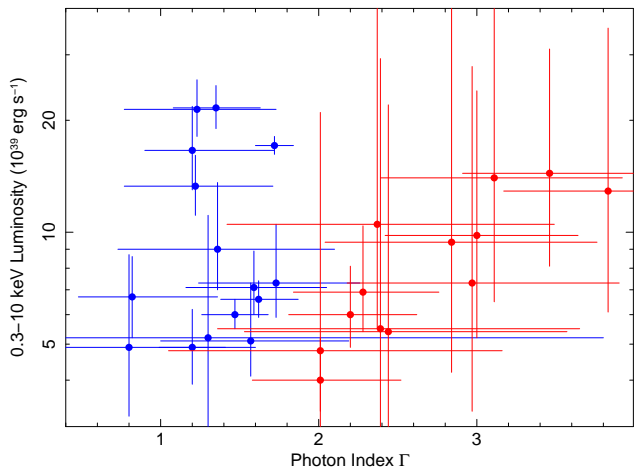}
 \caption{Unabsorbed 0.3--10 keV luminosity of the most luminous ULXs with individual spectral modelling (Table 3) as a function of their spectral hardness (power-law photon index). Blue datapoints correspond to ULXs in the hard ultraluminous state ($\Gamma < 2$); red datapoints correspond to the soft ultraluminous state ($\Gamma > 2$). We omitted from the plot the unusual thermal source (young SN candidate) for which we cannot define a photon index (Section 4.2.2). }
  \label{hardness}
\end{figure}

\subsection{ULX population}

\subsubsection{Spectral modelling for the ULXs}
In the first step of our ULX search (Section 3.2), we had assumed a fixed spectral model (power-law with photon index $\Gamma = 1.8$) and Galactic line-of-sight column density to each host galaxy (typically, $N_{\rm H} \approx 2$--$3 \times 10^{20}$ cm$^{-2}$; values from \citealt{hi4pi16} and \citealt{kalberla05}). Under those two assumptions, we identified 85 off-nuclear sources that exceed a 0.3--10 keV unabsorbed luminosity of $10^{39}$ erg s$^{-1}$ in at least one observation. We report their positions, observed 0.5--8 keV fluxes and unabsorbed 0.3--10 keV luminosities in Table 2, including the fluxes and luminosities in the observations in which the same sources were below the ULX threshold. Table 2 provides a useful working list of interesting point-like sources; however, all luminosity values are likely to be under-estimated, because we have not yet accounted for the loss of X-ray photons due to intrinsic absorption in the host galaxies and around the compact objects. We will now assess the effect of intrinsic absorption, and then improve on our ULX catalogue by using individual spectral modelling at least for the brightest sources.

We obtained reliable fits with free photon index and column density for all the sources with unabsorbed luminosities $\gtrsim$4 $\times 10^{39}$ erg s$^{-1}$ (Table 3). Instead, the majority of sources with $L_{\rm X} \sim$ 1--4 $\times 10^{39}$ erg s$^{-1}$ do not have enough counts below 1 keV to constrain $N_{\rm H}$ and $\Gamma$ (and therefore the unabsorbed luminosity) independently. This is particularly the case for sources observed in later {\it Chandra} cycles, with a much degraded ACIS sensitivity in the soft band. Thus, we limit our spectral analysis to the most luminous ULXs.
At the signal-to-noise available for the majority of our sources, an absorbed power-law model provides an adequate fit, with plausible photon indices $1 \lesssim \Gamma \lesssim 3$. The only exception is a ULX (CXOU\,J122541.70$+$071338.9) in the disk plane of the perfectly edge-on galaxy IC\,3322A, which appears dominated by thermal plasma emission, as we discuss later (Section 4.2.2).

Based on our spectral modelling results, we found at least 25 ULXs that reach a 0.3--10 keV luminosity of $\gtrsim$4 $\times 10^{39}$ erg s$^{-1}$ at least in one observation (Table 3); 9 of them reached a luminosity of $10^{40}$ erg s$^{-1}$. In addition, a 10th ULX exceeded $10^{40}$ erg s$^{-1}$, in NGC\,4480, if its Tully-Fisher distance of 42 Mpc is correct; however, in that case, NGC\,4480 would most likely not be a member of the Virgo cluster. Thus, we omitted that source from subsequent plots and modelling of the cumulative and differential luminosity distributions. For consistency, we also subtracted the SFR and stellar mass of NGC\,4480 when we calculated the predicted contribution of HMXBs and LMXBs, and subtracted the projected area of its $D_{25}$ when we calculated the predicted contribution of foreground and background sources. Moreover, Table 3 includes three sources whose identification as stellar-mass accreting compact objects is less clear: the peculiar thermal-plasma source in IC\,3322A (Section 4.2.2), and two sources that may be the nuclei of smaller galaxies behind or near NGC\,4492 and NGC\,4567 (Section 4.2.3).

We examined the hardness distribution as a function of unabsorbed luminosity (Figure 12), for 28 individual observations (as listed in Table 3) of 25 different ULXs above a luminosity of 4 $\times 10^{39}$ erg s$^{-1}$. We omitted from that plot the four above-mentioned sources with less clear identification, in the $D_{25}$ of IC\,3322A, NGC\,4480, NGC\,4492 and NGC\,4567. In 15 out of 28 cases, the sources can be classified in the hard ultraluminous state (using the classification scheme of \citealt{sutton13}), with a photon index $\Gamma < 2$ (which means that the spectral luminosity $EL_E$ rises with energy, in the {\it Chandra} band), and the other 13 are in the soft ultraluminous state, with $\Gamma > 2$. The hardness-luminosity distribution (Figure 12) is also consistent with the alternative ULX classification of \cite{pintore14}, into group 1 (harder) and group 2 (softer) sources. In both ULX classification schemes, the difference between between harder and softer sources is mostly attributed to the scattering optical depth of the super-critical disk outflow along our line of sight. In addition, a harder ULX spectrum may be the signature of a neutron star rather than BH accretor \citep{pintore17,walton18}.

In our sample, there is no statistical difference (at least below $10^{40}$ erg s$^{-1}$) between the luminosity of hard and soft sources, although the four most luminous sources (with $L_{\rm X} > 1.5 \times 10^{40}$ erg s$^{-1}$) are all in the hard state. The median photon index is $\approx$1.8, which also justifies our initial assumption of the photon index. The median total absorption column density is $\approx$3.0 $\times 10^{21}$ cm$^{-2}$, that is a factor of 10 higher than what we had assumed in our preliminary luminosity estimates based on {\it srcflux} results (Col.~6 in Table 2). 

To correct (at least on a statistical level) the likely under-estimation of the preliminary luminosity estimates for the sources that are too faint for individual spectral modelling, we re-estimated all their fluxes again with {\it srcflux}, and $\Gamma = 1.8$, but this time with the assumption of $N_{\rm{H,tot}} = 3.0 \times 10^{21}$ cm$^{-2}$, for all the sources not already included in Table 3. In summary, the list of ``corrected'' 0.3--10 keV luminosities provided in Table 2, Col.~(7), is our best estimate, combining the spectral fitting results for the most luminous sources and the revised {\it srcflux} estimates for all the others. On average, the corrected unabsorbed luminosities for sources observed in earlier {\it Chandra} cycles (before the soft-energy degradation) are a factor of 1.6--1.7 times higher than estimated when only Galactic line-of-sight absorption is assumed. In Cycle 18, the difference between $N_{\rm {H}} = 3.0 \times 10^{21}$ cm$^{-2}$ and $N_{\rm {H}} = 3.0 \times 10^{20}$ cm$^{-2}$ was reduced to a factor of $\approx$1.3.

\subsubsection{Candidate Type IIn SN in IC\,3322A}
CXOU\,J122541.70$+$071338.9, detected in IC\,3322A (distance of $\approx$25 Mpc) on 2018 April 24, is the most luminous ULX in the whole sample. Its X-ray spectrum (Table 3) suggests multi-temperature thermal plasma emission and an intrinsic column density of $\approx$2 $\times 10^{22}$ erg s$^{-1}$, for an intrinsic luminosity of $\approx$6 $\times 10^{40}$ erg s$^{-1}$. A spectrum dominated by X-ray thermal plasma at such a high luminosity is very unusual for a ULX. The X-ray spectrum is consistent with that of a young SN; more specifically, of a Type IIn, which explodes in a denser circumstellar medium. Based on the Supernova X-ray Database online catalog\footnote{https://kronos.uchicago.edu/snax/} \citep{ross17}, there are at least five core-collpase SNe (SN 1998Z, SN 1995N, SN 2005ip, SN 2005kd, SN 2006jd and SN 2010jl), detected at X-ray luminosities of a few $10^{40}$ erg s$^{-1}$, even up to a few $10^{41}$ erg s$^{-1}$, for several years after the explosion. See also \cite{chandra12,chandra15} and \cite{chandra18} for detailed discussions of the X-ray and multi-band properties of Type IIn SNe. The possibility that some of the brightest ULXs may be young SNe in a dense environment was discussed in other cases, for example for the brightest ULX in the Cartwheel galaxy \citep{pizzolato10}.

No core-collapse SNe have ever been reported in IC\,3322A, but that is not strong evidence against our suggested interpretation. The peak absolute optical brightness of Type IIn SNe is broadly distributed between about $-17$ mag and $-21$ mag, with a median value around $-19$ mag \citep{nyholm20}. The distance modulus of IC\,3322A is $\approx$32 mag. To that, we may add 10 mag of visual extinction, based on the best-fitting value of $N_{\rm H} \approx 2 \times 10^{22}$ cm$^{-2}$ \citep{predehl95,guver09,willingale13}. Thus, it is likely that the presumed SN was fainter than 20 mag in the optical, and would have been easily missed by large-area sky searches.

We used the Open SN Catalog\footnote{https://sne.space/} catalogs \citep{guillochon17} to check whether any of the point-like sources (even those fainter than $10^{39}$ erg s$^{-1}$, not listed in Table 2) may correspond to a historical SN. There are 39 optically-identified SNe reported to occur in the 75 sample galaxies prior to at least one of the {\it Chandra} observations used for this study. Only three of them are detected in X-rays (Table 4). The most luminous of them is SN 1979C (a Type II L) in NGC\,4321, for which we infer an X-ray luminosity of $\approx$6 $\times 10^{38}$ erg s$^{-1}$ when its spectrum (stacked over the six {\it Chandra} observations of this galaxy) is fitted with a two-temperature thermal plasma model. The best-fitting temperatures are $kT_1 = 0.27^{+0.20}_{-0.08}$ keV and $kT_2 = 1.3^{+0.2}_{-0.2}$ keV, and the best-fitting intrinsic column density is  $\left(5.8^{+2.7}_{-2.9}\right)\times 10^{21}$ cm$^{-2}$.

\begin{figure}
\hspace{-0.5cm}
\includegraphics[height=0.49\textwidth, angle=270]{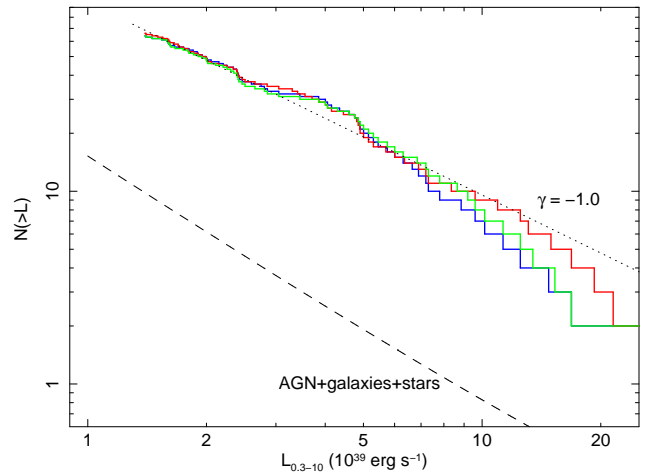}
 \caption{Cumulative LF in the 0.3--10 keV band. See Col.~(7), Table 2 for the data values. The datapoints include a subsample of ULXs modelled with free $N_{\rm H}$ and $\Gamma$ (Table 3), while for the fainter sources we fixed $\Gamma = 1.8$ and $N_{\rm{H,tot}} = 3.0 \times 10^{21}$ cm$^{-2}$. The red histogram represents the luminosities inferred from the earliest exposure available for each galaxy with multiple observations; the green histogram shows the luminosities inferred from the latest available exposures for each galaxy; the blue histogram is for the longest available exposures. The small excess at the higher end of the earliest-look LF is the result of small-number statistics (namely, two ULXs with $L_{\rm X} > 10^{40}$ erg s$^{-1}$ in NGC\,4527, detected only in the first observation of that galaxy). For graphical purposes only, the most luminous source (candidate SN) in the Virgo sample (Section 4.2.2) is falling outside the plotted region, at $L_{\rm X} \approx 6 \times 10^{40}$ erg s$^{-1}$. The black dashed line is the predicted contribution from the cosmic background plus foreground stars \citep{lehmer12}. The black dotted line shows that the lower end of the cumulative LF is consistent with a power-law of index $-1$, with a downturn at the high-luminosity end.}
  \label{cumullumf}
\end{figure}

\begin{figure}
\hspace{-0.5cm}
\includegraphics[height=0.49\textwidth, angle=270]{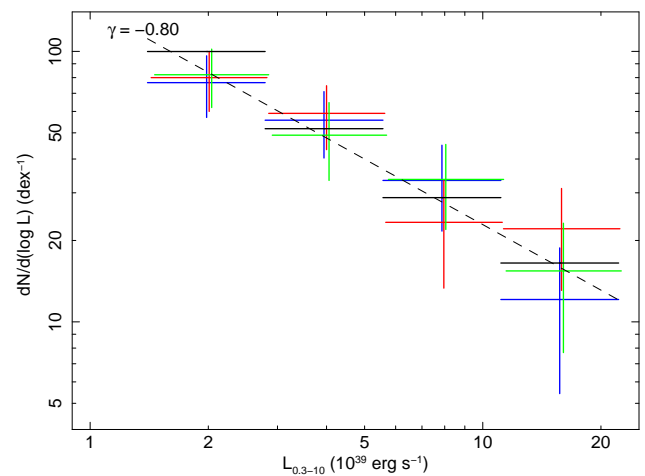}
 \caption{Differential LF in the 0.3--10 keV band. For each of the four luminosity bins, of equal width $\Delta \log \left(L_{\rm X}/{\rm {erg~s}}^{-1}\right) = 0.3$, the colours are defined as in Figure 13 (red, green and blue for the earliest, latest and longest observation, respectively). The relative position of the three coloured datapoints in each bin is offset by a very small amount along the X axis for visual clarity.
 The predicted background and foreground contributions have already been subtracted from each luminosity bin. The dashed black line is an indicative power-law model, with the form $dN/d(\log L) \propto L_{39}^{\gamma}$ with $\gamma = -0.80$; values of $\gamma \approx -0.8 \pm 0.2$ are consistent with the observed datapoints. 
 For each luminosity bin, a black line marks the value predicted for that bin by the LF models of \citet{lehmer19} (LMXBs plus HMXBs).}
  \label{difflumf}
\end{figure}

\subsubsection{Likely interlopers from other projected galaxies}
Two bright sources (Table 3) are associated with the nuclei of (apparently) small galaxies projected within the $D_{25}$ of a larger spiral (NGC\,4492 and NGC\,4567). We found no measurements for the redshifts of the two small companion galaxies. If the host galaxies are dwarf satellites at approximately the same distance as NGC\,4492 and NGC\,4567, respectively, the two X-ray sources would have a luminosity of several times $10^{39}$ erg s$^{-1}$ (Table 3). However, they would not be classified as ULXs in the most common definition of this class, because they are nuclear sources. They would still be interesting as candidate intermediate-mass BHs in the nuclei of dwarf satellites. Instead, the bright source projected inside the $D_{25}$ of NGC\,4698 (Tables 2, 3), known in the literature as XMMU J124825.9$+$083020 or WISE J124825.85$+$083020.4, is a BL Lac at redshift 0.43, confirmed by optical spectroscopy \citep{foschini02}. 

Another curious source (Table 3) is CXOU J122538.98$+$124000.8, projected inside the $D_{25}$ of NGC\,4388, but also projected onto the spiral arm (not the nucleus) of an anonymous background galaxy at redshift $z \approx 0.156$, seen behind the Virgo galaxy. If this transient source belongs to NGC\,4388, it is a run-of-the-mill ULX with $L_{\rm X} \approx 1.5 \times 10^{39}$ erg s$^{-1}$; instead, if it belongs to the background spiral, it is a hyperluminous X-ray source with $L_{\rm X} \approx 2.6 \times 10^{42}$ erg s$^{-1}$ (in the 2001 June 8 observation), a strong intermediate-mass BH candidate. In a subsequent observation of the same galaxy a decade later (2011 December 7), the source was no longer detected (Table 2), which implies a decline of a factor of at least 40.

\begin{figure}
\hspace{-0.5cm}
\includegraphics[height=0.49\textwidth, angle=270]{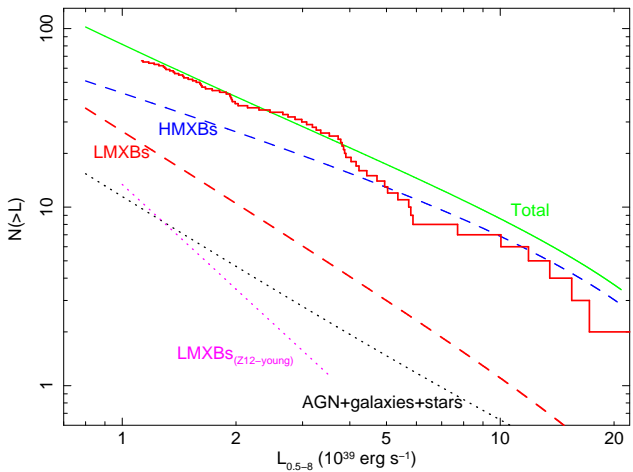}
 \caption{Comparison between the empirical cumulative LF from our survey, and the model components from \citet{lehmer19}. Here, we have plotted all functions in the 0.5--8 keV band, to facilitate the comparison. Conversions between 0.3--10 keV and 0.5--8 keV bands were done according to the best-fitting spectral models of the sources. Red histogram: cumulative ULX LF for the Virgo spiral sample (here we chose the distribution derived from the earliest observations). Dashed blue line: predicted contribution from HMXBs, for a star formation rate of $70 M_{\odot}$ yr$^{-1}$ (similar to the SFR estimated by B15). Dashed red line: predicted contribution from LMXBs in spiral disks, for a stellar mass of $1.5 \times 10^{12} M_{\odot}$. Dotted black line: predicted contribution from background galaxies, AGN, and foreground stars. Solid green line: sum of the previous three terms (HMXBs, LMXBs and bg/fg sources), which represents the expected observed luminosity function from our Virgo sample. Dotted magenta line: predicted contribution from LMXBs also for a stellar mass of $1.5 \times 10^{12} M_{\odot}$, but calibrated on the younger population of elliptical galaxies \citep{zhang12}. For graphical purposes only, the most luminous source (candidate SN) in the Virgo sample (Section 4.2.2) is falling outside the plotted region, at $L_{\rm X} \approx 6 \times 10^{40}$ erg s$^{-1}$. }
  \label{lumf_models}
\end{figure}

\begin{table*}
\caption{Spectral parameters of the most luminous ULXs in Virgo ($L_{\rm X} \gtrsim 4 \times 10^{39}$ erg s$^{-1}$)}
\vspace{-0.4cm}
\scriptsize{
\begin{center}
\begin{tabular}{lccccccccccc}  
\hline \hline\\[-5pt]    
Host Galaxy  &    R.A.(J2000) & Dec.(J2000) & ObsID & Obs.~Date & 
$N_{\rm {H,Gal}}$ & $N_{\rm {H,int}}$ & $\Gamma$ & $N_{\rm{pl}}^a$ & C-stat/dof &
$F_{0.5-8}^b$ & $L_{0.3-10}^c$  \\
   &   & & & & 
   ($10^{22}$ cm$^{-2}$) & ($10^{22}$ cm$^{-2}$) &  & & &
   ($10^{-14}$ erg cm$^{-2}$ s$^{-1}$) & ($10^{39}$ erg s$^{-1}$)  \\
\hline  \\[-6pt]
\multicolumn{12}{c}{Most likely ULXs} \\[4pt]
NGC\,4178  &  12 12 44.51  & $+$10 51 13.6  &  12748  & 2011-02-19 &
    [0.026]  & $0.24^{+0.09}_{-0.09}$ & $1.47^{+0.21}_{-0.21}$ &
    $3.3^{+0.9}_{-0.7}$  &  $206.9/235$  &
    $19.7^{+2.2}_{-2.0}$  &  $6.0^{+0.6}_{-0.5}$ \\[4pt]
NGC\,4197  &   12 14 40.29 &  $+$05 49 00.0  &  19420  & 2017-07-26  & 
    [0.015] & $0.23^{+1.21}_{-0.23}$ & $1.3^{+2.5}_{-\ast}$ &
    $0.6^{+5.8}_{-0.4}$ & $11.5/14$  &
    $4.3^{+5.8}_{-2.6}$  &  $5.2^{+5.9}_{-2.4}$  \\[4pt]
%
    & 12 14 41.31 & $+$05 49 03.9 &  19420  & 2017-07-26  & 
    [0.015]  &  $<$0.8  &  $0.8^{+0.8}_{-0.6}$ & 
    $0.30^{+0.44}_{-0.16}$  & $30.8/17$ &
    $4.6^{+2.6}_{-2.0}$  &  $4.9^{+3.8}_{-1.7}$  \\[4pt]
%
NGC\,4212   &  12 15 39.59  &  $+$13 53 42.2  &  19395  & 2017-02-14 &
    [0.026]   &  $0.19^{+0.25}_{-0.19}$ & $1.59^{+0.46}_{-0.43}$ &
    $2.7^{+1.8}_{-1.0}$    &  $92.3/104$  &
    $14.0^{+3.1}_{-2.5}$   &  $7.1^{+1.8}_{-1.1}$  \\[4pt]
NGC\,4216 &  12 15 50.12  & $+$13 06 18.3 & 19391 & 2017-07-24 & 
   [0.028]  & $0.81^{+1.13}_{-0.77}$  &   $2.01^{+1.15}_{-0.96}$  &
   $3.0^{+10.2}_{-2.2}$  & $26.8/36$ &
   $7.4^{+3.0}_{-3.1}$  &  $4.8^{+16.2}_{-2.0}$  \\[4pt]
%
NGC\,4254 &  12 18 56.10  & $+$14 24 19.3  &  7863 & 2007-11-21 &
    [0.026] & $0.10^{+0.10}_{-0.09}$  & $1.35^{+0.28}_{-0.27}$   & 
    $8.8^{+3.1}_{-2.2}$  & $137.4/170$ &
    $64.3^{+10.3}_{-8.8}$  & $21.6^{+3.2}_{-2.6}$ \\[4pt]
    &   &    & 17462   & 2015-02-16   &
    [0.026]  & $0.27^{+0.16}_{-0.14}$   & $1.62^{+0.25}_{-0.24}$  &
    $3.6^{+1.3}_{-0.9}$   & $185.4/237$  &
    $17.3^{+1.9}_{-1.7}$  &   $6.6^{+0.8}_{-0.7}$ \\[4pt]
NGC\,4316  & 12 22 39.54  & $+$09 20 21.0 & 19400  & 2018-04-20 &
    [0.014]  &   $0.27^{+0.53}_{-0.27}$ & $2.44^{+1.13}_{-0.91}$ &
    $1.3^{+2.7}_{-0.8}$  & $22.9/34$  &
    $2.9^{+1.3}_{-0.9}$  &  $5.4^{+16.6}_{-2.4}$ \\[4pt]
NGC\,4321  & 12 22 54.14 &  $+$15 49 12.6 &  9121  & 2008-04-20 &
    [0.018] & $0.28^{+0.12}_{-0.11}$  & $3.46^{+0.61}_{-0.55}$  &
    $7.3^{+3.6}_{-2.3}$  &  $101.5/112$ &
    $8.8^{+1.2}_{-1.1}$    &  $14.4_{-6.3}^{+16.7}$    \\[4pt]
    &   &   & 23141 &  2020-03-13 &
    [0.018]  & $0.32^{+0.33}_{-0.27}$ & $3.11^{+0.81}_{-0.72}$ &
    $8.4^{+9.1}_{-4.1}$   &  $54.7/75$ &
    $10.9^{+3.0}_{-2.2}$ & $14.0^{+30.6}_{-7.5}$   \\[4pt]
    &  12 22 54.78 &  $+$15 49 16.3 &  9121  & 2008-04-20 &
    [0.018]  &  $0.31^{+0.16}_{-0.14}$  &  $2.28^{+0.48}_{-0.44}$  &
    $4.5^{+2.4}_{-1.5}$  &  $87.2/115$  &
    $10.6^{+2.3}_{-1.8}$  & $6.9^{+3.5}_{-1.5}$  \\[4pt]
IC\,3322A  &  12 25 41.70 & $+$07 13 38.9 & 19401   &  2018-04-24  &
    [0.017]  & $2.10^{+1.27}_{-0.74}$   &  
    \multicolumn{2}{c}{(NA--thermal source)$^d$}   & $30.9/46$ &
    $3.9^{+1.3}_{-0.9}$   &  $62.2^{+69.8}_{-38.8}$ \\[4pt]
NGC\,4396  &  12 26 01.81 & $+$15 39 41.8 & 19417  & 2018-05-25 &
    [0.020]  &  $0.73^{+0.61}_{-0.48}$  &   $2.84^{+0.92}_{-0.80}$ &
    $9.6^{+17.3}_{-5.7}$  &  $55.0/61$  &  
    $10.2^{+2.8}_{-2.2}$  & $9.4^{+31.0}_{-5.2}$ \\[4pt]
NGC\,4411B  &  12 26 49.48 &  $+$08 52 18.9 & 19429 & 2017-02-09 &
    [0.013] & $0.54^{+0.72}_{-0.51}$  & $2.39^{+1.26}_{-1.03}$ &
    $1.9^{+5.2}_{-1.3}$  &  $31.6/28$ &
    $3.5^{+1.7}_{-1.2}$ &   $5.5^{+23.8}_{-2.7}$  \\[4pt]
%
NGC\,4496A &  12 31 38.08  & $+$03 56 42.2 &  16995 & 2015-07-26 &
    [0.023] & $<0.20$  & $1.20^{+0.52}_{-0.30}$   & 
    $4.8^{+3.5}_{-1.1}$  & $61.4/80$ &
    $44.2^{+11.8}_{-10.3}$  & $16.6^{+5.2}_{-3.6}$ \\[4pt]
NGC\,4501  & 12 32 00.39  &  $+$14 24 42.4  & 2922 & 2002-12-09 &
    [0.027]  & $0.31^{+0.14}_{-0.12}$  &   $3.83^{+0.77}_{-0.66}$ &
    $4.7^{+2.7}_{-1.6}$  &   $91.5/91$ &
    $4.7^{+0.7}_{-0.7}$  &  $12.9^{+22.5}_{-6.8}$ \\[4pt]
    &  12 32 00.94  & $+$14 25 02.6 & 2922 & 2002-12-09 &
    [0.027]  & $0.26^{+0.12}_{-0.11}$ &  $2.20^{+0.42}_{-0.39}$ &
    $3.3^{+1.5}_{-1.0}$  &  $130.2/127$  &
    $8.9^{+1.8}_{-1.5}$ & $6.0^{+2.1}_{-1.1}$ \\[4pt]
NGC\,4527  &  12 34 10.94  &  $+$02 39 25.2  &  4017 & 2003-03-09 &
    [0.020]  &   $0.84^{+0.46}_{-0.35}$   &  $1.23^{+0.50}_{-0.46}$ &
    $7.9^{+7.6}_{-3.6}$  &     $99.1/126$ &
    $55.3^{+22.1}_{-19.9}$ &  $21.4^{+4.3}_{-3.3}$\\[4pt]
          &  12 34 11.43 &  $+$02 39 28.9  &  4017 & 2003-03-09 &
    [0.020]  & $0.55^{+0.35}_{-0.27}$  &  $1.22^{+0.49}_{-0.45}$ &
    $4.9^{+4.2}_{-2.1}$  &   $71.0/99$  &
    $36.4^{+8.8}_{-7.2}$ &  $13.3^{+2.8}_{-2.2}$\\[4pt]
NGC\,4532  &  12 34 18.89  & $+$06 28 15.4  & 19407 & 2018-05-15 &
    [0.016]  & $0.40^{+0.37}_{-0.30}$ &  $1.73^{+0.53}_{-0.49}$ &
    $5.4^{+4.8}_{-2.4}$ & $89.3/99$ &
    $21.7^{+4.9}_{-4.1}$ & $7.3^{+3.2}_{-1.4}$ \\[4pt]
    &  12 34 20.34  & $+$06 27 37.5  & 19407 & 2018-05-15 &
    [0.016]  & $0.43^{+0.43}_{-0.33}$  & $2.97^{+0.93}_{-0.79}$ &
    $7.0^{+10.1}_{-3.8}$ & $37.7/55$ & 
    $8.7^{+3.5}_{-2.0}$  &  $7.3^{+20.6}_{-4.0}$ \\[4pt]
NGC\,4536  &  12 34 26.16  & $+$02 11 22.3  &  19387 & 2017-07-01 &
    [0.016]  & $0.52^{+0.49}_{-0.39}$ &   $1.57^{+0.62}_{-0.57}$  & 
    $2.4^{+2.8}_{-1.2}$ & $86.9/89$ &
    $11.4^{+3.0}_{-2.3}$  &  $5.1^{+2.2}_{-1.0}$  \\ [4pt]
    &  12 34 36.93  & $+$02 08 51.4  & 19387 & 2017-07-01 &
    [0.016]  &  $0.29^{+0.28}_{-0.23}$  &  $3.00^{+0.64}_{-0.58}$  &
    $7.1^{+5.9}_{-3.0}$  &  $64.9/92$  &
    $10.2^{+2.3}_{-8.4}$  &  $9.8^{+14.2}_{-4.6}$  \\[4pt]
%
%
NGC\,4568 &  12 36 33.04 & $+$11 14 05.6  & 19389 & 2018-03-01 &
    [0.024]  & $0.05^{+0.37}_{-0.05}$  & $0.82^{+0.54}_{-0.34}$ &
    $0.88^{+0.87}_{-0.30}$  & $68.6/75$ &
    $13.0^{+3.2}_{-2.9}$  & $6.7^{+1.9}_{-1.5}$
    \\[4pt]
    & & & 19396 & 2018-05-10 &
    [0.024] & $0.47^{+0.72}_{-0.47}$  & $1.36^{+0.74}_{-0.63}$   & 
    $2.5^{+4.2}_{-1.5}$  & $52.0/66$ &
    $15.5^{+4.6}_{-3.6}$  & $9.0^{+4.6}_{-2.0}$
    \\[4pt]
    & 12 36 33.22 &  $+$11 14 37.1 & 19389 & 2018-03-01 &
    [0.024]  &  $<$0.16  &   $1.20^{+0.21}_{-0.21}$ &
    $1.09^{+0.36}_{-0.29}$  &      $71.4/78$  &
    $10.2^{+2.4}_{-2.0}$   & $4.9^{+1.3}_{-1.0}$ \\[4pt]   
NGC\,4579 &  12 37 40.31  & $+$11 47 27.8  &  807 & 2000-05-02 &
    [0.031] & $0.11^{+0.03}_{-0.03}$  & $1.72^{+0.12}_{-0.12}$   & 
    $7.4^{+0.9}_{-0.8}$  & $275.4/300$ &
    $35.7^{+2.5}_{-2.4}$  & $17.1^{+1.0}_{-0.9}$
    \\[4pt]
    &  12 37 53.88  & $+$11 50 20.4  &  807 & 2000-05-02 &
    [0.031] & $0.10^{+0.11}_{-0.09}$  & $2.01^{+0.51}_{-0.43}$   & 
    $2.1^{+1.1}_{-0.7}$  & $92.5/92$ &
    $7.8^{+2.0}_{-1.6}$  & $4.0^{+1.2}_{-0.7}$
    \\[4pt]
NGC\,4607  &  12 41 12.61  &  $+$11 53 50.2  &  19403 & 2018-05-09 &
    [0.025] & $0.82^{+0.88}_{-0.62}$  & $2.37^{+1.12}_{-0.95}$ &
    $4.6^{+12.7}_{-3.1}$ &  $36.9/40$ &
    $7.6^{+3.0}_{-2.1}$  & $10.5^{+40.1}_{-5.0}$ \\ [4pt]
%
\hline\\[-6pt]
\multicolumn{12}{c}{Likely ULX but very uncertain distance} \\[4pt]
NGC\,4480 &  12 30 28.14  & $+$04 14 40.9  &  19423 & 2016-12-06 &
    [0.021] & $0.17^{+0.58}_{-0.17}$  & $1.8^{+1.2}_{-0.8}$   & 
    $0.96^{+2.32}_{-0.55}$  & $26.3/19$ &
    $4.1^{+2.3}_{-1.5}$  & $13.0^{+22.4}_{-4.7}$
    \\[4pt]
\hline\\[-6pt]
\multicolumn{12}{c}{Nuclei of apparently small galaxies: either dwarf satellites in Virgo, or unrelated background galaxies at unknown redshift} \\[4pt]
NGC\,4492 &  12 30 57.83 & $+$08 04 35.2 & 15759 & 2014-04-25 &
    [0.014] & $0.96^{+0.54}_{-0.45}$  & $1.87^{+0.57}_{-0.53}$   & 
    $3.1^{+3.4}_{-1.5}$  & $89.6/108$ &
    $8.7^{+1.8}_{-1.5}$  & $\left(8.3^{+5.8}_{-2.1}\right)^e$
    \\[4pt]
NGC\,4567  &  12 36 33.75 & $+$11 16 10.6 & 19396 & 2018-05-10 &
    [0.024] &    $<0.67$  &   $1.22^{+0.94}_{-0.51}$    &
     $2.5^{+4.2}_{-1.4}$     &  $29.6/28$   &
      $0.70^{+1.57}_{-0.29}$     &  $\left(4.9^{+2.5}_{-1.5}\right)^f$
\\[4pt]
\hline\\[-6pt]
\multicolumn{12}{c}{Either a ULX in Virgo, or an off-nuclear hyperluminous X-ray source in a background galaxy at $z \approx 0.155$} \\[4pt]
NGC\,4388   & 12 25 38.98 &  $+$12 40 00.8  & 1619 & 2001-06-08 &
    [0.026]  & $<$0.08   &   $1.62^{+0.28}_{-0.27}$  & 
    $0.57^{+0.11}_{-0.09}$   &  $51.8/71$  &
    $3.4^{+1.0}_{-0.8}$  & $\left(1.67^{+0.51}_{-0.38}\right)^g$  \\[4pt]
     & &  &  &  &
     &   &   & 
      &   &
     & $\left(2652^{+743}_{-565}\right)^h$  \\[4pt]  %
\hline\\[-6pt]
\multicolumn{12}{c}{Background BL Lac at $z \approx 0.43$} \\[4pt]
NGC\,4698  & 12 48 25.87 &  $+$08 30 20.8  &  3008 & 2002-06-16 &
     [0.016] &   $0.07^{+0.04}_{-0.04}$  &  $2.05^{+0.21}_{-0.19}$  &
     $3.8^{+0.7}_{-0.5}$   &  $164.5/210$  &
      $14.2^{+1.5}_{-1.3}$  &  $\left(1.4^{+0.2}_{-0.1} \times 10^5\right)$   \\[3pt]
\hline\\[-17pt] 
\end{tabular} 
\label{tab3}
\end{center}
\begin{flushleft} 
{\footnotesize{
$^a$: units of $10^{-5}$ photons keV$^{-1}$ cm$^{-2}$ s$^{-1}$ at 1 keV.\\
$^b$: observed fluxes in the 0.5--8 keV band.\\
$^c$: isotropic de-absorbed luminosities in the 0.3--10 keV band, defined as $4\pi d^2$ times the de-absorbed 0.3--10 keV fluxes.\\
$^d$: thermal-plasma spectrum, fitted with 2-temperature {\it apec} model; 
$kT_1 = 0.64_{-0.32}^{+0.47}$ keV; $kT_2 = 2.0$ keV (frozen).\\
$^e$: if located at the distance of NGC\,4492.\\
$^f$: if located at the distance of NGC\,4567.\\
$^g$: if located inside NGC\,4388.\\
$^h$: if projected onto a spiral arm of a background galaxy ($z \approx 0.155$: 
\citealt{alam15}) seen through the $D_{25}$ of NGC\,4388.\\
}}
\end{flushleft}
}
\end{table*}

\begin{table*}
\caption{Historical SNe observed in the sample}
\vspace{-0.2cm}
\scriptsize{
\begin{center}
\begin{tabular}{lccccccc}  
\hline \hline\\[-5pt]    
Host Galaxy  &  SN & R.A.(J2000) & Dec.(J2000) & Type & Peak Date &
$F_{0.5-8}^a$ & $L_{0.5-8}^b$  \\[3pt]
   &   & & & &  &
   ($10^{-16}$ erg cm$^{-2}$ s$^{-1}$) & ($10^{37}$ erg s$^{-1}$)  \\
\hline  \\[-5pt]
%
NGC\,4178  &  1963I  &  12 12 45.61  & $+$10 51 31.0  &  
     Ia	&  1963-05-05   &  $<7$   &   $<1.5$\\[4pt]
NGC\,4254  &  1967H  & 12 18 55.12 &	$+$14 24 40.3  &  
    II  &	1967-07-01  &	$<8$  &		$<2.0$\\[4pt]
     & 1972Q   &   12 18 50.64 & $+$14 26 36.3 &
     II P &	1972-12-17	&  $<8$	  &	$<2.0$\\[4pt]
     &  1986I   &   12 18 52.04 & $+$14 24 44.1  &
     II P &	1986-05-17	&  $<8$	 &	$<2.0$\\[4pt]
     & 2014L   &   12 18 48.68 & $+$14 24 43.5  &	
     Ic	 &  2014-01-26	&  $\left(20^{+20}_{-15}\right)^c$ &  $\left(5^{+5}_{-3}\right)^c$ \\[4pt]
%
NGC\,4302  & 1986E   &   12 21 41.21 & $+$14 37 55.0	&
    II L  &	1986-04-13	&  $38^{+26}_{-17}$ &  $16^{+12}_{-7}$\\[4pt]
NGC\,4303	&  1926A   &   12 21 54.15 & $+$04 29 34.3	&
    II L	& 1926-05-09  &	$<10$	& $<2$\\[4pt]
    &   1961I   &   12 22 00.45 & $+$04 28 13.5	 &
    II	&  1961-06-03	&  $<10$	& 	$<2$\\[4pt]
    &   1964F   &   12 21 53.02 & $+$04 28 24.3	 &
    II	& 1964-06-13	& $<10$  & $<2$\\[4pt]
    &   1999gn   &   12 21 57.02 & $+$04 27 45.6	&
    II P	& 1999-12-17   & $<10$  & $<2$\\[4pt]
    &   2006ov   &   12 21 55.30 & $+$04 29 16.7	& 
    II P	&  2006-11-24   & $<15$  & $<3$\\[4pt]
    &   2008in   &   12 22 01.77 & $+$04 28 47.5	&
    II P	& 2008-12-26   & $<15$  & $<3$\\[4pt]
    &   2014dt   &   12 21 57.57 & $+$04 28 18.5	&
    Ia pec	& 2014-10-29   & $<15$  & $<3$\\[4pt]
NGC\,4316 & 2003bk   &   12 22 42.14 & $+$09 20 01.2	&
    II	& 2003-02-28  &	$<37$  &	 $<33$\\[4pt]
NGC\,4321  &     1901B   &   12 22 47.60 & $+$15 49 25.0	&
    I	&  1901-02-10	& $<10$		&   $<1.4$\\[4pt]
    &  1914A   &   12 22 57.00 & $+$15 47 30.0	 &
    ?   & 1914-03-02	&  $<7$		& $<1$\\[4pt]
    &   1959E   &   12 22 58.91 & $+$15 49 01.3	   &
    I	&  1959-02-21	&  $<7$		& $<1$\\[4pt]
    &    1979C   &   12 22 58.63 & $+$15 47 51.7	&
    II L	& 1979-04-15	&  $160^{+20}_{-20}$  & $63^{+13}_{-11}$\\[4pt]
    &   2006X   &   12 22 53.99 & $+$15 48 33.1	  &
    Ia	&  2006-02-19  &	$<7$	&	 $<1$\\[4pt]
    &   2020oi   &   12 22 54.96 & $+$15 49 25.0	&
    Ic	& 2020-01-07  &	$<100^d$	&	 $<15^d$\\[4pt]
NGC\,4402	& 1976B   &   12 26 11.91 & $+$13 06 44.8	&
    I	&  1976-04-05	&  $<17$  &	 $<5$\\[4pt]
NGC\,4411b	& 1992ad   &   12 26 49.59 & $+$08 52 38.2	&
    II	& 1992-07-01 &	$<30$	&	 $<7$\\[4pt]
NGC\,4419	& 1984A   &   12 26 55.71 & $+$15 03 17.7	&
    Ia	& 1984-01-17  &	$<30$   &	$<10$\\[4pt]
    &   2012cc   &   12 26 56.81 & $+$15 02 45.5	&
    II	&  2012-04-29	&  $50^{+50}_{-30}$	&  $17^{+17}_{-10}$\\[4pt] 
NGC\,4424  &  1895A$^e$   &   12 27 16.90 & $+$09 25 05.0	&
    ?   &   1895-03-16	&  $<25$  &	 $<8$\\[4pt]
    &   2012cg   &   12 27 12.83 & $+$09 25 13.2	&
    Ia	&  2012-05-17	&  $<25$	& $<8$\\[4pt]
NGC\,4451   &  1985G   &   12 28 40.40 & $+$09 15 40.0	&
    II P  &	1985-03-17	&  $<20$   &	$<17$\\[4pt]
NGC\,4496a	& 1960F   &   12 31 42.06 & $+$03 56 47.9	&
    Ia	& 1960-04-20	& $<80$  &	$<25$\\[4pt]
NGC\,4501	& 1999cl   &   12 31 56.01 & $+$14 25 35.3	&
    Ia	& 1999-06-15   &   $<15$  &	 $<6$\\[4pt]
NGC\,4527	& 1915A   &   12 34 10.95 & $+$02 39 03.5	&
    ?   & 1915-03-20  &	 $<20$  &	$<5$\\[4pt]
    &   1991T   &   12 34 10.17 & $+$02 39 56.4	  &
    Ia pec	&  1991-04-28   &	 $<20$  &	$<5$\\[4pt]
    &   2004gn   &   12 34 12.10 & $+$02 39 34.4	&
    Ic	&  2004-12-01  &	 $<30$  &	$<8$\\[4pt]
NGC\,4536	& 1981B   &   12 34 29.58 & $+$02 11 59.3	&
    Ia	&  1981-03-09  &	 $<30$  &	$<8$\\[4pt]
NGC\,4568	& 1990B   &   12 36 33.83 & $+$11 14 29.8	&
    Ic	& 1990-01-18  &	 $<30$  &	$<10$\\[4pt]
    &   2004cc   &   12 36 34.40 & $+$11 14 32.8	&
    Ic	&  2004-06-10  &	 $<30$  &	$<10$\\[4pt]
NGC\,4579	& 1988A   &   12 37 43.54 & $+$11 48 19.4	&
    II P  &	1988-01-19  &	 $<5$  &	$<2$\\[4pt]
    &   1989M   &   12 37 40.72 & $+$11 49 26.0	  &
    Ia  &	1989-06-28 &	 $<5$  &	$<2$\\[4pt]
NGC\,4639	& 1990N   &   12 42 56.70 & $+$13 15 23.7	&
    Ia	& 1990-07-10 &	 $<20$  &	$<10$\\[4pt]
NGC\,4647   &   1979A   &   12 43 29.14 & $+$11 35 27.1	 &
    I &	1979-01-15 &	 $<3$  &	$<1$\\[4pt]
\hline\\[-5pt] 
\end{tabular} 
\label{tab4}
\end{center}
\begin{flushleft} 
{\footnotesize{
$^a$: observed fluxes in the 0.5--8 keV band. To estimate the non-detection upper limits, we assumed a thermal-plasma {\it apec} model, with fixed temperature $kT = 1$ keV and line-of-sight Galactic $N_{\rm H}$. For SN 2014L, SN 1986E, SN 2012cc, {\it apec} temperatures and column densities were left as free parameters in the fit, with Cash statistics. For SN 1979C, a two-temperature {\it apec} model was used (Section 4.2.2).\\
$^b$: de-absorbed luminosities in the 0.5--8 keV band, based on the models mentioned above.\\
$^c$: likely due to extended thermal emission\\
$^d$: located in a region of strong diffuse emission\\
$^e$: first historical SN discovered outside the Local Group\\
}}
\end{flushleft}
}
\end{table*}

\subsection{Properties of the ULX luminosity function}
LFs are ideally defined and quantified as instantaneous snapshots of an X-ray population. Collecting and including all sources (including transient ones) detected in several successive observations of the same galaxy leads to an over-estimate of the luminosity distribution. For this reason, when plotting the 0.3--10 keV cumulative LF (Figure 13), we included only sources from one observation per galaxy. We compared the distribution obtained by selecting the earliest observation for each galaxy, the latest observation, and the longest observation. The three alternative histograms are consistent with each other at the low luminosity end. At the high end, the small excess of the earliest-look LF is the result of small-number statistics: two ULXs with $L_{\rm X} > 10^{40}$ erg s$^{-1}$ in NGC\,4527, detected only in the first observation of that galaxy. 

We used the $\log N$-$\log S$ curves in the {\it Chandra} Deep Field study of \cite{lehmer12} to estimate the number of contaminating sources (background galaxies, AGN, quasars, and foreground stars) in our survey. To do this, we calculated the expected number of such sources projected inside the $D_{25}$ of each of our sample galaxies, above a flux corresponding to a 0.3--10 keV luminosity of 10$^{39}$ erg s$^{-1}$ for that galaxy (at its median NED distance, given in Table 1, Column 8). For most of the galaxies, the expected number is $<$1. Adding those numbers for all the galaxies, we estimate a total of $\approx$15 contaminating sources above $10^{39}$ erg s$^{-1}$ (Figure 13). In fact, the total number of 15 is likely to be an over-estimate, because it assumes that the count rates and fluxes of background AGN suffer no attenuation from photoelectric absorption passing through the halo or disk of the Virgo galaxies. With a similar method, we estimate the expected contamination at other luminosity levels; for example, only about 2 contaminating sources are expected above $\approx$5 $\times 10^{39}$ erg s$^{-1}$. (A few candidate background/foreground sources have already been flagged in Table 2 and Table 3).


We divided the energy range between $1.4 \times 10^{39}$ erg s$^{-1}$ and $2.2 \times 10^{40}$ erg s$^{-1}$ (0.3--10 keV band) into four equally spaced bins (in log scale), and computed the contribution of the three alternative LFs (earliest, latest and longest) in each bin. We then subtracted the expected background and foreground contributions in each bin from each of three observed distributions. This gives three alternative realizations of the differential LF of the point-like sources, corresponding to three different snapshots in time (Figure 14). As expected, there is a lot of scatter at the high-luminosity end, due to small-number statistics. A power-law with an index $\gamma = -0.80 \pm 0.20$ (defined as $dN/d(\log L) \propto L^{\gamma}$) is a good approximation of the slope of the differential LF.
This is intermediate (as expected, given the mix of the two populations) between the slopes of $\approx$ $-0.6$ for a pure HMXB population, and $\approx$ $-1.3$ for LMXBs in that energy range \citep{lehmer19}. The normalization is also consistent (Figure 14) with the spiral galaxy models of \cite{lehmer19}, for a total SFR $\approx$70 $M_{\odot}$ yr$^{-1}$ (B15 value) and a total stellar mass $M_{\ast,{\rm{tot}}} = 1.5 \times 10^{12} M_{\odot}$. This does not imply that the SFR of $\approx$100 $M_{\odot}$ yr$^{-1}$ inferred from the {\it WISE} photometry is wrong. It simply means that the SFR proxies (far-UV {\it GALEX} plus 24$\mu$m {\it Spitzer} luminosity) used by \cite{lehmer19} to calibrate the normalization of the HMXB population provide SFR values more similar to the B15 values than to the {\it WISE}-derived values.

Our empirical differential LF falls slightly below the model predictions at the low-luminosity end (Figure 14). This suggests that we are missing a few sources. This is expected. Sources with a 0.3--10 keV luminosity of $\sim$1.5 $\times 10^{39}$ erg s$^{-1}$ and an absorbing column $N_{\rm H} \gtrsim$ a few times $10^{21}$ cm$^{-2}$ have a substantial chance of missing our first selection cut (Table 2, Col.~6), depending on their photon index and on the {\it Chandra} observing cycle. For this work, we are not particularly concerned by this issue: ULXs in the luminosity range $\sim$(1--1.5) $\times 10^{39}$ erg s$^{-1}$ are the upper end of the ordinary X-ray binary population; we are more interested in the high-luminosity end of the ULX population. 

The total 0.3--10 keV luminosity of the resolved $D_{25}$ population above $1.4 \times 10^{39}$ erg s$^{-1}$ is $\sim$(3.5--3.7) $\times 10^{41}$ erg s$^{-1}$ (depending on the choice of observation for galaxies with multiple exposures). About 10\% of this luminosity comes from background AGN projected inside the $D_{25}$ regions. An extrapolation of the differential LF (Figure 14) with a power-law slope of $\gamma = -0.8 \pm 0.2$ down to $10^{36}$ erg s$^{-1}$ predicts an additional contribution of $\approx$2--3 $\times 10^{41}$ erg s$^{-1}$ from unresolved point sources. This value is not very sensitive to the choice of the lower limit, as most of the integrated emission comes from the high-luminosity sources. Thus, our best estimate for the background-subtracted point-source 0.3--10 keV luminosity in the Virgo sample is $\sim$5--6 $\times 10^{41}$ erg s$^{-1}$, corresponding to $\sim$4.5--5 $\times 10^{41}$ erg s$^{-1}$ in the 0.5--8 keV band.

First, we compare this extrapolated value with the integrated background-subtracted point-source galaxy luminosities estimated in Section 4.1. There, we had obtained a luminosity of $\approx$3.6 $\times 10^{41}$ erg s$^{-1}$ in the 0.5--8 keV band. However, for many of the galaxies with moderate or low emission (or short exposures) we had simply assumed a Galactic line-of-sight absorption in the conversion from count rates to fluxes and luminosities. Consequently, that value must be considered a lower limit. We have already seen (Section 4.2.1) that increasing the column density from $3 \times 10^{20}$ cm$^{-2}$ to $3 \times 10^{21}$ cm$^{-2}$ increases the inferred luminosity by 20 to 50 per cent (for a given count rate). Therefore, we conclude that the total luminosity estimates from the integration of the $D_{25}$ source regions and from the (extrapolated) point-source LF are consistent with each other. We had already noted (Section 4.1) that the value of $\approx$3.6 $\times 10^{41}$ erg s$^{-1}$ appeared under-estimated for the given mass and SFR of the sample, and suggested that this under-estimation particularly affected galaxies with dusty disks, seen at high inclination. 

Second, we compare the measured point-source luminosity with the model predictions \citep{lehmer19}, in the energy range where we have better constraints ($L_{\rm X} \gtrsim 1.2 \times 10^{39}$ erg s$^{-1}$ in the 0.5--8 keV band). Direct sum of the ULX luminosities (minus the predicted AGN and foreground star fraction) gives a net 0.5--8 keV luminosity of $\approx$2.7 $\times 10^{41}$ erg s$^{-1}$, but largely dependent on the small-number statistics of the most luminous sources. Over the same energy range, for an SFR of 70 $M_{\odot}$ yr$^{-1}$ and a stellar mass of $1.5 \times 10^{12} M_{\odot}$, model predictions are $\approx$3.2 $\times 10^{41}$ erg s$^{-1}$, of which $\approx$2.5 $\times 10^{41}$ erg s$^{-1}$ from HMXBs and $\approx$0.7 $\times 10^{41}$ erg s$^{-1}$ from LMXBs. (The number of HMXBs per unit luminosity interval starts to dominate over the number of LMXBs at $L_{\rm X} \approx 1.2 \times 10^{39}$ erg s$^{-1}$). As we noted earlier, the observed ULX luminosities are more consistent with the lower end of the published SFR values (B15 values) rather than the {\it WISE}-derived values. 

We analyzed the 0.5--8 keV cumulative luminosity distributions of observed sources and model predictions (Figure 15) to determine whether the small excess luminosity in the model of \cite{lehmer19} is significant. For sources below $\approx$4 $\times 10^{39}$ erg s$^{-1}$, the fiducial model follows our empirical distribution extremely well. Above that energy bin, the model is significantly above the observed LF. 
We suggest several alternative reasons for the discrepancy. 
The first possibility is that there is an imperfect match between the sector of the LF for which most of the sources have an individual spectral fit, and the sector where the luminosity was estimated with a fixed $N_{\rm H}$ and $\Gamma$. Splicing the two sub-samples together may create a normalization jump between them. The second possibility is that our simple (power-law) spectral models are still underestimating the luminosities of a few of the most luminous sources (more than for the fainter ones). For example, the most luminous ULXs might have a stronger thermal component below 1 keV, in addition to the power-law component, which was not included in the spectral models. More likely, our luminosity function may include a few ULXs with an intrinsic luminosity $L_{\rm X} \gtrsim 4 \times 10^{39}$ erg s$^{-1}$ but an absorption column $>$10$^{22}$ cm$^{-2}$ ({\it e.g.}, in dense star-forming regions or inside an edge-on galactic disk), which have been incorrectly placed in lower-luminosity bins. Finally, and perhaps more interestingly, the assumed shape of the model functions may not be correct. In \cite{lehmer19}, both the LMXB and HMXB differential LFs in the ULX range were assumed to be unbroken power-laws (with different slopes) up to a cut-off at $5 \times 10^{40}$ erg s$^{-1}$. While this may be justified for HMXBs, there is no sufficient evidence to say that it is also a good model for LMXBs. In fact, in the ULX sample studied by \cite{lehmer19}, there are no LMXBs above a luminosity of $\approx$2 $\times 10^{39}$ erg s$^{-1}$. If the LMXB distribution has a cut-off at $L_{\rm X} \approx$ a few $\times 10^{39}$ erg s$^{-1}$, it would help explain why our observed ULX distribution is consistent with the LMXB plus HMXB model function at the low luminosity end but follows the HMXB-only function at the high luminosity end (Figure 15).

\subsection{ULXs in early-type and late-type spiral galaxies}
Let us now compare the total number of ULXs found in our study with the numbers predicted by the models of \cite{kovlakas20} for spiral galaxies (valid from S0/a to Sm). The energy band used to define the ULX luminosity in their study is 0.5--8 keV. In the same band, we observed $\approx$60 ULXs, after removing the expected background and foreground contributions. (We used the first observation of each galaxy for this estimate, but the number is approximately the same if we choose the longest observation or the latest observation.) Considering a mild effect of incompleteness around 1.0--1.3 $\times 10^{39}$ erg s$^{-1}$, hinted at in Section 4.3, the corrected number of ULXs at any given time is probably $\approx$65--70, but we keep the estimate of 60 as a safe lower limit. The model of \cite{kovlakas20} predicts $N_{\rm ULX} \approx 0.45^{+0.06}_{-0.09} \times$ SFR $+  3.3^{+3.8}_{-3.2} \times M_{\ast,12}$, where $M_{\ast,12}$ is in units of $10^{12} M_{\odot}$. For a meaningful comparison with our numbers, we need to consider the definition of SFR used by \cite{kovlakas20} for their calibration of the ULX scaling relation. In their case, the SFR was derived from the total infrared luminosity, primarily from the {\it IRAS} photometric measurements (calibration of \citealt{dale02,kennicutt12}) supplemented for some galaxies by the 12-$\mu$m and 22-$\mu$m {\it WISE} measurements. Therefore, for consistency, we use the {\it WISE} measurements of SFR for our galaxy sample rather than the B15 values (the latter being more biased in favour of UV- and H$\alpha$-based SFR measurements). Namely, we take a total SFR $\approx 86$--104 $M_{\odot}$ yr$^{-1}$, where the lower and higher values corresponds to the 22-$\mu$m and 12-$\mu$m estimates, respectively (Table 1).
For this SFR range, the scaling relation of \cite{kovlakas20} predicts a best-fitting $N_{\rm ULX} \approx 44$--52. If we consider also the error range in the coefficients of the scaling relation, the predicted number of ULXs in our sample is $\approx$36--58, lower than observed. \cite{kovlakas20} did point out that their ULX luminosities and hence ULX numbers are lower than those found in other surveys, including that of \cite{swartz11} (which is the most similar to the present work, for methodology). The difference was attributed to a different morphological composition of the samples, and a different conversion procedure between {\it Chandra}/ACIS count rates, fluxes and luminosities (Appendix C1 in \citealt{kovlakas20}).

More importantly, a comparison with the scaling of \cite{kovlakas20} is interesting because of their claimed strong difference between ULX rates in early-type spirals (S0/a to Sbc) and late-type spirals (Sc or later). For the former, $N_{\rm ULX} \approx 0.16^{+0.08}_{-0.08} \times$ SFR $+  11.2^{+5.2}_{-5.6} \times M_{\ast,12}$; for the latter, $N_{\rm ULX} \approx 0.98^{+0.11}_{-0.20} \times$ SFR. To test those relations, we split the Virgo spiral sample into two corresponding early-type and late-type subsamples, based on the NED classification (see also Table 1 and Figure 4). The early-type-spiral sample includes 45 galaxies, with a total stellar mass of $\approx$1.2 $\times 10^{12} M_{\odot}$ and a {\it WISE}-derived SFR of $\approx$55--66 $M_{\odot}$ yr$^{-1}$; the late-type-spiral sample has 30 galaxies, with a stellar mass of $\approx$0.3 $\times 10^{12} M_{\odot}$ and SFR  $\approx$30--38 $M_{\odot}$ yr$^{-1}$. For the parameters of the early-type-spiral sample, \cite{kovlakas20} predict a central value of $\approx$23--24 ULXs ($\approx$16--31 when we include the uncertainty on the scaling coefficients); our observed number is 39. For the late-type-spiral sample, the predicted central value is $\approx$29--37 ULXs ($\approx$24--42 with full uncertainty range); our observed number is 21. We conclude that the scaling models of \cite{kovlakas20} underpredict the number of ULXs in early-type spirals and over-predict those in late-type ones, at least when applied to the Virgo sample.

\section{Discussion and summary}


We are conducting the first large-scale {\it Chandra}/ACIS survey of the spiral galaxy population in the Virgo cluster. This project complements and integrates the AMUSE-Virgo survey, which studied the early-type galaxy population of Virgo, a decade ago. We selected a sample of 75 galaxies, including all the largest and most active spirals ($\approx$20 late-type galaxies with SFR $\gtrsim$1 $M_{\odot}$), plus a selection of fainter ones. 
As detailed in Section~\ref{sub_select}, our sample is skewed towards brighter objects, in
the sense that about 60 of our 75 galaxies have SFR $\gtrsim$0.3 M$_\odot$
yr$^{-1}$. This was done to provide a contrast with the early-type galaxy
sample \citep{gallo08,plotkin14,gra-sor19} and to aid the discovery of luminous XRBs.

The total observing time was about 1.95 Ms (combining new and archival observations), and for all galaxies in our sample we reached a point-source detection limit of $\approx$3 $\times 10^{38}$ erg s$^{-1}$ (or deeper, in a few cases). 

In this first paper, we have outlined the general properties of the sample: the distance, morphological type, stellar mass and SFR. We have then presented the X-ray population properties of the sample galaxies, in relation to their stellar mass (for which we estimate a total of $\approx$1.5 $\times 10^{12} M_{\odot}$) and SFR. For the SFR, we used the {\it WISE} 12$\mu$m and 22$\mu$m proxies and compared them with a collection of other SFR proxies (in particular, based on H$\alpha$ and the far-UV) in the literature. Estimates vary between $\approx$70 $M_{\odot}$ yr$^{-1}$ and $\approx$100 $M_{\odot}$ yr$^{-1}$. Two-thirds of the galaxies can be described as active star-forming disks (dominated in X-rays by the HMXB population), while the others are early-type spirals, with low sSFR and an X-ray luminosity dominated by LMXBs. 

We identified a sample of 85 ULXs that exceeded a 0.5--8 keV luminosity of $10^{39}$ erg s$^{-1}$ in at least one observation, and determined their coordinates. We fitted the spectral parameters for the most luminous sources (25 ULXs with $L_{\rm X} \gtrsim 4 \times 10^{39}$ erg s$^{-1}$) and used their median power-law index and column density to convert count rates to luminosities for the rest of the ULXs, when they were too faint for individual fitting. We showed that the ULX luminosity distribution from our Virgo study is broadly consistent with the LF models from \cite{lehmer19}, for a SFR $\sim 70 M_{\odot}$ yr$^{-1}$ and a total stellar mass of $\approx$1.5 $\times 10^{12} M_{\odot}$. 

The model functions of \cite{lehmer19} predict a significant contribution (in number) from LMXBs: about half of the sources in the $\approx$(1--1.5) $\times 10^{39}$ erg s$^{-1}$ range, and about one third of our total number of ULXs. 
At first sight, this appears to contradict the estimate from \cite{plotkin14} (based on their AMUSE-Virgo study), of only about 1 old-population ULX per $M_{\ast} \approx 1.6 \times 10^{11} M_{\odot}$, all of them with $L_{\rm X} \lesssim 2 \times 10^{39}$ erg s$^{-1}$. (In our case, this would correspond to only about 10 old-population ULXs rather than about 30). In fact, there is no contradiction. The term ``LMXB'' is generally applied to a variety of X-ray binary populations that are proportional to stellar masses, but the normalization, break and slope of their LFs are substantially different, depending on age and environment \citep{zhang12}. In the disk of spiral galaxies, especially for sSFR $\lesssim$10$^{-10}$ yr$^{-1}$, the number and luminosity of bright LMXBs (above a few times 10$^{37}$ erg s$^{-1}$) per unit mass is a few times higher than in old ellipticals. This is evident for example in the properties of the luminosity function in the disk of the nearby disk-dominated spiral M\,83, where the LMXB component is at least as important as the HMXB component \citep{long14}. Most of the stellar mass in the AMUSE-Virgo survey is in massive elliptical galaxies (Table 1 in \citealt{gallo10}). Therefore, the scaling relations of \citep{zhang12} are suitable to model the LMXB population mapped by that survey. Instead, the LMXB population in our sample includes a mix of spheroidal (spiral bulge) population for the early-type spirals, and pure disk population. Thus, we expect it to follow the scaling relation derived by \cite{lehmer19}, at least up to a few times $10^{39}$ erg s$^{-1}$.

The contribution of LMXBs to the higher-luminosity ULXs remains an open question. From the models of \cite{lehmer19} we expect four old-population sources above $L_{\rm X} \approx 4 \times 10^{39}$ erg s$^{-1}$ (0.5--8 keV band), compared with about 16 HMXBs. Instead, our Virgo ULX LF is consistent with the HMXB component alone. A cut-off of the LMXB population at $L_{\rm X} \approx 4 \times 10^{39}$ erg s$^{-1}$ is one of the possible explanations we have proposed for the slight mismatch. Some theoretical population synthesis models \citep{wiktorowicz17} do predict the formation of neutron star ULXs at least as luminous as $10^{40}$ erg s$^{-1}$ during phases of steadily super-critical mass transfer from low-mass donors (main sequence, Hertzsprung-gap stars, red giants, asymptotic giants, or white dwarfs), long after the end of star formation. The predicted number of old-population ULXs does, however, drop rapidly with age, decreasing by a factor of 20 from 1 to 5 Gyr after a star formation burst, and by another order of magnitude from 5 to 10 Gyr \citep{wiktorowicz17}. In addition to those steady mass-transfer systems, old-population ULXs should include also transient systems with a large (outer radius $\gtrsim$10$^{12}$ cm) accretion disk, subject to the thermal-viscous disk instability, which may exceed the Eddington limit during outburst \citep{hameury20}.

In further work currently in preparation, we will illustrate and discuss the optical counterparts and stellar environments of the ULXs in this Virgo galaxy sample, based on the Next Generation Virgo Cluster Survey maps \citep{ferrarese12} from the Canada-France-Hawaii Telescope. We will distinguish between younger and older populations, for the lower-luminosity and higher-luminosity subsets of ULXs, and for the early-type-spiral and late-type-spiral subsamples of galaxies. We will also discuss the metallicity effect on the high-luminosity end of the younger population.

Individual spectral studies of the most interesting {\it Chandra} sources will also be presented in follow-up work. Here, we have simply mentioned a few highlights. The most luminous ULX in the sample (a source in IC\,3322A with $L_{\rm X} \approx 6 \times 10^{40}$ erg s$^{-1}$) may not be an accreting compact object, as its X-ray spectrum is more consistent with that of a young SN (which could have been missed by optical/IR searches). Apart from that mysterious source, the three most luminous ULXs (in NGC\,4254, NGC\,4496A and NGC\,4579) all reached a luminosity $L_{\rm X} \approx 2 \times 10^{40}$ erg s$^{-1}$, which is the well-known characteristic threshold above which ULXs become much rarer \citep{grimm03,swartz04,swartz11,mineo12a}. At least three {\it Chandra} sources seen inside the $D_{25}$ of sample galaxies are likely to be the nuclear sources of smaller galaxies: further investigation is needed to determine whether those galaxies are also in Virgo (possibly satellites of their larger companions) or are a more distant background.
 



\section*{Acknowledgements}
RS acknowledges grant number 12073029 from the National Science Foundation of China. He is also grateful for support and hospitality from the Curtin Institute of Radio Astronomy (Perth, Australia) and from the Observatoire de Strasbourg during part of this work. MK acknowledges support from the French Centre National d'\'Etudes Spatiales (CNES). We are extremely grateful to the WXSC team (Team directors: Thomas Jarrett and Russ Taylor) for their support and their sharing of results in advance of their catalog publication. We thank Fabien Gris\'e, Manfred W. Pakull, Pierre-Alain Duc, Kinwah Wu, for their comments and suggestions, which improved the quality and the chances of success of our {\it Chandra} Large Program proposal, and Benjamin L. Davis for his expert advice on galaxy structures and scaling relations. We also thank Luca Cortese, Pat Cot\'e, Jean-Charles Cuillandre and Stephen Gwyn, who helped us access the NGVS dataset. Finally, we thank the anonymous referee for their careful reading of the first version of this paper, and their insightful comments and suggestions.

\section*{Data Availability}
The {\it Chandra} datasets used for this work are all available for download from the public archives. Reprocessed data can be provided upon request.


\bsp	
\label{lastpage}

\end{document}